\documentclass[12pt,preprint]{aastex}

\newcommand{\dm}{$\Delta m_{15}(B)$}

\begin{document}

\shortauthors{Hicken et al.}
\shorttitle{185 Type Ia Supernova Light Curves}

\title{CfA3:  185 Type Ia Supernova Light Curves from the CfA}

\author{Malcolm Hicken\altaffilmark{1,2},
 Peter Challis\altaffilmark{1},
 Saurabh Jha\altaffilmark{3},
 Robert P. Kirshner\altaffilmark{1},
 Tom Matheson\altaffilmark{4},
 Maryam Modjaz\altaffilmark{5},
 Armin Rest\altaffilmark{2},
 W. Michael Wood-Vasey\altaffilmark{6},
 Gaspar Bakos\altaffilmark{1},
 Elizabeth J. Barton\altaffilmark{7},
 Perry Berlind\altaffilmark{1},
 Ann Bragg\altaffilmark{8},
 Cesar Brice\~no\altaffilmark{9},
 Warren R. Brown\altaffilmark{1},
 Nelson Caldwell\altaffilmark{1},
 Mike Calkins\altaffilmark{1},
 Richard Cho\altaffilmark{1},
 Larry Ciupik\altaffilmark{10},
 Maria Contreras\altaffilmark{1},
 Kristi-Concannon Dendy\altaffilmark{11},
 Anil Dosaj\altaffilmark{1},
 Nick Durham\altaffilmark{1},
 Kris Eriksen\altaffilmark{12},
 Gil Esquerdo\altaffilmark{13},
 Mark Everett\altaffilmark{13},
 Emilio Falco\altaffilmark{1},
 Jose Fernandez\altaffilmark{1},
 Alejandro Gaba\altaffilmark{11},
 Peter Garnavich\altaffilmark{14},
 Genevieve Graves\altaffilmark{15},
 Paul Green\altaffilmark{1},
 Ted Groner\altaffilmark{1},
 Carl Hergenrother,\altaffilmark{12},
 Matthew J. Holman\altaffilmark{1},
 Vit Hradecky\altaffilmark{1},
 John Huchra\altaffilmark{1},
 Bob Hutchison\altaffilmark{1},
 Diab Jerius\altaffilmark{1},
 Andres Jordan\altaffilmark{1},
 Roy Kilgard\altaffilmark{1},
 Miriam Krauss\altaffilmark{16},
 Kevin Luhman\altaffilmark{17},
 Lucas Macri\altaffilmark{18},
 Daniel Marrone\altaffilmark{1},
 Jonathan McDowell\altaffilmark{1},
 Daniel McIntosh\altaffilmark{19},
 Brian McNamara\altaffilmark{20},
 Tom Megeath\altaffilmark{1},
 Barbara Mochejska\altaffilmark{21},
 Diego Munoz\altaffilmark{1},
 James Muzerolle\altaffilmark{12},
 Orlando Naranjo\altaffilmark{1},
 Gautham Narayan\altaffilmark{1},
 Michael Pahre\altaffilmark{1},
 Wayne Peters\altaffilmark{1},
 Dawn Peterson\altaffilmark{1},
 Ken Rines \altaffilmark{1},
 Ben Ripman \altaffilmark{1},
 Anna Roussanova\altaffilmark{16},
 Rudolph Schild\altaffilmark{1},
 Aurora Sicilia-Aguilar\altaffilmark {1},
 Jennifer Sokoloski\altaffilmark{1},
 Kyle Smalley\altaffilmark{1},
 Andy Smith\altaffilmark{1},
 Tim Spahr\altaffilmark{1},
 K. Z. Stanek\altaffilmark{22},
 Pauline Barmby\altaffilmark{23},
 St\'ephane Blondin\altaffilmark{24},
 Christopher W. Stubbs\altaffilmark{1,2},
 Andrew Szentgyorgyi\altaffilmark{1},
 Manuel A. P. Torres\altaffilmark{1},
 Amili Vaz\altaffilmark{16},
 Alexey Vikhlinin\altaffilmark{1},
 Zhong Wang\altaffilmark{1},
 Mike Westover\altaffilmark{1},
 Deborah Woods\altaffilmark{1},
 Ping Zhao\altaffilmark{1}
}

\altaffiltext{1}{Harvard-Smithsonian Center for
 Astrophysics,Cambridge, MA 02138; mhicken@cfa.harvard.edu,
kirshner@cfa.harvard.edu}
\altaffiltext{2}{Department of Physics, Harvard University, Cambridge, MA
02138}
\altaffiltext{3}{Department of Physics and Astronomy, Rutgers, the
 State University of New Jersey, Piscataway, NJ 08854}
\altaffiltext{4}{
National Optical Astronomy Observatory,
NOAO Gemini Science Center, 950 North,
Cherry Avenue, Tucson, AZ
85719 }
\altaffiltext{5}{
UC Berkeley, Astronomy Department, Campbell Hall,
Berkeley, CA 94720 }
\altaffiltext{6}{Department of Physics and Astronomy, University of
 Pittsburgh, Pittsburgh, PA 15260}
\altaffiltext{7}{
 University of California, Irvine,
 Department of Physics and Astronomy,
 2158 Frederick Reines Hall Irvine, CA 92697
}
\altaffiltext{8}{
Physics Department, Marietta College, 215 Firth Street, Marietta, OH
45750
}
\altaffiltext{9}{
 Centro de Investigaciones de Astronomia (CIDA),
 Apdo Postal 264 VE - Merida 5101-A, Bolivarian Republic of Venezuela 
}
\altaffiltext{10}{
Adler Planetarium \& Astronomy Museum,
1300 S. Lake Shore Drive,
Chicago, Illinois 60605 }
\altaffiltext{11}{
  University of North Carolina at Chapel Hill,
Department of Physics and Astronomy, Chapel Hill,
 NC 27599-3255
}
\altaffiltext{12}{
 The University of Arizona,
 Tucson AZ 85721
}
\altaffiltext{13}{
Planetary Science
 Institute, 1700 E. Fort Lowell Rd., Tucson, AZ 85719
}
\altaffiltext{14}{
   University of Notre Dame,
 25 Nieuwland Science Hall,
 Notre Dame, IN 46556-5670 USA
}
\altaffiltext{15}{
   University of Caifornia - Santa Cruz,
 Astronomy \& Astrophysics,
 201 Interdisciplinary Sciences Building (ISB),
 Santa Cruz, CA 95064
}
\altaffiltext{16}{
   Massachusetts Institute of Technology,
 77 massachusetts avenue,
 cambridge, ma 02139-4307
}
\altaffiltext{17}{
Department of Astronomy and Astrophysics,
525 Davey Lab,
The Pennsylvania State University,
University Park, PA 16802 }
\altaffiltext{18}{
Department of Physics and Astronomy, Texas A\&M University, College
Station, TX 77843-4242}
\altaffiltext{19}{
   University of Massachusetts, Amherst,
Department of Astronomy, 
Amherst, MA 01003
}
\altaffiltext{20}{
Department of Physics \& Astronomy, 
University of Waterloo, 
200 University Avenue,
West Waterloo, Ontario,
Canada N2L 3G1
}
\altaffiltext{21}{
 Copernicus Astronomical Center in Warsaw, Poland Bartycka 18, 00-716 Warszawa
}
\altaffiltext{22}{
 OSU Department of Astronomy, McPherson Laboratory 140 W 18th Avenue,
Columbus,
 Ohio 43210 }
\altaffiltext{23}{Department of Physics \& Astronomy, University
of Western Ontario, London, ON N6A 3K7, Canada}
\altaffiltext{24}{European Southern Observatory, D-85748 Garching, Germany}



\begin{abstract}

We present multi-band photometry of 185 type-Ia supernovae (SN Ia), with over
11500 observations.  These were acquired between 2001 and 2008 at the F. L.
Whipple Observatory of the Harvard-Smithsonian Center for Astrophysics (CfA).
This sample contains the largest number of homogeneously-observed and reduced
nearby SN Ia ($z\lesssim0.08$) published to date.  It more than doubles the
nearby sample, bringing SN Ia cosmology to the point where systematic
uncertainties dominate.  Our natural system photometry has a precision of
$\lesssim0.02$ mag in \emph{BVRIr'i'} and $\lesssim0.04$ mag in $U$ for points
brighter than 17.5 mag.  We also estimate a systematic uncertainty of 0.03 mag
in our SN Ia standard system \emph{BVRIr'i'} photometry and 0.07 mag for $U$.
Comparisons of our standard system photometry with published SN Ia light curves
and comparison stars, where available for the same SN, reveal agreement at the
level of a few hundredths mag in most cases.  We find that 1991bg-like SN Ia
are sufficiently distinct from other SN Ia in their color and
light-curve-shape/luminosity relation that they should be treated separately in
light-curve/distance fitter training samples.  The CfA3 sample will contribute
to the development of better light-curve/distance fitters, particularly in the
few dozen cases where near-infrared photometry has been obtained and, together,
can help disentangle host-galaxy reddening from intrinsic supernova color,
reducing the systematic uncertainty in SN Ia distances due to dust.  

\end{abstract}


\section{Introduction}

SN Ia are standardizable candles ($\sigma\lesssim0.2$ mag after correction
for light-curve shape) and have been used
to measure the expansion history of the universe  \citep[e.g.,][]{phillips93,
riess96, hamuy96a, goldhaber01, jha07}, giving rise to the startling conclusion that the
universe is accelerating \citep[e.g.,][]{riess98, perlmutter99}.  

Some 30 years prior to the discovery of the accelerating universe,
\citet{kowal68} found a dispersion of $\sim0.6$ mag in the SN Ia
redshift-magnitude relation.  Reasons for the high dispersion are that many of
the objects in his sample were not in the Hubble flow, corrections for
light-curve shape and absorption were not made, and not all objects were SN Ia.
He predicted that SN Ia might become distance estimators with better than
$10\%$ precision and enable measurement of the curvature of the Hubble diagram
at greater redshifts.  Nearly 20 years later, \citet{norgaard89} made a valiant
attempt to measure SN Ia at $z\sim0.3$, using methods similar to modern SN
searches.  They had the right idea but their telescope and detector were too
small and they only reported one SN Ia in two years of searching.  The advent
of significantly larger detectors, mounted on larger telescopes, provided the
higher discovery rates needed and was one of the main factors in the discovery
of the accelerating universe.

SN Ia studies can be divided into two broad groups:  low redshift and high
redshift.  For our purposes, the dividing line between the two groups is at
$z\approx0.15$.  Low redshift SN Ia are easier to study to higher precision and
can generally be observed over a greater range in phase.  They map out the
recent expansion of the universe and can be used to study the local bulk flows
and peculiar velocities of galaxies in the nearby universe
\citep[e.g.][]{tammann90, riess95, jha07, neill07, haugbolle07}.  They also serve as the template
against which the high redshift SN Ia are compared.  Having, and understanding,
a nearby sample that fills out the phase space of SN Ia properties is vital to
the use of SN Ia as precise distance indicators at greater redshifts.  High
redshift SN Ia allow measurements of the change in the expansion rate of the
universe over time, as well as in any presumed underlying models, such as dark
energy.  Up to a point, more data at both high and low redshift decreases the
statistical uncertainty in the derived cosmological parameters.  It can also
help refine our understanding of possible systematic uncertainties such as host
galaxy reddening and intrinsic color variation of SN Ia.

On the nearby front, the Calan-Tololo survey produced the first large,
multi-band, CCD sample of SN Ia photometry, publishing 29 light curves
\citep{hamuy96b}.  This was followed by 22 \emph{BVRI} SN Ia light curves from the
CfA in 1999 \citep{riess99} and a further 44 \emph{UBVRI} in 2006 \citep[hereafter,
J06]{jha06} (these two samples will be referred to as CfA1 and CfA2,
respectively).  Additionally, Krisciunas and his collaborators have published a
significant number \citep{krisc00, krisc01, krisc03, krisc04a, krisc04b,
krisc06}, the European Supernova Collaboration has published photometry of
several nearby SN Ia \citep[see][and references therein]{stanishev07}, and
\citet{kowalski08} recently published eight nearby SN Ia.

Other groups that are working on significant nearby samples are
KAIT \footnote{http://astro.berkeley.edu/$\sim$bait/kait.html}, in conjunction
with the LOTOSS/LOSS SN searches, CSP \citep{hamuy06}
and the Nearby Supernova Factory \citep{aldering02}.  The 2004 SDSS SN Survey
\citep{sako05} found 16 spectroscopically-confirmed SN Ia as a preparation run
for the SDSS-II SN Survey \citep{frieman08}.  In its first two years, SDSS-II
observed over 300 spectroscopically-confirmed and $\sim100$
photometrically-identified SN Ia in $ugriz$ and in the redshift range 
$0.05<z<0.35$.  This survey has
good control of systematics in their photometry and will be very useful in
calculating SN rates in the nearby to intermediate redshift range.
\citet{dilday08} present a calculation of nearby SN Ia rates based on 17 SN Ia
at $z\le0.12$ from the 2005 season of SDSS-II.

Systematic differences often exist between different groups' photometry of the
same SN Ia at low redshift, typically at the level of a few hundredths mag and
sometimes larger.  These differences are mainly due to difficulties in
transforming to the standard system and, to a lesser extent, the use of
different photometry pipelines.  A large, homogeneously-observed and reduced
nearby sample does not internally suffer from these two problems and can help
reduce systematic uncertainties in dark energy measurements.  However, there is
still the issue of ensuring that the nearby and faraway samples are
photometrically consistent.

The goal of our research was to produce a large sample
of homogeneously-observed and reduced SN Ia light curves that fills out the
sampling of the whole range of SN Ia properties and can be used to reduce the
statistical and systematic uncertainties in SN Ia cosmology.   
Here we publish 185 multi-band optical SN Ia light curves, with data taken
between the years 2001 and 2008.  This is the third sample of nearby CfA SN Ia
photometry (CfA3 sample).  This is the largest set of 
nearby SN Ia photometry to date, more than doubling the literature sample.  It
consists of over 11500 observations.  For comparison, the CfA1 and CfA2 samples
consist of 1210 and 2190 observations.  

A better understanding of the nature and range of SN Ia properties improves
their use as standardizable candles and may reveal that certain subsamples are
more useful than others.  We intentionally built up the sample of slow (more
luminous) and fast (less luminous)
decliners.  The slow decliners are particularly helpful for improving SN Ia
cosmology since they are found more often at high redshift.  The study of
peculiar SN Ia also deepens our understanding of what physical mechanisms might
be at work and large samples are more likely to include rare types.  One such
object that the CfA Supernova
Group\footnote{http://www.cfa.harvard.edu/supernova/} observed was SN 2006gz
\citep{hicken07, maeda09}, a very slow-declining and bright SN Ia that may have
come from a double-degenerate merger and/or a Super-Chandresekhar progenitor.
With the larger sample, light-curve fitters can be trained better and a proper
prediction error can be calculated by excluding individual objects (or groups
of objects) from the training sample one at a time.  \citet{mandel09} has
developed the machinery for this in the near infrared and will be including the
optical bands shortly.  Combining the optical and near-infrared photometry
should help disentangle host-galaxy reddening from intrinsic SN Ia color.

The impact of adding additional nearby SN Ia can be seen in \citet{kowalski08},
who take 49 nearby and 250 faraway SN Ia from the literature and add eight of
their own, using the light-curve fitter SALT \citep{guy05}.  
These additional eight reduce the statistical uncertainty on
the dark energy parameter, $w$, by a factor of 1.04, when the intrinsic or
additional uncertainty of $\sim0.1$ mag that they discuss is added.  If the intrinsic uncertainty is not added then their
eight reduce the statistical uncertainty in $w$ by a factor of 1.07.  The
application of the CfA3 data set to studying dark energy is presented in
\citet[][hereafter, H09]{hicken09}, where 90 of the 185 objects presented here
pass the quality cuts (on such things as redshift and phase of first observation) of
\citet{kowalski08}.  These 90 are added to their ``Union" set to form the
``Constitution" set (a more perfect union) with a total of 157 nearby and 250
faraway objects.  The Constitution set produces a value of
$1+w=0.013^{+0.066}_{-0.068} (0.11 \rm~syst)$, consistent with the cosmological
constant.  The uncertainty on $w$ for the Constitution set is found to be 1.2--1.3
times smaller than the comparable Union value (1.3 
when the $\sim0.1$ mag intrinsic
uncertainty is included and 1.2 when it is not), in line with approximate statistical expectations.
The systematic uncertainty is estimated to be $\sim65\%$ larger than the
statistical uncertainty.  The other fitters, SALT2 \citep{guy07}, MLCS2k2
\citep{jha07} with $R_V=3.1$ and MLCS2k2 with $R_V=1.7$ were found to reduce
the statistical uncertainty by a factor of $\sim1.2-1.3$, slightly less than
the rough statistical expectation.  The addition of the CfA3 sample achieves
its goal of reducing the statistical uncertainty of $w$.  Both the good and the
bad news is that systematic errors are now the main limit for making further
progress in better understanding dark energy with SN Ia!  Improvements in
systematic uncertainties are needed to maximize the contributions of future SN
Ia surveys, such as the Joint Dark Energy Mission, that aim to place tight
constraints on the time dependence of dark energy.

In this paper, we first show that the CfA3 sample is consistent with previous
nearby samples in its color and host-galaxy reddening distributions.  By
design, the CfA3 sample has a wider distribution of light-curve shapes than
earlier work since we gave the fast and slow decliners higher priority in
deciding which objects to follow most thoroughly.  However, the range of
decline rates covered is the same.  We also show that the agreement of our
photometry with that of other groups, for the same objects, is as good as the
agreement between other groups, typically at the level of a few hundredths mag.
These facts give us confidence that the CfA3 sample can be used by current
light curve fitters developed primarily from the Calan-Tololo, CfA1, and CfA2
surveys.  We invite people to combine the CfA3 sample with previous samples to
retrain existing fitters or invent new ones.  The CfA3 sample itself was not
part of the training sample for any of the light-curve fitters in H09 and so
the good agreement found there of the CfA3 results with previous samples is
encouraging.  

The CfA3 sample shares many of its methods with ESSENCE
\citep{miknaitis07}.  The same data-reduction pipeline was used, minimizing the
introduction of systematic effects due to different reduction methods.  The
CfA3 sample also helps reduce the systematic uncertainty in $w$ because it was
reduced and mostly observed in a homogeneous fashion (the use of two different
cameras and changing from \emph{RI} filters to $r'i'$ being the exceptions
to completely homogeneous observation). 

In conjunction with this optical photometry, the CfA Supernova Group has taken
spectra \citep{matheson08, blondin09} of many of these SN using the FAST
spectrograph \citep{fabricant98} and, starting in 2004, began use of the
PAIRITEL near-infrared telescope\footnote{http://www.pairitel.org/} to acquire
valuable $JHK$-band data for the brighter SN Ia in the sample.  These
near-infrared SN Ia light curves stand on their own as standard candles
\citep{woodvasey08} and, when combined with the optical data, will help clarify
the properties of host-galaxy dust and intrinsic color variation of SN Ia
\citep{friedman09}.  This should help decrease the systematic uncertainties due
to these intertwined phenomena.  

In \S2, we describe our observing strategy, explain our data reduction choices,
and present the CfA3 light curves.  We also show that our photometry is
internally consistent and agrees well externally (to a few hundredths mag,
roughly) in cases where others have published light curves for the same
objects.  Typical uncertainties in our \emph{V}-band SN photometry are 0.015 mag
around maximum light.  We estimate a systematic uncertainty of 0.03 mag in
\emph{BVRIr'i'} and 0.07 mag in $U$.  In \S3, we examine the decline rates,
intrinsic colors, and intrinsic absolute magnitudes.  We confirm many of the
relations seen before.  The one main new insight is that the fast decliners
have a range in intrinsic magnitude of 1.0-1.5 mag, with the 1991bg-like
objects signficantly fainter and not forming part of the otherwise tight locus
of SN Ia points, suggesting that they should be treated separately 
in light-curve fitter training samples.  We present our conclusions in \S4. 

The CfA3 light curves, comparison star magnitudes and passbands can be found at
our website\footnote{http://www.cfa.harvard.edu/supernova/CfA3} and are
archived with the journal.  Luminosity distances from H09 can also be found
at our website.
 
\section{Data and Reduction}

The instruments, data aquisition and data reduction are described here.  The
data reduction consists of three stages:  reduction, calibration and
host-galaxy subtraction (where necessary).  The reduction and subtraction
stages are carried out by a version of the ESSENCE and SuperMACHO pipeline
\citep{miknaitis07, rest05, garg07}, modified for use on the CfA3 data.  The
calibration was carried out very similarly to the calibration in the CfA1 and
CfA2 samples.  We made use of differential photometry by calibrating the field
or comparison stars surrounding the SN on photometric nights and then measuring
the flux of the SN relative to the comparison stars in each image, on both
photometric and non-photometric nights.  In most cases, the underlying
host-galaxy light had to be subtracted, using reference images taken after the
SN had faded.

\subsection{Instruments}

The 1.2m telescope at the F. L. Whipple Observatory (FLWO) was used to obtain
the photometry presented here.  The 4Shooter 2x2 CCD
mosaic\footnote{http://linmax.sao.arizona.edu/FLWO/48/OLD/4shccd.html} was
employed up until 2004 August for 64 objects of the CfA3 sample.  From 2004
September to 2005 July, the 2x1 CCD mosaic
Minicam\footnote{http://linmax.sao.arizona.edu/FLWO/48/OLD/miniccd.html} was
used to observe five SN Ia.  The single-chip CCD
Keplercam\footnote{http://linmax.sao.arizona.edu/FLWO/48/kepccd.html} was used
for the remaining 116 SN Ia beginning in 2005 September.  

The 4Shooter camera uses four thinned, backside-illuminated, anti-reflective
coated Loral 2048x2048 CCD detectors.  Our 4Shooter observations were always on
chip three (read out by a single amplifier) in bin-by-2 mode such that the
binned pixel scale is $0".674$ per pixel and the field of view was 11.5'x11.5'.
The typical image quality was $1".5$ to $3"$ FWHM.  The Minicam chips are
thinned, backside-illuminated Marconi (ex-EEV) 2248x4640 CCD detectors with two
long-rectangular shaped amplifiers per chip.  In bin-by-2 mode, the pixel scale
is $0".600$ per pixel.  Our observations were always on amplifier three with an
approximate field of view of 5.1'x23.1'.  The Keplercam uses a Fairchild ``CCD
486."  It is read out by four amplifiers, each covering a region of 2048x2048
pixels.  Our observations were always on amplifier 2.  In bin-by-2 mode, the
pixel scale is $0".672$ per pixel, resulting in an amplifier-2 field of view of
approximately 11'.5x11'.5.

All three instruments have good response in the red while the 4Shooter was
superior in the near ultraviolet.  The 4Shooter had a significant number of bad
pixels that required masking, the Minicam had few and the Keplercam had
virtually none.  A bad-pixel mask was not required or used for the Minicam and
Keplercam.  The Johnson \emph{UBV} passbands were used with all three detectors.
The Krons-Cousins \emph{RI} passbands were used on the 4Shooter.  In order to
cooperate better with other FLWO observing programs, SDSS $r'i'$ filters were
used on the Minicam and Keplercam.  The ``Harris" set of \emph{BVRI} filters and a
$U$ filter with a $\rm CuS0_4$ cell for red blocking were used for all CfA3
4Shooter observations.  The same \emph{UBV} filters, and SDSS $r'i'$ filters were
used on the Minicam and Keplercam.  The $U$ filter broke in January, 2007 and
was replaced in June, 2007.  A liquid leak was discovered in the $\rm CuS0_4$
cell of the $U$ filter in November, 2007 and after repair and testing it was
installed in February, 2008.  These problems with the $U$ filter account for
missing $U$-band photometry in 2007-2008.

The 64 4Shooter objects are all observed with the same camera and filters and
reduced with the same pipeline, constituting one homogeneously-observed and
reduced sample.  The 116 Keplercam objects also represent a
homogeneously-observed and reduced sample.  The use of three different 
cameras and changing from \emph{RI} filters to $r'i'$ limits us from calling
the entire CfA3 sample homogeneously observed and reduced.  However, 
its acquisition and reduction can be called quasi-homogeneous, since
the \emph{UBV} filters were used on all three cameras, the detector responses are
similar, and the same reduction pipeline was used.

\subsection{Observations}

Nearby SN are discovered by both amateur and professional astronomers.  Many of
the discoverers promptly report their findings to the SN community via email.
The IAU's Central Bureau for Astronomical Telegrams, the IAU Circulars and The
Astronomer's Telegram are commonly used to disseminate information.  Usually
the initial discovery does not include spectroscopic confirmation and typing.
The CfA Supernova Group depends on these discoveries, north of declination
$-20^\circ$, for the SN it studies.  The CfA3 discovery data is displayed in
Table \ref{table_sndiscovery}.  Roughly two thirds of the CfA3 sample were
discovered by professional observers.  Roughly one third was discovered by
amateurs, demonstrating their valuable contribution to nearby SN science.  In
first place, KAIT/LOTOSS/LOSS discovered $46\%$ of the CfA3 sample.  In second
place, the Puckett Observatory Supernova
Search\footnote{http://www.cometwatch.com/search.html} discovered $18\%$.  Most
of these search surveys had typical limiting magnitudes of 19.5 mag.  SDSS-II 
is the most obvious exception. 

The CfA Supernova Group rapidly responds to new objects, acquiring spectra and
optical and $JHK$ light curves.  This allows for a deeper investigation into
individual SN.  For the CfA3 sample, we would sometimes initiate photometric
observations of untyped SN candidates, depending on their brightness and any
additional properties provided in the email circulars, such as color or when
the last non-detection of the SN candidate was made.  If the SN candidate was
brighter than 18 to 18.5 mag and north of $-20^\circ$ then we would take
spectra with the FAST spectograph.  Our efforts have contributed roughly $40\%$
of the reported identifications of SN type over the last six years.  We did not
follow any SN that had peak magnitudes fainter than $\sim18.5$ mag, making this
the effective limiting magnitude for the CfA3 sample.  However, this does not
mean we observed every SN brighter than $\sim18.5$ mag.

With the information on type, age and any peculiar features in hand, either
from our own spectra or from others' reports, a decision on whether to begin or
continue observing the SN candidate was made.  As one of our goals was to fill
out the sampling across the whole range of SN Ia (to provide a more
complete training set for light curve fitters), highest
priority was given to SN Ia that were young, slow-declining, fast-declining, or
otherwise peculiar.  Another reason to prioritize slow decliners is that
these are preferentially found at high redshift.  Our program also observes
core collapse SN and high priority was given to stripped-envelope SN IIb/Ib/c.
Lower priority was given to SN II and older SN Ia.  If a SN Ia was found to be
older than $\sim$14 days after $B$-band maximum at time of our first
observation then it was usually removed from our list.  

We emphasize that the CfA3 sample distribution is not representative of the
abundances of SN Ia type or host galaxies.  Objects announced during the bright
phase of the moon were also less likely to be included since spectroscopic
identification was less likely to be obtained.  The Keplercam and Minicam were
usually mounted on the telescope at all phases of the moon while the 4Shooter
was often taken off for several days around full moon.  Our preference for
young and more extreme events makes the CfA3 sample distribution less
representative of the underlying population but does ensure that the wide range
is being amply sampled.  Finally, the limiting magnitude of both the searches
and our follow up mean that highly-reddened or intrinsically less-luminous SN
Ia are only observed in a small volume:  they are severly under-represented
in this sample compared to the cosmic rate.

\begin{figure}
\scalebox{0.90}[0.90]{
\plotone{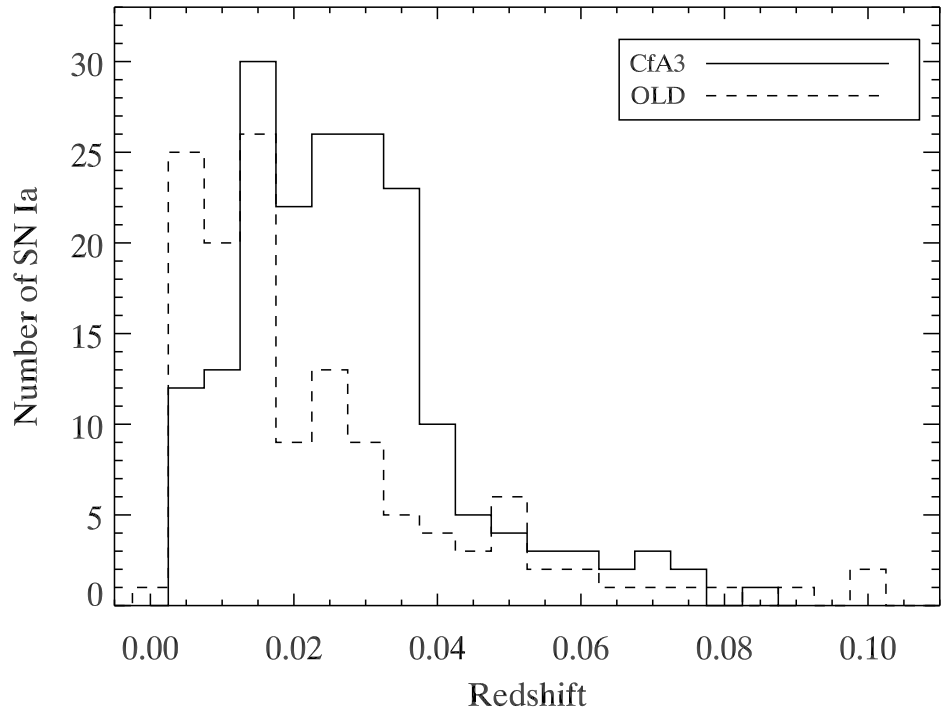}
}
\caption{Histograms of redshift ($z_{CMB}$) for both CfA3 and OLD SN Ia.  The mean redshifts are, respectively, 0.027 and 0.024.  There is one OLD SN above
$z=0.12$ not shown.
}
\label{fig_zhist}
\end{figure}

\begin{figure}
\scalebox{0.90}[0.90]{
\plotone{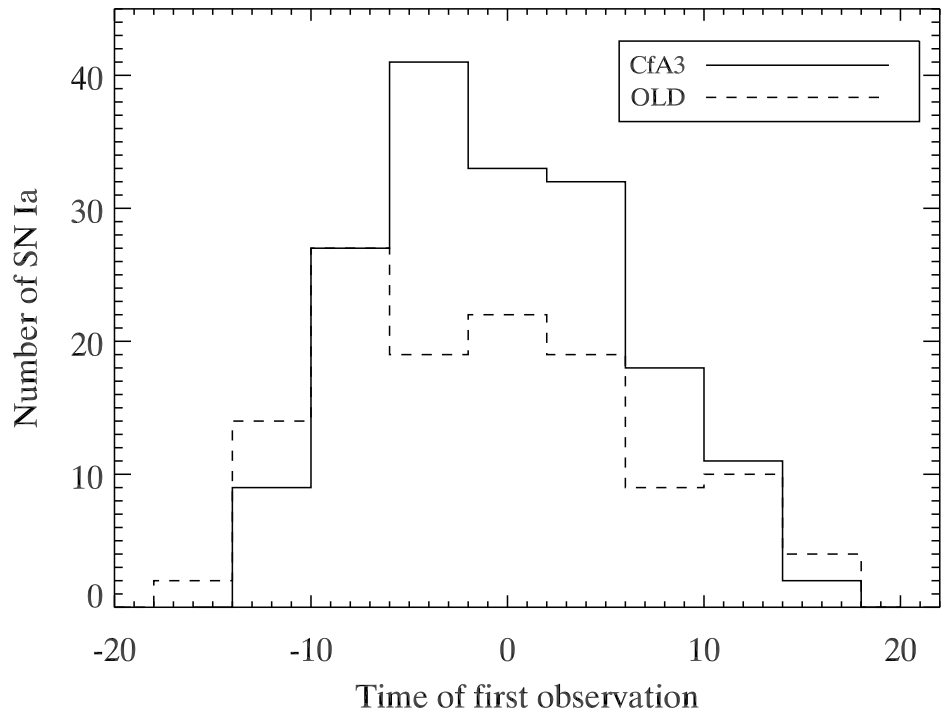}
}
\caption{Histograms of time of first observation in the rest frame, relative 
to maximum light in $B$, as calculated by MLCS2k2.  OLD has more objects
with very early measurements which are useful for constraining the rise time
and better understanding the explosion mechanism.
}
\label{fig_t1sthist}
\end{figure}

In Figure \ref{fig_zhist}, we plot redshift histograms of the CfA3 and OLD
samples.  The OLD sample is the nearby SN Ia sample as compiled in 
\citet{jha07}.
The CfA3 sample is primarily in the $0.02<z_{CMB}<0.04$ region, where
$z_{CMB}$ is the redshift in the cosmic microwave background reference frame.
The OLD sample is primarily below $z_{CMB}\approx0.03$.  Above $z_{CMB}=0.01$,
the median CfA3 and OLD redshifts are, respectively, 0.027 and 0.025.  Figure
\ref{fig_t1sthist} shows the time of first observation, relative to $B$-band
maximum light, with median values of -0.8 and -1.5 for CfA3 and OLD,
respectively.  The OLD sample has a higher percentage with very early
observations.  Respectively, the CfA3 and OLD samples have 48 and 47 objects
with time of first observation beginning sooner than five days before maximmum,
and 90 and 76 objects beginning before maximum.

\begin{figure}
\scalebox{0.90}[0.90]{
\plotone{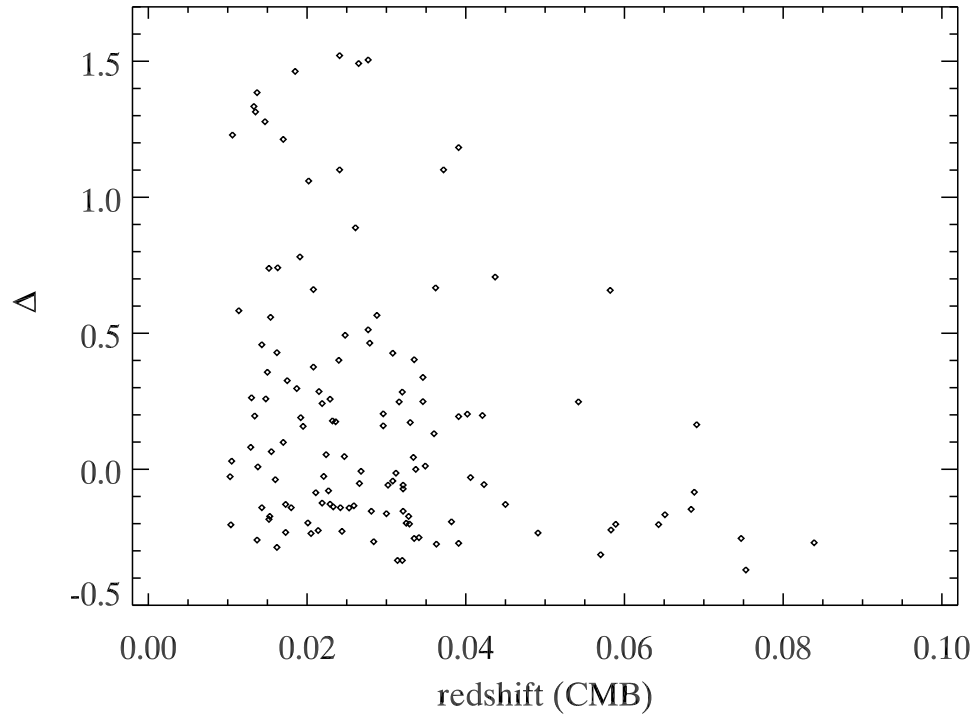}
}
\caption{Plot of MLCS2k2 $\Delta$ versus redshift (CMB) for the CfA3 sample.  
Objects closer than
$z=0.01$ are not shown.  The highest redshift for a given $\Delta$ is 
consistent with an approximate, effective peak limiting magnitude of 18.5 mag. 
At high redshift, in H09, no objects with $\Delta > 0.75$ are found in the
ESSENCE, SNLS, and Higher-Z samples used.
}
\label{fig_delta_z}
\end{figure}

The MLCS2k2 \citep{jha07} light-curve parameter, $\Delta$, is roughly a measure
of the relative \emph{V}-band brightness compared to the $\Delta=0$ model light
curve.  Negative $\Delta$ means greater intrinsic luminosity and broader light
curves and positive $\Delta$ means fainter luminosity and narrower light
curves.  Figure \ref{fig_delta_z} shows the CfA3 distribution of $\Delta$
versus redshift above $z_{CMB}=0.01$.  The whole range of $\Delta$ is present
out to $z_{CMB}\approx0.03$ and then the magnitude limits of discovery,
spectroscopic identification, and photometric-follow-up decisions discriminate
against
fainter objects which are not present in our sample at higher redshifts.  The roughly diagonal
slope in the right edge of the $\Delta$-versus-redshift distribution is
consistent with a limiting peak magnitude of $\sim18.5$ mag in the CfA3
objects.

\begin{figure}
\scalebox{0.90}[0.90]{
\plotone{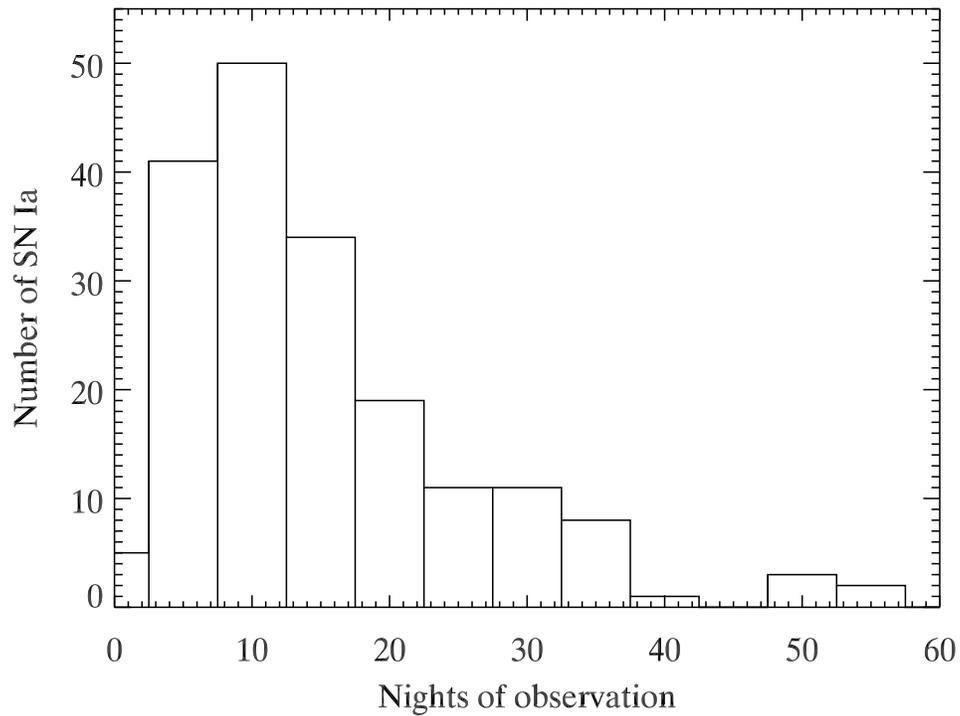}
}
\caption{Histogram of the number of nights each CfA3 SN Ia was observed
in $R/r'$ band, representative of \emph{V} and $i'$ also.  $B$ is slightly
less.  $U$ is often much less as it fades first, or nonexistent for 
when the filter was broken.  The mean is 15 nights and the median is 
12.  There are 121 objects with 10 or more nights and 45 with 20 or more.
}
\label{fig_nightshist}
\end{figure}

The FLWO 1.2m telescope has its time allocated to a specific observing program
each night with the requirement that roughly $10\%$ of the night be devoted to
other programs' observations.  From 2001 to summer, 2005, our typical time
allocation was one night per month with a few months of multiple nights to
aquire calibration and host-galaxy reference images.  Nightly requests of two
SN to other observers was typical during this period.  Beginning in the fall of
2005, two changes significantly increased both the number of SN we observed and
the sampling per object.  Instead of a single night per month, we received
roughly seven nights per month.  Additionally, several other observing programs
made significant numbers of SN observations for us in time they could not use.
The most notable group was the CfA component of the Kepler
Mission\footnote{http://kepler.nasa.gov/}.  We tried to observe new,
high-priority SN every one or two nights until $\sim$10 days past maximum light
and less frequently thereafter.  Weather and competing targets sometimes
reduced the actual cadence.  Secondary standards from \citet{landolt92} and
\citet{smith02} were observed on photometric nights and reference images for
host-galaxy subtraction were obtained after the SN had faded sufficiently,
usually a year after maximum light.  Figure \ref{fig_nightshist} shows a
histogram of the number of nights observed for each SN in the CfA3 sample.  The
mean number is 15 and the median is 12.  The number of objects with 20-or-more
nights of observation is 45 and the number with 10-or-more nights is 121.

\subsection{Pipeline:  Reduction Stage}

In this stage, raw images are processed to the point where all their star-like
objects have had their flux measured, but not yet calibrated.  Images first
undergo bias subtraction and flat fielding.  Dome-screen flats were used for
\emph{BVRIr'i'} while twilight flats were used for $U$.  The 4Shooter images
had their bad pixels masked out while the Minicam and Keplercam images did not
require this.  The small, but non-negligible, $I$-band fringes on the 4Shooter
were removed to the extent possible by subtracting fringe frames created from
several nights of $I$-band images.  The $i'$-band fringes on the Minicam and
Keplercam were much smaller in amplitude, making fringe correction
unnecessary.

The cosmic-ray removal algorithm, la\_cosmic \citep{vandokkum01}, in the form
of the IDL code, la\_cosmic.pro, by Joshua Bloom, was then applied to the
flat-fielded images to remove most of the cosmic rays.  It uses a 2-dimensional
Laplacian algorithm to detect cosmic rays.  Although removing the cosmic rays
did not have a significant effect on the photometry and reference-image
subtraction, this step was applied to each image.  

A linear astrometric solution was calculated for each image.  We
used astrometric solutions based on an external astrometric catalog for a
handful of good-seeing images of a single field.  We then ran SWarp
\citep{bertin02} on these images to properly scale and align them, and
center them on the SN position.  DoPHOT \citep{schechter93} was used to
get the field star positions to make an ``internal" astrometic catalog from our
own images.  We then reran these same images through these same stages with the
internal astrometric catalog and recalculated the field star positions to make
our final internal astrometric catalog.  This was done for each SN field.

The UCAC2 catalog \citep{ucac2} was our preferred external catalog but it does
not extend above declinations of roughly $+45^\circ$.
Where the UCAC2 catalog was sparsely populated, we used either the
USNO-B1.0 \citep{monet03} or USNO-A2.0 catalogs \citep{monet98}.
UCAC2
has an accuracy of around $0".03$ while USNO-B1.0 and USNO-A2.0 have poorer
accuracies of roughly $0".20$ and $0".25$ respectively.  The resulting average
standard deviation and relative accuracy of the star coordinates in our
internal astrometric catalogs did not depend significantly on which external
catalog was used.  The absolute accuracy of our internal catalogs will be
better in those that used UCAC2 for the initial solution but our positions
will generally be better than those reported at discovery.  Since we are
primarily interested in relative accuracy, though, all our internal astrometric
catalogs are adequate.  The typical standard deviation of a star's position in
our internal astrometric catalogs is $0".06$. 

We then used our internal astrometric catalogs to create a linear
astrometric solution for all of the flat-fielded images.  A linear solution was
adequate for the small field of view of the 1.2m images.  The astrometric
solution was used in SWarp to align the images to a common pixel system so that
host-galaxy reference images can be subtracted.  DoPHOT was run on the SWarped
images to calculate fluxes for all stellar-shaped objects.

DoPHOT uses a parameterized point-spread function (PSF) model.  A range of
functions can be effectively chosen by setting different values of the DoPHOT
PSF-shape parameters $\alpha$ and $\beta$.  With the PSF function set, DoPHOT
first fits for a single PSF shape and size over the whole image.  The high
signal-to-noise ratio (SNR) stars most heavily influence the best-fit PSF in
DoPHOT.  Then it fits this PSF to each star-like detection, calculating a
best-fit position, sky value and flux amplitude.  It is important that the PSF
model be capable of fitting the actual PSF shape of the data.  We found that an
order-2 Moffat fit our stars' PSF well while the default, truncated Gaussian
underestimated the flux in the wings of the stars.


A mismatched PSF function will do a better job of fitting low-SNR stars than of
fitting high-SNR stars (since low-SNR data is less contraining), possibly
introducing relative inaccuracies between the faint and bright stars.  We
compared our DoPHOT truncated-Gaussian-PSF magnitudes with aperture-photometry
magnitudes and found that the DoPHOT magnitudes differed from the
aperture-photometry magnitudes by about 0.01-0.02 mag per mag.  The fainter
stars were being interpreted as fainter relative to the aperture photometry
magnitudes than were the bright stars.  When we used the well-matching, order-2
Moffat function for our PSF, this effect was drastically diminished.

In $\S$2.5, we describe the calibration process to generate photometric
catalogs for the comparison stars in the SN fields.  To calculate a photometric
zero-point
for each SN image, we took a weighted mean of the differences between our
catalog magnitudes (in the natural system) and the DoPHOT measurements of the
comparison stars.  In the cases where the SN is sufficiently distant from 
any underlying structure (such as host-galaxy light or neighboring stars) the
DoPHOT magnitudes of the SN can be combined with their respective image 
zero-points to produce a calibrated light curve in the natural system.  

\subsection{Pipeline:  Host-Galaxy Subtraction}

Most of the SN in our sample required host-galaxy subtraction.  Reference
images were acquired on clear nights with good seeing and little or no moon so
as to maximize their SNR.  We also took reference images of SN that did not
need host subtraction as a way to test the host-subtraction
process.

Accurately subtracting the reference image from the SN image which was
obtained under different seeing conditions requires a convolution kernel that
can transform the PSF of one image to the PSF of the other.  The convolution
kernel is calculated using the algorithm of \citet{alard98} and
\citet{alard00} with slight improvements as in \citet{becker04} and
\citet{miknaitis07}.  The two images are each divided into stamps and
substamps and the best-fit convolution kernel is determined.  The image with a
narrower PSF is convolved to the other image.  Usually the reference image was
convolved but sometimes the SN image was.  The SN flux in the difference image
is measured with the DoPHOT PSF from the stars of the (wider) unconvolved
image.  

All of the reference images for the Keplercam SN Ia were obtained with the
Keplercam, resulting in ``same-camera" subtractions.  Some of the reference
images for the 4Shooter and Minicam SN Ia were taken with the Keplercam,
resulting in ``cross-camera" subtractions.  The responsivity of the 
different cameras are similar enough in a given passband so there is no
problem in using the Keplercam reference images for 4Shooter and Minicam
SN images.  The flux normalization for the
difference image can be chosen from either the SN image or the reference image.
In the case of the same-camera subtractions, we chose to use the flux
normalization from the reference image so that this would be used for every
observation of that SN in a given band.  In the cross-camera subtractions, the
flux normalization from the SN image was used in order to stay in the natural
system of the camera in which the SN data was observed.  If the unconvolved
image happens to be the one chosen for the flux normalization of the difference
image then its zeropoint magnitude can be directly applied to the DoPHOT SN
magnitude to achieve the calibrated natural-system SN magnitude.  If the
zeropoint of the image-that-got-convolved is used for the flux normalization
then the flux of the SN in the difference image must be divided by the
normalization (sum) of the convolution kernel to preserve the 
pre-convolution flux scale.

Noise maps are propagated for both images and are used to calculate a noise
map for the difference image.  Information from the noise image is combined
with the DoPHOT uncertainty and calibration uncertainty to produce the
uncertainty of the natural system SN measurement.

\subsection{Calibration}

On photometric nights, we observed one or two fields of secondary standards
every hour, over a range in airmass that matched the SN observations.  For the
\emph{UBVRI} bands used on the 4Shooter, we used secondary standards from
\citet{landolt92}.  \citet{smith02} establish the photometric system for the
SDSS passbands, $u'g'r'i'z'$.  They use many of the fields from
\citet{landolt92} but much fewer stars.  For the Minicam and Keplercam, where
we are using \emph{UBVr'i'}, we chose our secondary standards from
\citet{smith02} to ensure that we have stars with $r'i'$ calibration and used
the \emph{UBV} magnitudes from \citet{landolt92}.  

We performed aperture photometry on the Landolt/Smith standard stars and on our
SN-field comparison stars using the NOAO/DIGIPHOT/APPHOT package in IRAF
\citep{tody93}.  The
comparison stars were chosen so that they were reasonably well isolated and
usually detected in all bands.  A few sparse fields required also using stars
that only had good detections in \emph{BVRI/r'i'} but not in $U$.  An aperture with
radius of 15 pixels was used on both the standard and comparison stars.  An
aperture correction was calculated from one or two bright, isolated, good
curve-of-growth stars by subtracting the 6-pixel-radius-aperture magnitude from
the 15-pixel-radius-aperture magnitude and applied to all of the stars in
the field.

A photometric transformation solution for a given night was calculated from our
Landolt/Smith stars using system of equations 1.  A linear dependence on
airmass and color was sufficient for our intended level of final
\emph{V}-band comparison star precision ($\sim 0.015$ mag).  Higher-order terms were found to be
consistent with zero and so we did not use them.  

\begin{eqnarray}
u-b &=& zp_{ub} + \alpha_{ub}x + \beta_{ub}(U-B) \nonumber \\
b-v &=& zp_{bv} + \alpha_{bv}x + \beta_{bv}(B-V)  \nonumber\\
v-V &=& zp_{v} + \alpha_{v}x + \beta_{v}(B-V)  \nonumber\\
v-r &=& zp_{vr} + \alpha_{vr}x + \beta_{vr}(V-R)  \nonumber\\
v-i &=& zp_{vi} + \alpha_{vi}x + \beta_{vi}(B-I)
\end{eqnarray}

The terms on the left side of the equations are the instrumental colors except
for the \emph{V}-band term.  The first term on the right side of each equation
is the zero-point, followed by the airmass coefficients, $\alpha$, times the
airmass, $x$.  The \emph{V}-band equation is unique in that it directly
relates the instrumental magnitude \emph{v} to the standard system magnitude
and color, \emph{V} and $B-V$.  The other four equations only relate the
instrumental and standard-system colors to each other.  The final term on the
right of the four color equations multiplies the standard-system color of the
standard stars by a coefficient, $\beta$, to convert the standard-system color
into the ``calibrated" natural-system color.  

Having solved for the zero-point, airmass and color coefficients by using the
Landolt/Smith standards, this photometric solution was then applied to the
comparison star measurements, producing tertiary standards that were used to
calibrate the SN measurements.  

Our goal was to observe each SN field on multiple photometric nights to ensure
more accurate calibration.  Sometimes this was not possible, but even in those
cases SN fields that produced consistent, multiple-night calibration were
observed on the same night, making us sufficiently confident that the SN fields
with a single night of calibration were accurate.  The uncertainties of the
comparison stars include the measurement uncertainties, the standard deviation
of measurements from multiple nights (for single nights, an appropriate error
floor was used instead) and the uncertainty of the transformation to the
standard system.  The typical uncertainty of our \emph{V}-band comparison star
measurements is 0.015 mag.  The average color coefficients are presented in
Table \ref{table_colorterms}.  

\begin{figure}
\scalebox{0.90}[0.90]{
\plotone{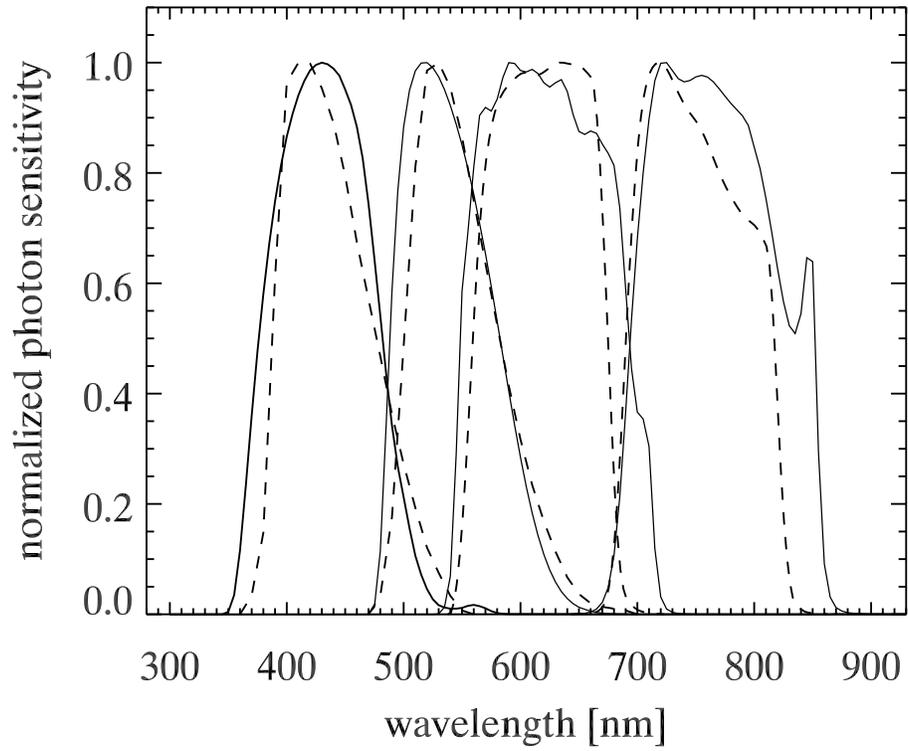}
}
\caption{Synthesized natural system Keplercam \emph{BVr'i'} passbands 
(solid curves) with \citet{bessell90} \emph{BV}
and SDSS \emph{r'i'} overplotted (dashed curves).
}
\label{fig_keppassband}
\end{figure}

\begin{figure}
\scalebox{0.85}[0.85]{
\plotone{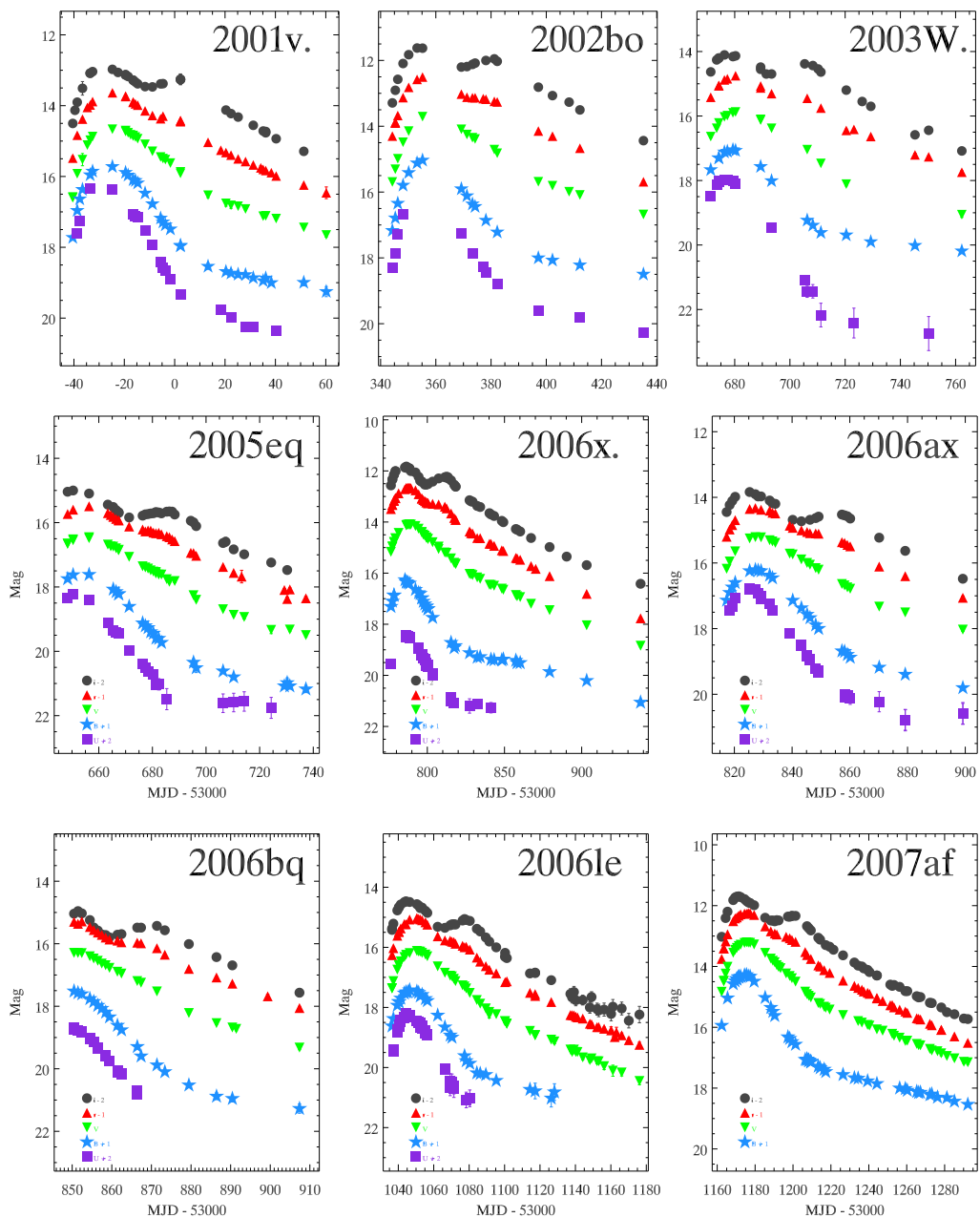}
}
\caption{Nine of the better CfA3 SN Ia light curves.  Error bars are smaller
than the symbols in most cases.  $U+2$, $B+1$, \emph{V}, $R/r'-1$ and $I/i'-2$
have violet, blue, green, red and black symbols, and are ordered from bottom
to top in each plot. 
}
\label{fig_9lc}
\end{figure}

We also synthesized natural system \emph{BVr'i'} passbands for the Keplercam by
combining the primary and secondary mirror reflectivities (taken as the square
of the measured reflectivity of the primary), the measured filter
transmissions, and the measured Keplercam quantum efficiencies.  No atmospheric
component is included.  We present these passbands as normalized photon
sensitivities.  A $U$-band filter transmission curve and the Minicam quantum
efficiency were not available so passbands were not made for Keplercam $U$ or
any of the Minicam bands.  The 4Shooter \emph{BVRI} passbands can be found in
J06 as the ``4Shooter/Harris" combination and we point out that they are
presented as normalized energy sensitivities.  To convert to normalized photon
sensitivities, the passbands should be divided by wavelength and renormalized.
See Figure \ref{fig_keppassband} for a visual representation of the Keplercam
\emph{BVRI} passbands.  Our light curves were produced in the natural system
and then converted to the standard system by using the color terms in Table
\ref{table_colorterms}.  The light curves and comparison stars, both natural
and standard system versions, can be found at our
website\footnote{http://www.cfa.harvard.edu/supernova/CfA3} and are archived
with the journal.  The Keplercam \emph{BVr'i'} passbands can also be found at
these two locations.  The natural system passbands and photometry can be used
together to avoid the uncertainty of using star-derived color terms but we do
not pursue this here.  Figure~\ref{fig_9lc} shows nine of our better light
curves.

\subsection{Internal Consistency Checks}

By choosing an appropriate shape for the PSF of the comparison stars we ensured
accurate flux measurements for well-isolated stars.  This also applies to cases
where the SN is well isolated, allowing for two tests of the image-subtraction
process:  comparing the unsubtracted light curve with the light curve produced
by subtracting a reference image taken with the same camera;  doing the same
procedure but with a reference image taken with a different camera.  A third
test involves comparing the light curve produced by subtracting a same-camera
reference image with the light curve produced by subtracting a reference image
from a different camera.  As described below, we have done these tests and 
find internal consistency at about the 0.01 mag level in most cases, 
when the SN is brighter than 17 mag.  

\subsubsection{Same-Camera Subtracted Versus Unsubtracted Light Curves}    

\begin{figure}
\begin{center}
\scalebox{0.80}[0.80]{
\plotone{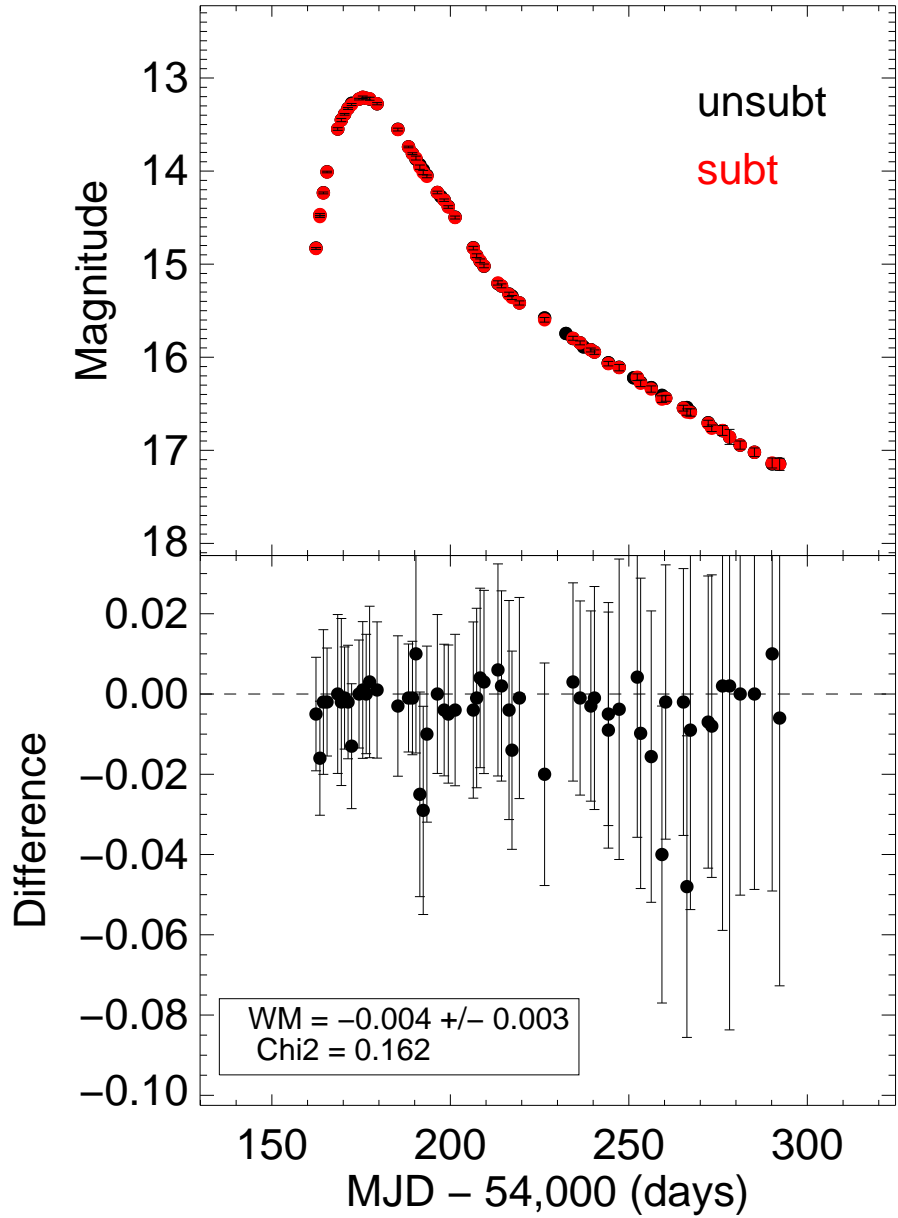}
}
\end{center}
\caption{Comparison of the \emph{V}-band unsubtracted and subtracted light curves
of the bright SN 2007af.  Most points agree to better than 0.01 mag.  The
weighted mean (WM) and $\chi^2$ (Chi2) of the differences are listed in 
the lower panel.
}
\label{fig_sn07af}
\end{figure}

SN 2007af was very bright compared to its underlying galaxy background and the
subtracted and unsubtracted light curves agree to better than 0.01 mag for most
points, as seen in Figure~\ref{fig_sn07af}, showing that the subtraction stage
of the pipeline works well.  The comparison plots also contain the weighted
mean (WM) and $\chi^2$ (Chi2) of the differences.

\begin{figure}
\begin{center}
\scalebox{0.80}[0.80]{
\plotone{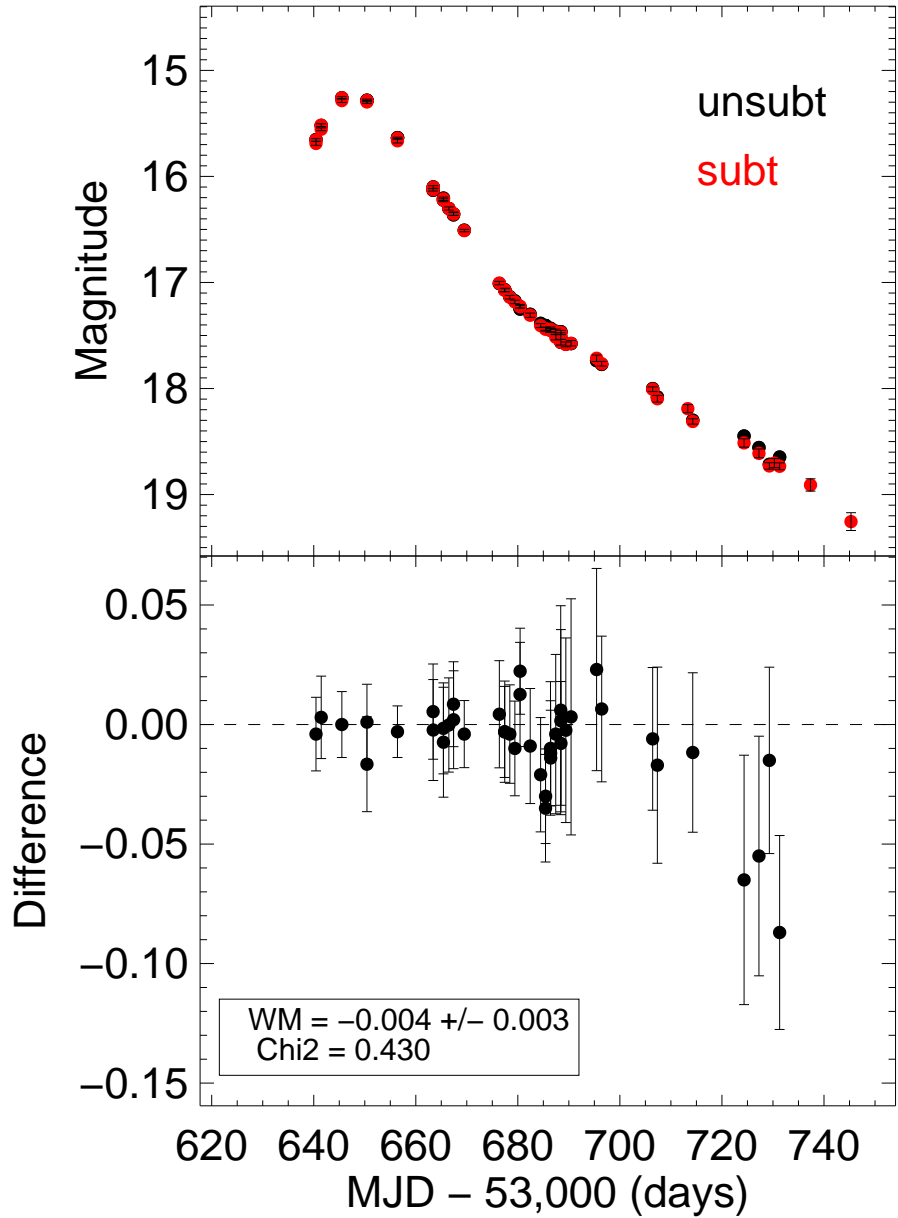}
}
\end{center}
\caption{Comparing the host-galaxy subtracted and unsubtracted \emph{V} light curves
of SN 2005el.  Most points agree to better than 0.02 mag. 
Some of the scatter is due to varying seeing.
The light curves diverge at very late times when the underlying
galaxy influences the unsubtracted light curve more. 
}
\label{fig_sn05el}
\end{figure}

SN 2005el is on the side of a smooth host galaxy.  Figure~\ref{fig_sn05el}
shows that the subtracted and unsubtracted \emph{V}-band light curves of SN 2005el
agree to better than 0.01 mag for all but one point brighter than 17 mag. 
Between 17 and 18 mag, most of the points agree within 0.02 mag.  The
underlying galaxy flux in the unsubtracted images begins to be picked up in the
last few points.  The agreement in the other bands was fairly similar.  

\begin{figure}
\begin{center}
\scalebox{0.80}[0.80]{
\plotone{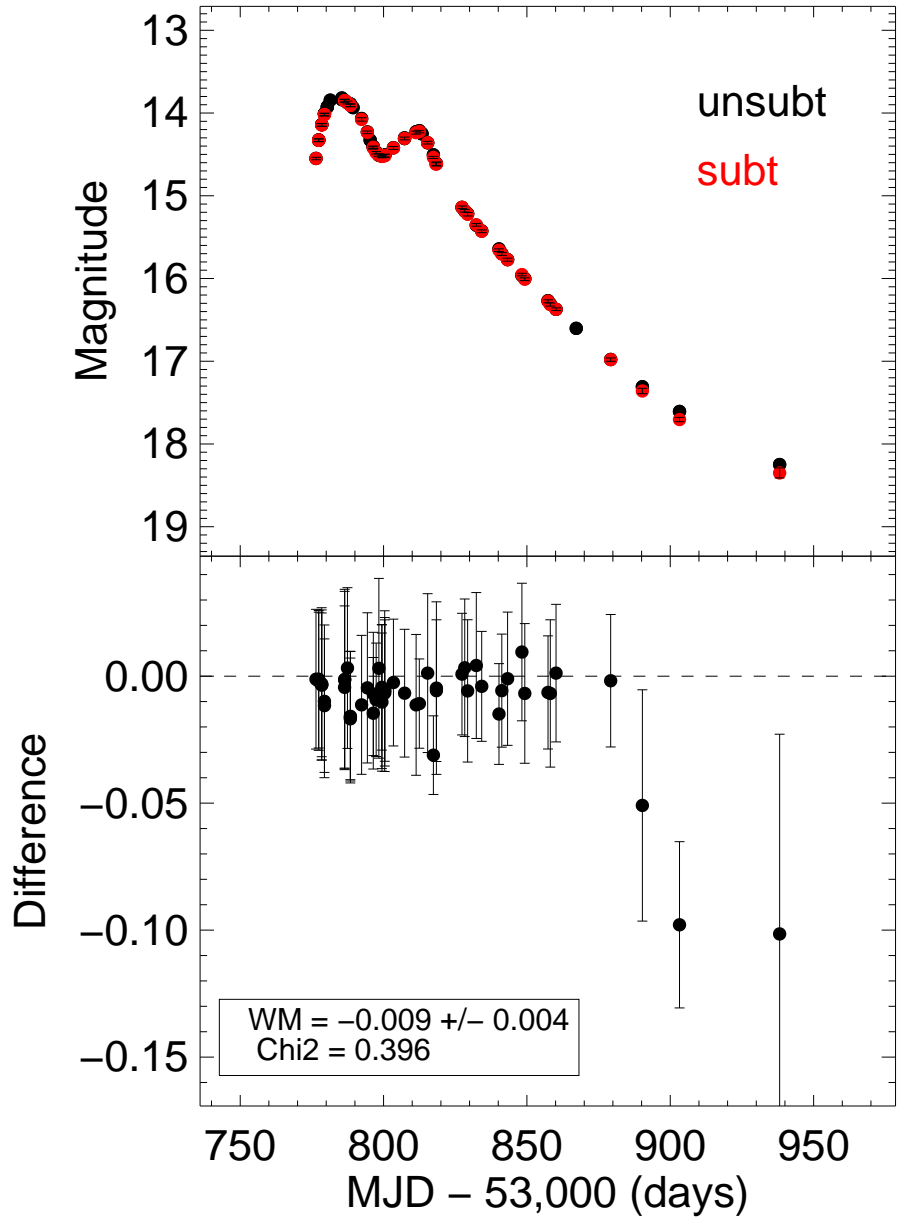}
}
\end{center}
\caption{Comparison of the subtracted and unsubtracted $i'$ light curves
of the bright SN 2006X.  Most points agree to better than 0.01 mag, suggesting 
that the reference-image subtraction is working well.  The underlying 
galaxy flux only becomes evident at late times. 
}
\label{fig_Inosubt06X}
\end{figure}

SN 2006X allows two comparisons.  Since it was fairly bright compared to its
underlying galaxy light, the subtracted and unsubtracted light curves can be
compared, especially in $i'$, where 
dust extinction is the least and the PSF is the narrowest.  Figure~\ref{fig_Inosubt06X} shows that the unsubtracted light curve is slightly
brighter at bright times, due to the small amount of underlying galaxy flux.
Nonetheless, most points agree to better than 0.01 mag.
At faint times, this galaxy flux becomes more significant.  However, the 
agreement at bright times is a good indication that the subtraction
stage is working well.  

\begin{figure}
\begin{center}
\scalebox{0.80}[0.80]{
\plotone{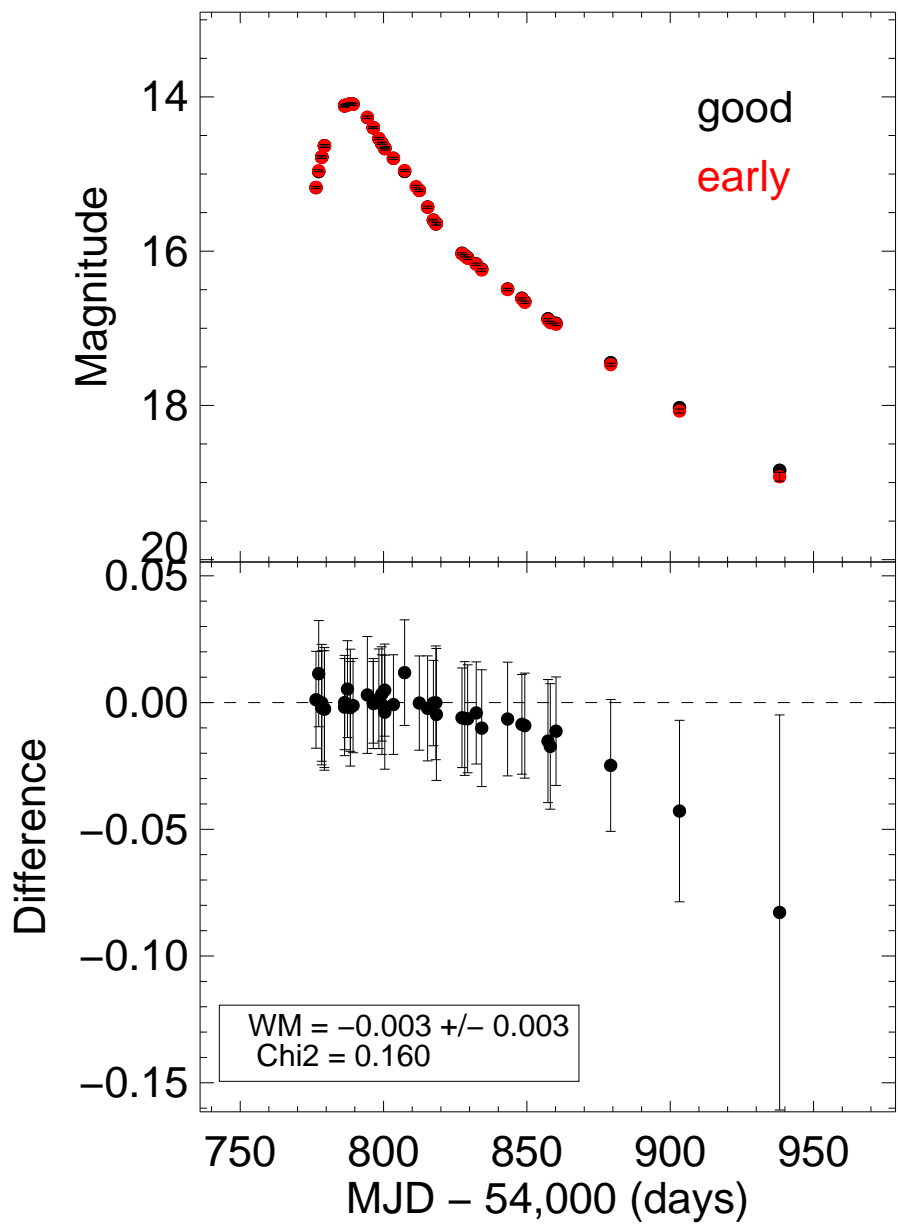}
}
\end{center}
\caption{Comparison of the subtracted \emph{V}-band light curves of SN 2006X using
two different reference images, one taken too early, with a small amount of SN 
flux in it still, and a later one with none.  Most points agree to better than
0.005 mag at bright times,
showing that the subtraction process is working well, while the later divergence
demonstrates the need for the SN to fade away before acquiring the reference
image.  The photometry we present used the later reference image, of course.
}
\label{fig_sn06Xearlylate}
\end{figure}

The other comparison for SN 2006X involves the subtracted light curves using
two different reference images, one that was taken early with some SN flux
still in it and one taken later when the SN had faded sufficiently (see
Figure~\ref{fig_sn06Xearlylate} for the \emph{V}-band comparison).  Most of the
points that are brighter than 16 mag agree to better than 0.005 mag, once again
showing that the subtraction pipeline introduces no spurious results.  Later
than this, the SN flux in the early reference image begins to make its light
curve fainter than it should be.  The later reference image was used for our
final light curve.  Waiting roughly one year (or more for the handful of SN
brighter than $\sim14$ mag at peak) to aquire the reference images for our SN
 ensured that the SN had faded sufficiently.

\begin{figure}
\begin{center}
\scalebox{0.80}[0.80]{
\plotone{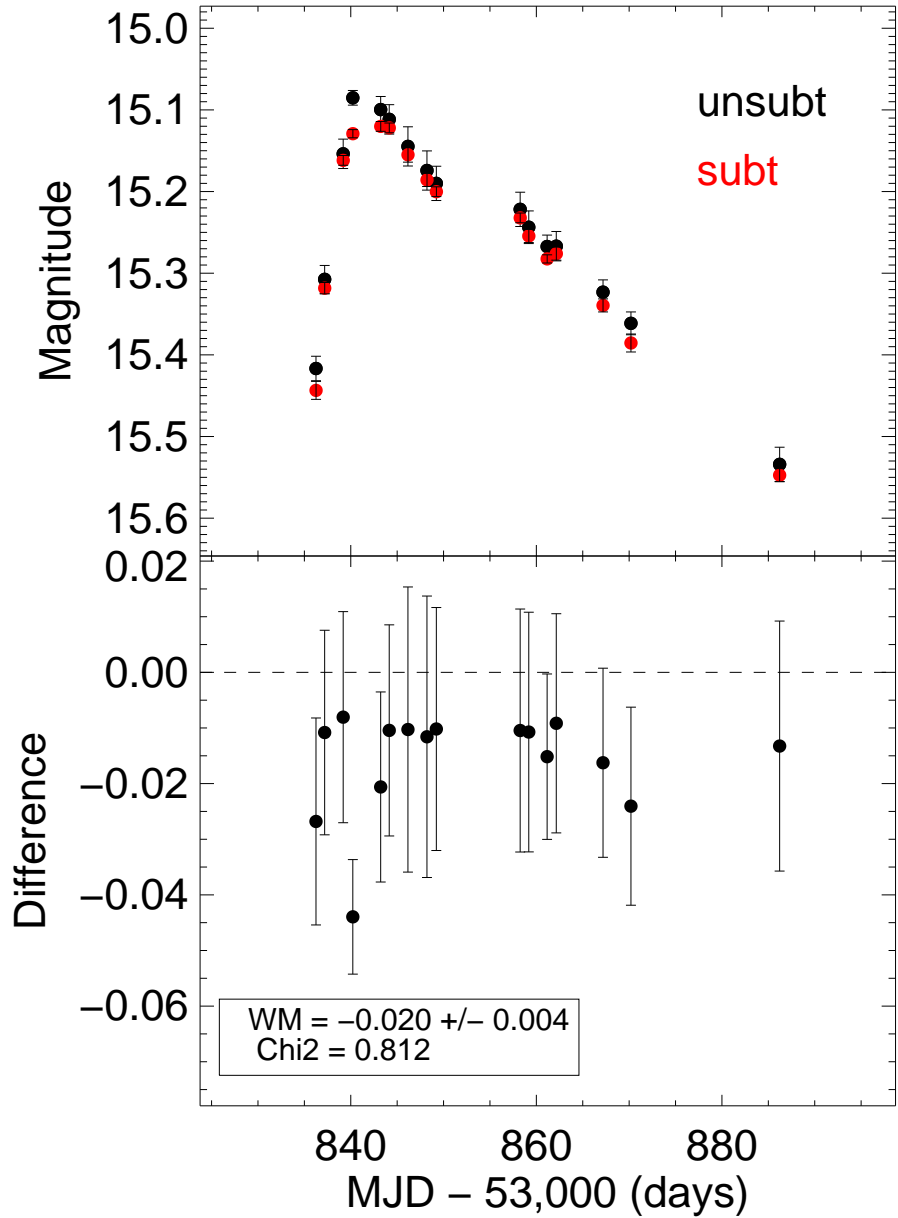}
}
\end{center}
\caption{Comparison of the \emph{V}-band unsubtracted and subtracted light curves
of the bright SN 2006bp.  Its position has some host-galaxy flux in it, giving
rise to the slightly brighter unsubtracted light curve but showing that the
subtraction process is working well.  Chi2 was calculated relative to the -0.01
mag offset.
}
\label{fig_sn06bp}
\end{figure}

The host-subtracted light curve for the type II SN 2006bp was presented in
\citet{dessart08}.  Underlying galaxy structure is definitely present at the SN
position but since the SN was bright, the unsubtracted light curve is only
slightly brighter (offset by $\sim0.01$ mag) than the host-subtracted light
curve, as seen in Figure~\ref{fig_sn06bp}, with only one point significantly
different from this offset.  The relatively constant offset between the 
two light curves serves as an indicator of the good pipeline performance. 

These four examples show that the reference-image subtraction process itself
does not introducing any significant offset into the final SN photometry.

\subsubsection{Cross-Camera Subtracted Versus Unsubtracted Light Curves}

\begin{figure}
\begin{center}
\scalebox{0.80}[0.80]{
\plotone{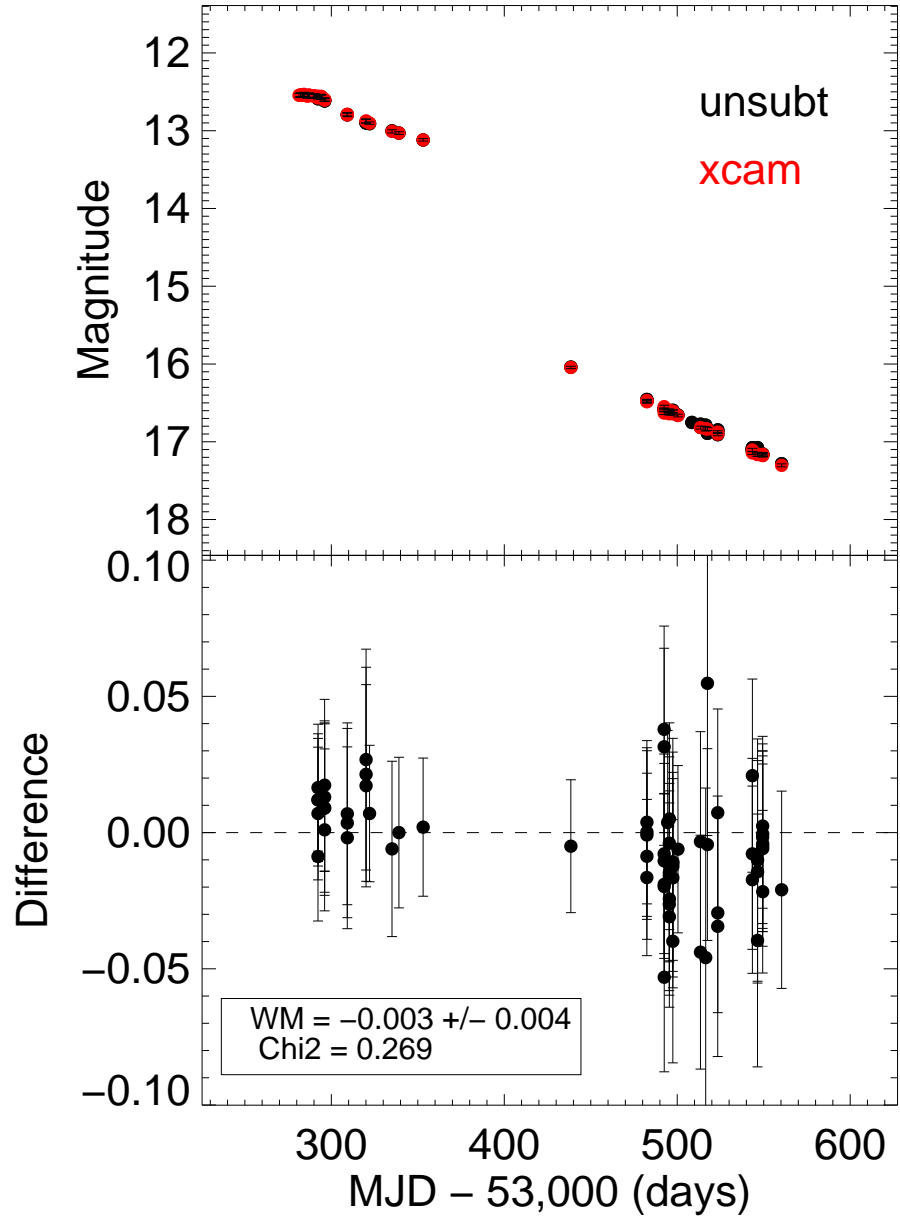}
}
\end{center}
\caption{Comparison of the unsubtracted and cross-camera subtracted \emph{V}-band
light curves of SN 2004et. 
}
\label{fig_sn04et}
\end{figure}

Many of the 4Shooter and Minicam SN Ia reference images were acquired with the
Keplercam and so it was important to see that the cross-camera subtraction
works well.  SN 2004et was a bright SN type II on a fairly smooth host-galaxy
background.  The SN data were taken on the Minicam while the reference image
was acquired on the Keplercam.  Figure~\ref{fig_sn04et} shows that the
cross-camera subtracted and the unsubtracted \emph{V}-band light curves agree within
the uncertainties.  At bright times, about two-thirds of the points agree to
better than 0.01 mag.  The largest discrepancy is 0.03 mag.  At faint times,
the galaxy light begins to contribute more, and the unsubtracted light curve
more is roughly 0.015 mag brighter with a scatter of 0.02 mag but when the SN
is bright, the cross-camera subtraction does not introduce any systematic
error.  

\subsubsection{Cross-Camera Subtraction Versus Same-Camera Subtraction} 

\begin{figure}
\begin{center}
\scalebox{0.80}[0.80]{
\plotone{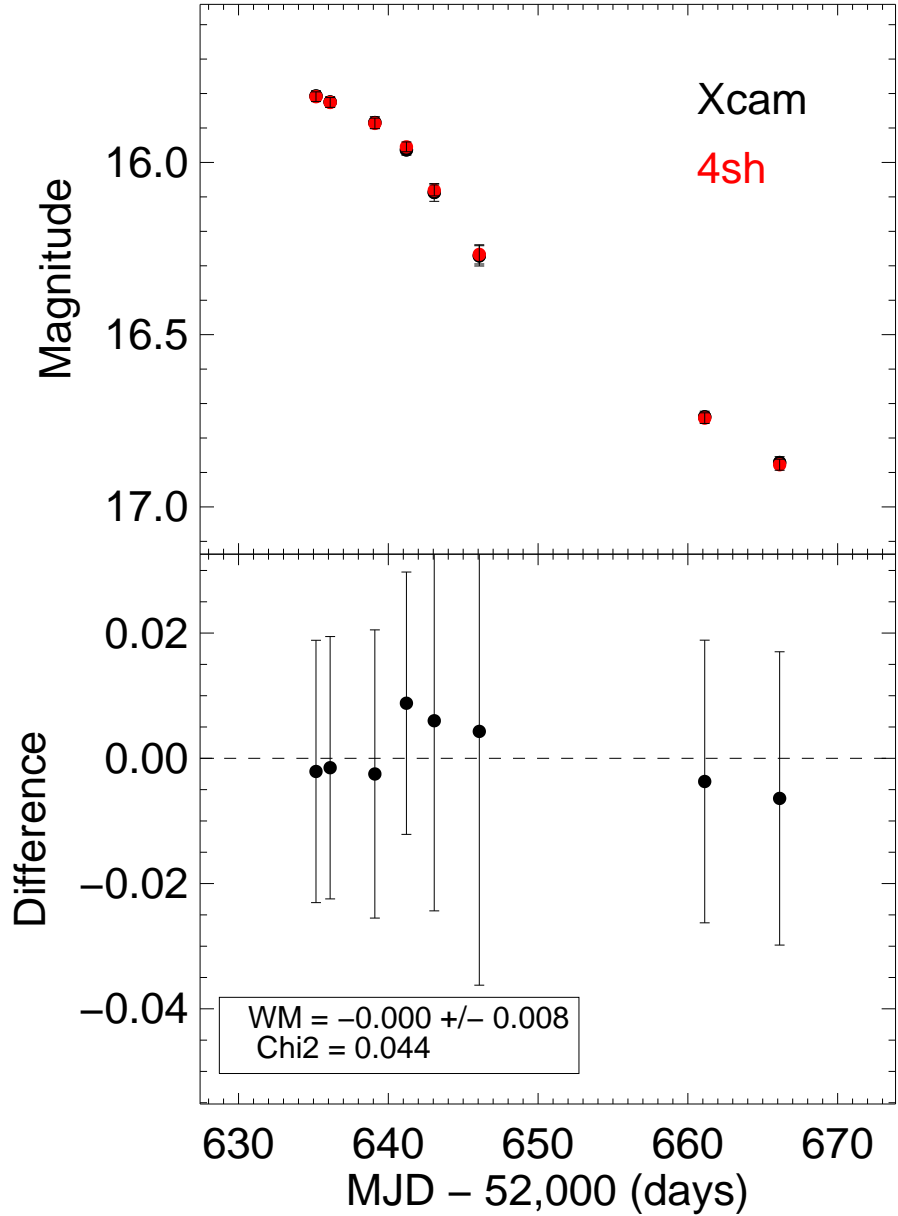}
}
\end{center}
\caption{Comparison of the $R$-band same-camera subtracted and cross-camera
subtracted light curves of SN 2002jy.  The agreement is good, bolstering 
confidence that
the cross-camera subtraction works reliably.
}
\label{fig_sn02jy}
\end{figure}

The SN data for SN 2002jy were obtained with the 4Shooter while reference
images were obtained with both the 4Shooter and the Keplercam.  There is
excellent agreement between the \emph{BVRI} same-camera and the cross-camera
subtracted light curves, with typical agreement at the 0.01 mag level or
better.  The 4Shooter $U$-band reference image was of inferior quality and
could not be used.  The scatter is much smaller than the error bars because the
only difference in the two light curves is the reference images, while the data
images are the same.  The $R$-band comparison is shown in
Figure~\ref{fig_sn02jy}, with all of the points agreeing to better than 0.01
mag.  The slight differences in the light curves may be due to slight flux and
seeing differences in the two reference images.  Other factors include poorer
4Shooter cosmetic properties and different responsivities between the cameras.
We also found good agreement in other SN, bolstering our confidence that the
cross-camera subtraction process was reliable.

\subsection{External Consistency Checks}

Comparisons with published photometry are made to check for consistency in
comparison star calibration and SN Ia light curves.  Differences in
instruments, reduction techniques and comparison star calibration are some of
the factors leading to disagreements in the photometry from different
telescopes of the same SN Ia.  Typical disagreement of SN Ia photometry is
roughly 0.02 to 0.05 mag in \emph{BVR} around maximum light with larger
discrepancies more common at later times and in $U$ and $I$ at all times.  J06
present photometry comparisons from different groups for several SN Ia and find
typical agreement of several hundredths mag in most cases but worse in others.  

SN 1999ee is an example where data was taken by two different telescopes on the
same mountain and reduced in the exact same fashion with the same comparison
star magnitudes \citep{stritzinger02}.  The only difference was in the two
telescopes/detectors.  The differences in the two \emph{UBVRI} light curves near
maximum light were -0.14, -0.01, -0.04, +0.04 and -0.03 mag, respectively and
slightly larger a month later.  S-corrections integrate the convolution of the
natural system passband and SN spectrum and subtract the convolution of the
standard system passband and SN spectrum.  Because of the non-stellar spectra
of SN Ia, especially at later times, they can be used instead of star-derived
color terms to more accurately place the SN photometry on the standard (or some
other) system.  
S-corrections were applied, resulting in partial improvement for some bands and
worsening in $R$, leading to the conclusion that accurate passbands must be
determined if S-corrections are to be of use.  Similarly to SN 1999ee,
\citet{suntzeff00} discusses the disagreement in the photometry of SN 1998bu
from two telescopes that he reduced in the same manner with the same comparison
stars.  He finds a color difference between the two telescopes of
$\delta(B-V)=0.12$ mag at late times, when the second telescope began
observing.  He finds that S-corrections would be able to correct this.

As another case, \citet{krisc03} applies S-corrections to SN 2001el.  These are
on the order of a few hundredths mag.  Most S-corrections in the literature are
roughly in the range 0.0 to $\pm0.1$ mag.  In general, S-corrections can be
large or small, depending on the mismatch between the natural system and
standard system passbands and the spectral properties of the SN.  SN 2005cf
\citep{wang08a} is an example where the disagreement between different
telescope's light curves is still 0.02 to 0.03 mag after S-corrections.  This
shows that differences of a few hundredths mag can occur even when many, but
not all, of the systematic differences are not present and great care is taken
in acquiring and processing the data.  


As a check on our photometry pipeline, in \S2.7.1, we first run the raw data of
17 SN Ia from J06 through our photometry pipeline and compare the results.  The
J06 photometry pipeline mainly differs from the CfA3 pipeline in the reference
image subtraction software.  Then, for six objects from the literature, in
\S2.7.2-2.7.7, we compare our CfA3 comparison stars and light curves with the
published values.  Of particular worth are the cases where values are presented
from two or more telescopes.  Overall, we find good consistency between
our comparison star calibration and light curves in comparison to those from
other groups for the same objects.  This is of great importance when combining
multiple data sets together to calculate dark energy properties.  For purposes
of comparing two sources of SN photometry, we define 'excellent' agreement for
all bands (except $U$) as better than 0.02 mag difference, 'good' as 0.02 to
0.04 mag, 'adequate' as 0.04 to 0.06, and 'poor' as greater than 0.06.  For
$U$, these values are doubled.  These labels will be used in the following
sections as a way of giving a measure of the agreement.
Applying one of these labels to a light curve comparison should be understood
to apply to each point.  For example, two light curves with excellent agreement
would have no points disagreeing by more than 0.02 mag.  Mostly-good agreement
would mean most of the points differ by no more than 0.04 mag.  These labels
are consistent with typical differences in published photometry of the same SN
Ia from different groups, using different instruments (see above).  

\subsubsection{Running J06 Data Through CfA3 Pipeline}

\begin{figure}
\begin{center}
\scalebox{0.80}[0.80]{
\plotone{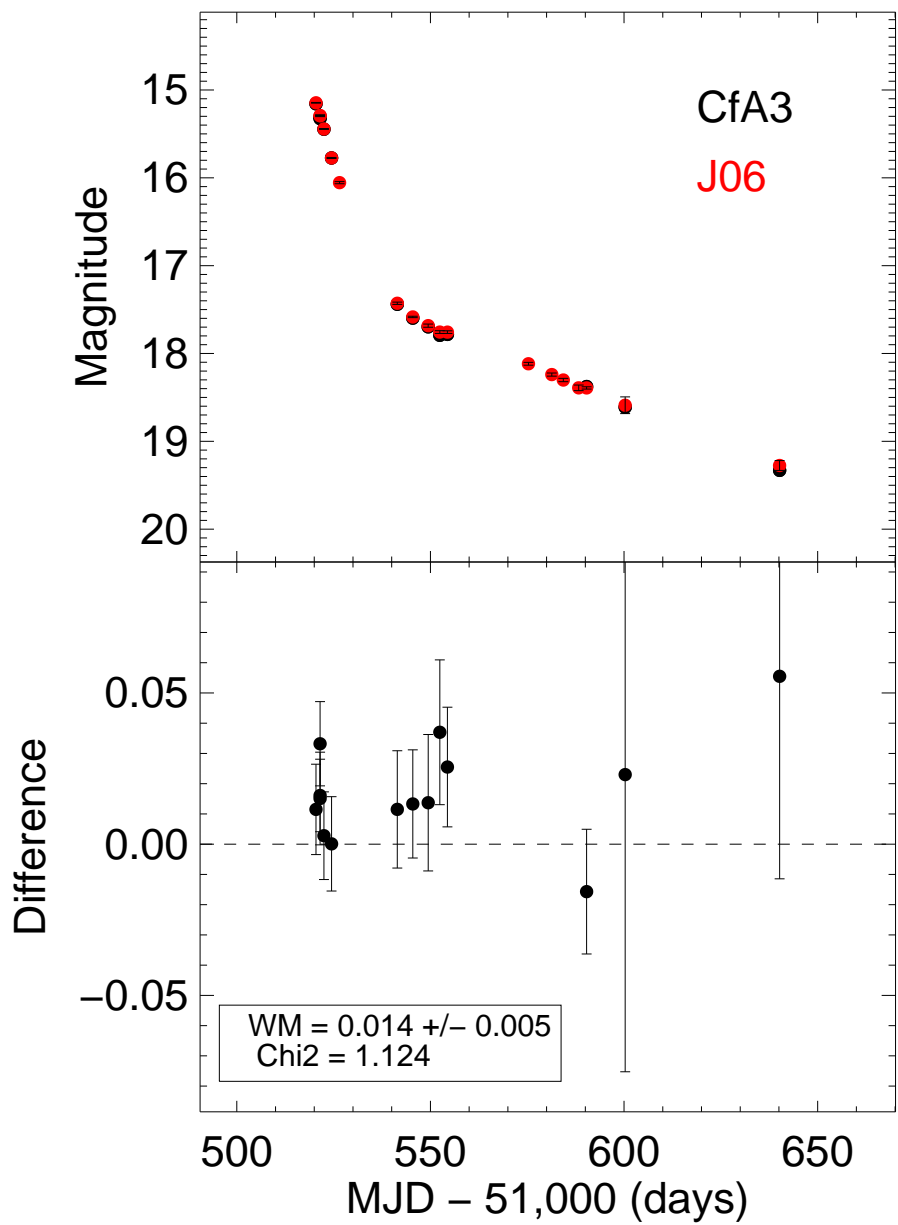}
}
\end{center}
\caption{Comparison of the CfA3 and J06 versions of SN 1999gh in $B$.  The
slight offset and scatter is typical of the 17 SN Ia from J06 that we ran
through the CfA3 pipeline, showing that there is a slight difference 
between the two pipelines.  However, there are both positive and negative
offsets and sometimes both at different phases of the light curves of the
17 objects, suggesting that the pipelines are not introducing a definite
positive or negative bias to all photometry.  
}
\label{fig_Bsn99gh}
\end{figure}

\begin{figure}
\begin{center}
\scalebox{0.80}[0.80]{
\plotone{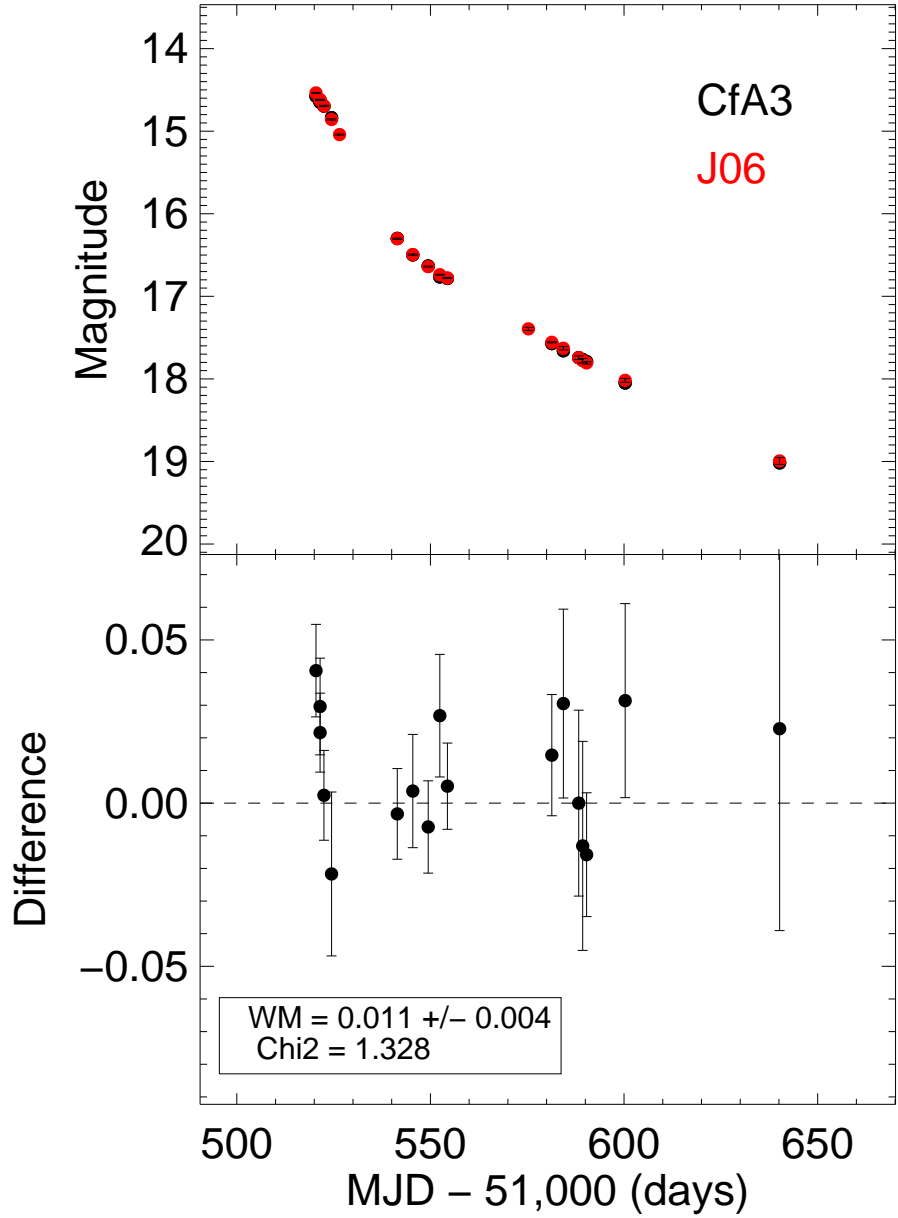}
}
\end{center}
\caption{Comparison of the CfA3 and J06 versions of SN 1999gh in \emph{V}, showing
generally good agreement but with some scatter.
}
\label{fig_Vsn99gh}
\end{figure}

To test our photometry pipeline, we ran the 4Shooter \emph{BVRI} raw data for 17 SN
Ia from J06 through our pipeline and compared the results with those obtained
by J06, whose reduction methods differed from ours in some ways.  The main
difference was in the reference image subtraction software.  J06 used the ISIS
subtraction package \citep{alard98} as modified by B. Schmidt.  We think that
our more modern subtraction software is an improvement but this needs an
empirical test.  Another difference was our use of DoPHOT while J06 used
aperture photometry.  We note that that we did not correct for fringing on the
$I$-band images for these 17 objects (not to be confused with the fact that we
did for the CfA3 $I$-band images).  Running the $U$-band data through
successfully requires more effort and does not add to determining the
consistency of the J06 photometry pipeline with ours so we did not do it.  The
agreement was typically good or excellent around maximum light with typically
good or adequate agreement at later times.  A weighted mean of the difference
between the two light curves was computed for each SN in each band.   The
average of these weighted means was $0.001\pm0.019$ for all bands while in each
separate band they were $0.010\pm0.015$ in $B$, $0.005\pm0.011$ in \emph{V},
$0.001\pm0.011$ in $R$ and $-0.012\pm0.029$ in $I$.  The larger disagreement in
$I$ is possibly due to our lack of fringe correction for these comparison
objects.  The two pipelines agree at $\lesssim~0.01$ mag in each band.
These differences make clear the advantages of using one large sample that has been
reduced by the same pipeline.  It also illustrates the advantage that the CfA3
sample has since one of the significant high-z samples, ESSENCE, uses the same
pipeline.  That each band's offset is consistent with zero is important since
no significant, net offset is being introduced relative to the CfA2 photometry,
much of which was used to train the various light curve fitters that are used
in H09.  We later will adopt 0.01 mag as the systematic uncertainty for our
pipeline.  As a representative example (neither the best nor the worst), we
show the \emph{BV} light curve comparisons of SN 1999gh in Figures~\ref{fig_Bsn99gh}
and~\ref{fig_Vsn99gh}.  

\subsubsection{SN 2003du}

SN 2003du has four sets of photometry:  CfA3 provides one and
\citet{stanishev07}, \citet{leonard05}, and \citet{anupama05} provide the other
three.  We will refer to these as CfA3, STAN, LEO and ANU.  Our comparison
stars agree to better than 0.02 mag with STAN in \emph{BVI}, with LEO in \emph{BVRI} and
with ANU in $I$.  ANU is fainter in \emph{BV}, by several hundredths magnitude
compared to the other three.  STAN is the most discrepant in $R$.  In $U$, CfA3
is fainter than STAN, but within the uncertainty.  CfA3 pairings with the other
groups are in as good or better agreement than most of the other non-CfA3
pairings.

\begin{figure}
\begin{center}
\scalebox{0.80}[0.80]{
\plotone{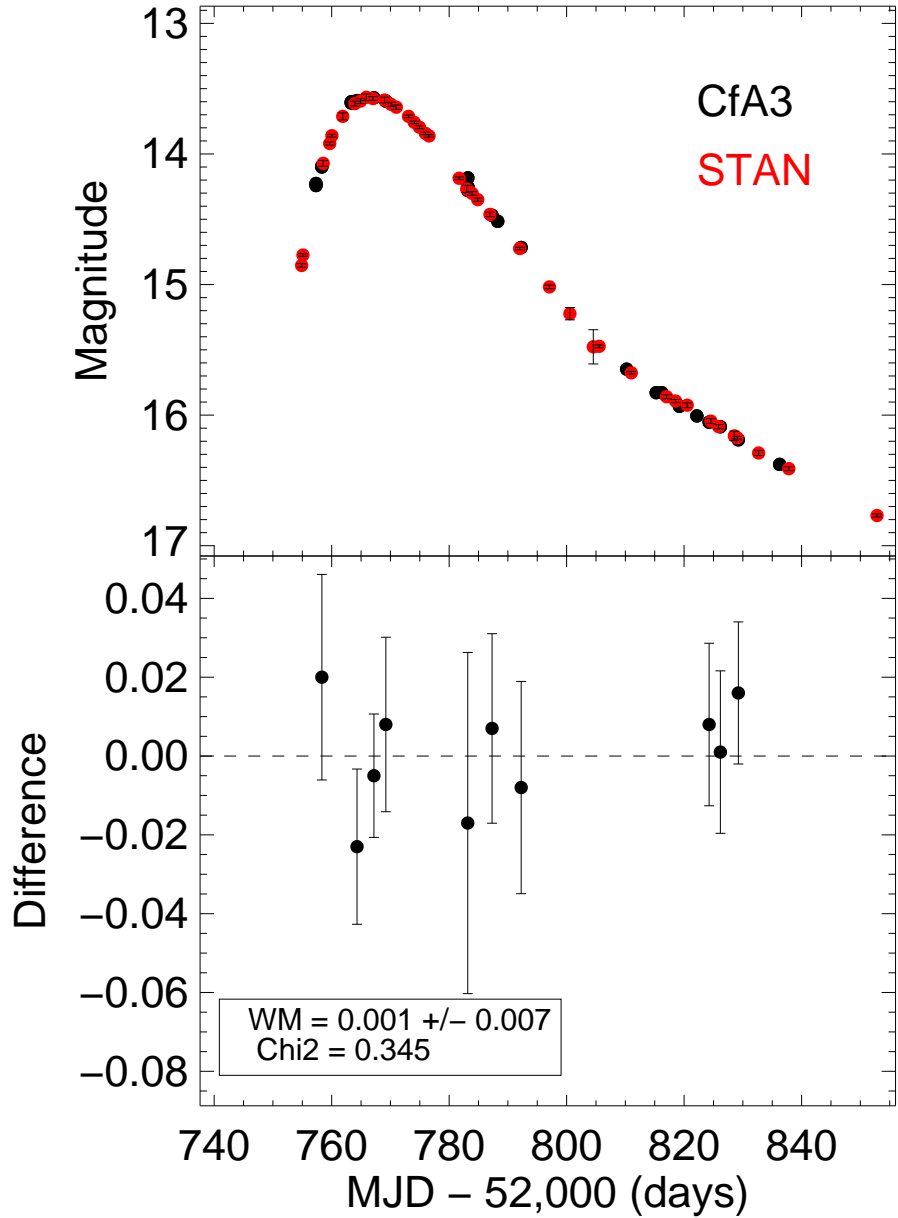}
}
\end{center}
\caption{Comparison of the CfA3 and STAN \emph{V}-band light curves of SN 2003du,
showing good agreement.  This is a good sign since the STAN light curve comes
from several telescopes and most of the points have been S-corrected while
the CfA3 light curve is from one detector and has not been S-corrected.  
}
\label{fig_sn03du}
\end{figure}

The agreement of the CfA3 light curve is mostly excellent with STAN in \emph{V}, as
seen in Figure~\ref{fig_sn03du}, and it is good in \emph{BRI}, with a few points
brighter by about 0.1 mag at late times in $I$.  STAN and LEO have good
agreement in $R$ but STAN is systematically brighter in \emph{BV} by a few
hundredths and in $I$ by 0.06 mag.  CfA3 has good and excellent agreement with
LEO in $R$ and $B$ but is brighter by about 0.03 mag in \emph{V} and 0.07 mag in
$I$.  ANU is fainter than CfA3 by around 0.04-0.05 mag in \emph{BV} around max, in
rough agreement with the offset in the three comparison stars in common, but
agrees better at late time.  CfA3 and ANU have good agreement in \emph{RI}.  It is
interesting to note that most of the STAN values we compare with were from
various telescopes and S-corrected, suggesting that the CfA3 color-term
transformations do a decent job of converting to the standard passbands.

\subsubsection{SN 2002bo}

\citet{krisc04b} (KRIS) and \cite{benetti04} (BEN) present optical light curves
of SN 2002bo.  In looking at the five comparison stars in common, those of KRIS
are brighter than those of BEN by several hundredths mag.  BEN only presents
\emph{BVRI} data while CfA3 and KRIS also present $U$-band data. For the three
comparison stars in common between KRIS and CfA3, excellent agreement is found
in \emph{VRI}, while KRIS is fainter by about 0.03 mag in $B$ and 0.02 mag in $U$,
with a large scatter in $U$.  CfA3 agrees excellently with BEN in $B$.  CfA3 is
brighter than BEN in \emph{VRI} by 0.03 to 0.05 mag, but KRIS is even brighter,
compared to BEN.  Overall, in the bands presented by all three groups, CfA3 and
KRIS agree best in \emph{VRI} while CfA3 and BEN agree best in $B$.

KRIS presents both S-corrected light curves and color-term-corrected light
curves.  We find that our light curves agree slightly better with the
S-corrected \emph{BV} light curves.  The $B$ difference, CfA3$-$KRIS, is 
$0.026\pm0.021$ and in \emph{V} it is $0.004\pm0.031$.  In $R$, there
is little difference in which KRIS light curve we compare to.  The agreement
is good except for one poor point.  In $I$, there is
one KRIS point--their minimum point between the two $I$-band peaks--that seems
excessively low compared to the shape of the neighboring points and so we
consider this to be a bad point and not really worth comparing.  It differs
from ours by about 0.25 mag.  Besides this point, our points around the first
peak agree have good agreement with the S-corrected KRIS version and
good-to-adequate agreement with the un-S-corrected version.  There are two
other KRIS $I$ points at later times where the un-S-corrected points are
fainter than CfA3 by about 0.1 mag but the (same) S-corrected points are only
$\sim$0.02 and 0.04 mag fainter.  Finally, in $U$, both KRIS light curves are
narrower and fainter than the CfA3 and there is very poor 
agreement--it is difficult to say which is right but
the CfA3 light curve has a smoother shape and smaller error bars so it may be
better.

KRIS and BEN generally have good-to-adequate agreement with a few poor points.
The bad KRIS $I$-band point mentioned above 
is about 0.35 mag fainter than BEN, confirming that it is likely an
aberrant point.  Similarly, BEN and CfA3 have good-to-adequate agreement with
a few poor points.

\subsubsection{SN 2002bf}

\citet{leonard05} (LEO) present \emph{BVRI} photometry of SN 2002bf.  We only have
two comparison stars in common but they are consistent with zero difference in
all bands except for one of the stars in $I$, where the LEO star is fainter by
slightly more than the $1\sigma$ level.  As LEO note, the SN is only $4".1$
from its host galaxy's center, resulting in subtraction difficulties.  As a
result, both CfA3 and LEO light curves are somewhat choppy and several points
have large error bars.  The agreement in photometry is good to excellent in the
two brightest $B$ points and in the brightest \emph{V} point, with adequate-to-poor
agreement in most of the other \emph{V} points.  The CfA3 light curves are mostly
fainter in $R$, and in $B$ after the two brightest points.  In $I$, there are
some phases of good agreement and some of large disagreement ($\sim0.2$ mag). 

\subsubsection{SN 2005hk}
 
\citet{phillips07} present optical comparison star photometry and light curves
for SN 2005hk from four different sources:  CSP, CTIO, KAIT and SDSS.  We make
no comparison with SDSS since their values are in $ri$ and not in $r'i'$.  We
compare the CfA3 comparisons stars with the other telescopes' values where the
stars and passbands are the same.  In \emph{BV}, CfA3, CSP and
CTIO all agree within 0.01 mag in the mean while KAIT is most different from
the other three but still in good-to-excellent agreement. In $U$, CfA3 and KAIT
show the best agreement (0.01 mag) and CfA3 and CTIO differ by $\sim0.03$ mag.
The CfA3 comparison star photometry was calibrated on three nights, with
excellent agreement, showing internal consistency.  The excellent agreement
with CSP and CTIO and the good-to-excellent agreement with KAIT suggest that
our comparison star calibration is reliably on the standard system.

\citet{phillips07} compare the \emph{BV} KAIT and CSP light curves of SN 2005hk.
The KAIT light curve is not reference-image subtracted and this is probably the
largest source of the discrepancies of several hundredths mag 
after S-correction.  The CSP light curve used a
$g'$ SDSS reference image for $B$ and a $g'+r'$ reference image for \emph{V} so this
might be a small source of inaccuracy.  They also compare the CSP and SDSS
light curves in $ri$.  S-corrections bring the two data sets into better 
agreement with some scatter at the level of a couple hundredths left over.

The CfA3 \emph{V} light curve show good-to-excellent agreement, and excellent
agreement in the mean, with CSP.  The CfA3 $B$ light curve mostly shows
good-to-excellent agreement with CSP, with a few poorly agreeing points, mostly
at late times.  The CfA3 $r'i'$ light curves are about 0.03 mag fainter, with a
few poorly-agreeing points at later times.  Possible sources of disagreement
are the reference images used, passband differences and no S-corrections for
the CfA3 light curves.

\subsubsection{SN 2005am and SN 2005cf}

\citet{li06} present \emph{UBV} comparison star calibration of SN 2005am and 2005cf.
Using the stars in common for SN 2005am, CfA3 is in excellent agreement with Li
in the mean in U ($\sim0$ mag), but with a scatter of 0.08 mag.  CfA3 is
fainter by about 0.03 and 0.02 mags in $B$ and \emph{V}, respectively, but
consistent with zero difference.  No comparisons of the SN 2005am light curves
is made since the Li light curve is not reference-image subtracted and the CfA3
light curve is.  There are three \emph{UBV} comparison stars in common for SN 2005cf
and there is excellent agreement in each band:  less than 0.01 mag mean 
difference for \emph{BV} and less than 0.025 mag in $U$.

\subsubsection{SN 2006X}

We compare our \emph{UBV} comparison stars and light curves with those from
\citet{wang08b}, who present data from KAIT, CTIO, and the Teramo-Normale
Telescope (TNT).  The comparison stars in common differ in \emph{UBV}, respectively,
as follows:  $-0.043\pm0.116$, $0.008\pm0.041$, $-0.013\pm0.032$.

The CfA3 $B$-band light curve agrees excellently in the mean with the composite
light curve from WANG but several of the individual points are only in 
adequate or poor agreement.  In \emph{V} band, CfA3 has excellent agreement
with KAIT and CTIO in eight points, adequate in another, and poor (0.07 to
0.09 mag) in three more.  TNT is systematically fainter than
CfA3 by about 0.06 mag.  The TNT data is the most discrepant of any of the four
groups.

\subsubsection{External Comparisons Summary}

Our external comparisons show that our light curves are consistent with those
from other groups at about the same level that other groups' light curves are
consistent with each other, typically at the ``good" level of a few hundredths
mag.  This is important so that multiple nearby samples can be combined, both
for training light-curve fitters and for calculating cosmological results.
The reduction pipeline and the calibration to the standard system are usually
the largest sources of systematic uncertainty between groups and so we
emphasize the value of one large, homogeneously-observed and reduced sample.
The CfA3 sample fits the bill, with the slight exception of the three cameras
used, and the change from \emph{RI} filters on the 4Shooter to $r'i'$ on the later
cameras.  The CfA3-4Shooter subsample, with 64 objects, and the CfA3-Keplercam
subsample, with 116 objects, each individually qualify as large,
homogeneously-observed and reduced samples, with no qualifications.

\subsection{Systematic Uncertainty}

The uncertainties of our comparison star photometry take into account both
statistical uncertainty and the uncertainty in the photometric transformation
and so no systematic error needs to be added.  However, in calculating the
differential photometry zeropoint to be added to the SN and its uncertainty (by
calculating the weighted mean of the difference of the instrumental and
calibrated magnitudes of the comparison stars) the individual star errors are
treated as if they were purely statistical.  The zeropoint error is roughly
equal to the typical individual comparison star uncertainty divided by the
square-root of the number of stars.  This usually lowers the
differential-photometry zeropoint uncertainty below the amount of systematic
uncertainty contained in the individual comparison star values and so our SN
photometry uncertainties can be considered accurate in the natural system.  A
systematic uncertainty, on the order of the photometric transformation
uncertainty (including photometric zeropoint and color term uncertainties),
should be added when comparing CfA3 standard system SN photometry with that of
other groups.  We estimate this systematic uncertainty to be 0.02 mag in
\emph{BVRIr'i'} and 0.06 mag in $U$.  This uncertainty was not added to the
comparisons of literature and CfA3 SN Ia light curves above.

The other main source of systematic uncertainty for the CfA3 light curves comes
from the photometry pipeline.  The CfA3 photometry pipeline does propagate all
of the uncertainties of the various stages to the final SN measurement and are
reliable in comparing with other CfA3 pipeline measurements.  However, there
may be differences compared to other photometry pipelines.  Based on the
scatter in the weighted means of the CfA3-versus-J06 comparisons of the 17 SN
Ia chosen from J06 (0.019 mag) and an estimated reference-image flux zeropoint
uncertainty of 0.005 mag, we attribute a systematic uncertainty of 0.02 mag to
our pipeline photometry in \emph{BVRIr'i'}.  For $U$, we estimate this to be 0.04
mag.  This is larger than the best agreement seen between our subtracted and
unsubtracted light curves for the same object and may be an overestimate if the
CfA3 pipeline is an improvement (especially in the subtraction stage) over the
J06 pipeline.  

Adding the photometric transformation and CfA3 pipeline systematics in
quadature gives a total systematic uncertainty of roughly 0.03 mag in
\emph{BVRIr'i'} and 0.07 mag in $U$.  The $U$ measurements should be
used with caution.  We emphasize that this level of systematic uncertainty is
typical of the literature SN Ia photometry as well and has the potential to
lead to systematic errors in derived cosmological quantities, such as the dark
energy equation of state parameter, $w$.  If there is a 0.03 mag offset from
the true distance modulus in the nearby sample then this would roughly give
rise to an error in $w$ of $\sim$0.06-0.08, roughly the same size as current
statistical uncertainties in $w$ (H09).  Using a homogeneously-observed and
reduced sample can reduce this systematic uncertainty.

As in H09, the ``OLD" sample refers to the list of SN Ia from
\citep{jha07}.  For $0.01\leq z_{CMB} \leq0.15$, the four different light curve
fitters, SALT, MLCS2k2 ($R_V=1.7$), MLCS2k2 ($R_V=3.1$), and SALT2, produce
CfA3 Hubble residuals that are fainter on average than the OLD sample by
$0.017\pm0.027$, $0.018\pm0.027$, $0.038\pm0.027$, and $0.056\pm0.032$ mag,
respectively.  The average of these ($\sim0.03\pm0.03$ mag) is similar to our
systematic uncertainty estimate, but consistent with no offset.  Part of the
difference in mean residuals is probably due to slightly different SN Ia
populations being sampled, and how the light-curve/distance fitters interacts
with them .  Another part is possibly due to some sort of systematic difference
in photometry.  It should be noted that the standard deviation of the Hubble
residuals is similar between the CfA3 and OLD samples, suggesting that the CfA3
photometry is about as intrinsically consistent as the OLD sample, although
other effects contribute to this as well.

\section{SN Ia Light Curve Properties}

\begin{figure}
\scalebox{1.00}[1.00]{
\plotone{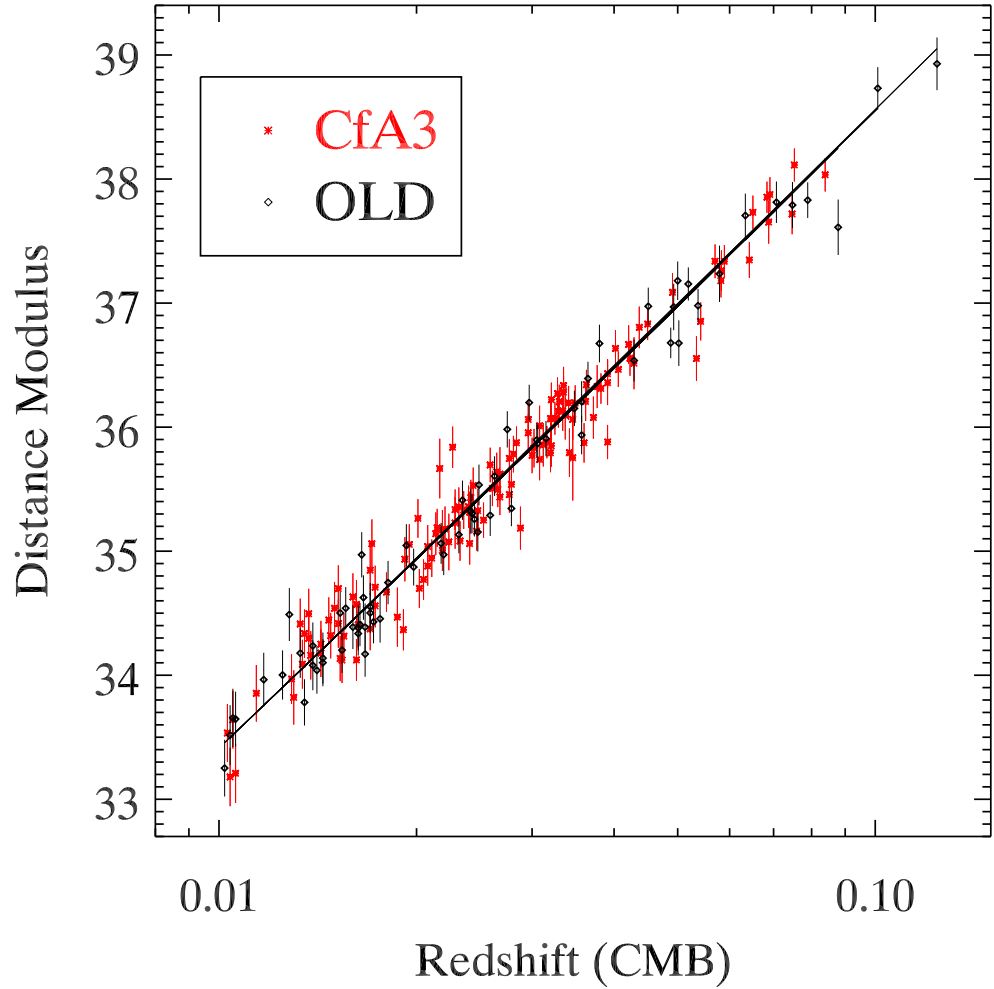}}
\caption{Hubble diagram of the CfA3 (red) and OLD (black) nearby SN Ia.
Distance moduli from H09 using MLCS2k2 ($R_V=1.7$).  The dispersion
is 0.20 mag and the solid line is the distance modulus for a 
($\Omega_M=0.27$, $\Omega_\Lambda=0.73$) universe.  
}
\label{fig_hubblediagram}
\end{figure}

We examine some of the SN Ia light curve properties of the CfA3 and OLD
samples.  In H09, the CfA3 and OLD samples are fit with MLCS2k2, using the
Galactic value of $R_V=3.1$ and $R_V=1.7$ (see \citet{conley07} for additional
discussion).  $R_V=3.1$ leads to an overestimate in the host extinction while
$R_V=1.7$, chosen to remove the trend in Hubble residuals versus $A_V$ for the
CfA3 sample, does not.  We believe that $R_V=1.7$ MLCS2k2 extinction estimates
are closer to reality so we choose here to use the light curve shape parameter,
$\Delta$, and the calculated host-galaxy extinction, $A_V$, from the $R_V=1.7$
MLCS2k2 fits.  A nearby Hubble diagram is presented in Figure
\ref{fig_hubblediagram} with the CfA3 points in red and the OLD points in
black.  This plot includes objects with a large range of extinction ($A_V<1.5$)
and light-curve shape ($-0.4\leq\Delta\leq1.6$), including the less-luminous,
1991bg-like objects.  The scatter for the CfA3+OLD sample is 0.20 mag, similar
to the scatter of CfA3 and OLD separately.  

We also take well-sampled light curves from the CfA3 and OLD samples that have
data before maximum light in $B$ as measured by MLCS2k2 and measure properties
directly from them, thus eliminating any dependency on a model or template
light curve (see Tables \ref{table_lcproperties1} and
\ref{table_lcproperties2}).  K-corrections are applied and Milky Way reddening
\citep{schlegel98} and time-dilation are removed, placing these in the rest
frame but without any host-galaxy reddening correction.  A high-order
polynomial (typically order 5 to provide enough flexibility to match the light
curve shape before and after maximum) is fit to the rest-frame \emph{UBV} light
curves, smoothly passing through a large majority of the light-curve points,
well inside the error bars, with only a few outliers in the more jagged light
curves.  

\subsection{\emph{UBV} Decline Rates, $\Delta m_{15}$}
 
Where possible, the time of maximum and the magnitude at those times 
are measured in each of the \emph{UBV} bands from the polynomial fits just 
described.  The $U$ and \emph{V} values are also
measured at the time of $B$ maximum, where possible, to give the color
at maximum light.
To measure the
peak magnitude for a light curve in a given band, we require a point before
and after maximum, with a separation of no more than $\Delta t = 9.2$ days.
Most of the
light curves have much smaller separations but we want to measure the
peak color for as many light curves as possible.  We choose to use
this wider separation limit because the peak magnitude
calculations are fairly insensitive to the calculated time of maximum,
as seen by removing points from light curves with many points around
maximum and recalculating the peak magnitude.
The uncertainty of each peak magnitude is taken to be the average of the
uncertainties of the nearest points, one before and one after, plus
$0.005\Delta t$, added in quadrature.

We also measure the decline rate, $\Delta m_{15}$,
in \emph{UBV} from the polynomial fits to the light curves.
For this measurement, we also require that there be a point before and
after $t=+15$ days, with a
separation no greater than 12.2 days.  The uncertainty in $\Delta m_{15}$ is
approximated by summing (in quadrature) the uncertainty at peak and the
uncertainty
at $t=+15$ days.
The magnitude at $t=+15$ days is sensitive to the time of maximum and we estimat
e
its uncertainty to be the sum in quadrature of the
average uncertainty of the surrounding points
and the product of the slope of the light curve at $t=+15$ days times
the uncertainty in $t_{max}$.
The slope of the light curve is roughly $0.1$ mag per
day since a typical SN Ia declines $\sim~1$ mag in 15 days with most of that
decline coming over the last 10 days.  The uncertainty in $t_{max}$ is
estimated to be roughly 0.15 times the separation in time between the
two surrounding points, determined by removing points from well-sampled
light curves and noting the effect.

Because we are mostly concerned with
looking for trends in the plots, we choose to include the light curves with
larger separations and lower precision measurements.
For more precise purposes we advise the reader to have caution in
using the peak magnitude and $\Delta m_{15}$ values derived from
light curves with time
separations between the two points near maximum of greater than six
days.
The time separations are
listed in Table \ref{table_lcproperties2} for this purpose.  We note that
only $13.7\%$ of the peak magnitudes and $12.5\%$ of the $\Delta m_{15}$
values come from light curves with separations of more than six days.

\subsection{Decline Rates, Colors at Maximum Light and $\Delta$}

\begin{figure}
\scalebox{0.90}[0.90]{
\plotone{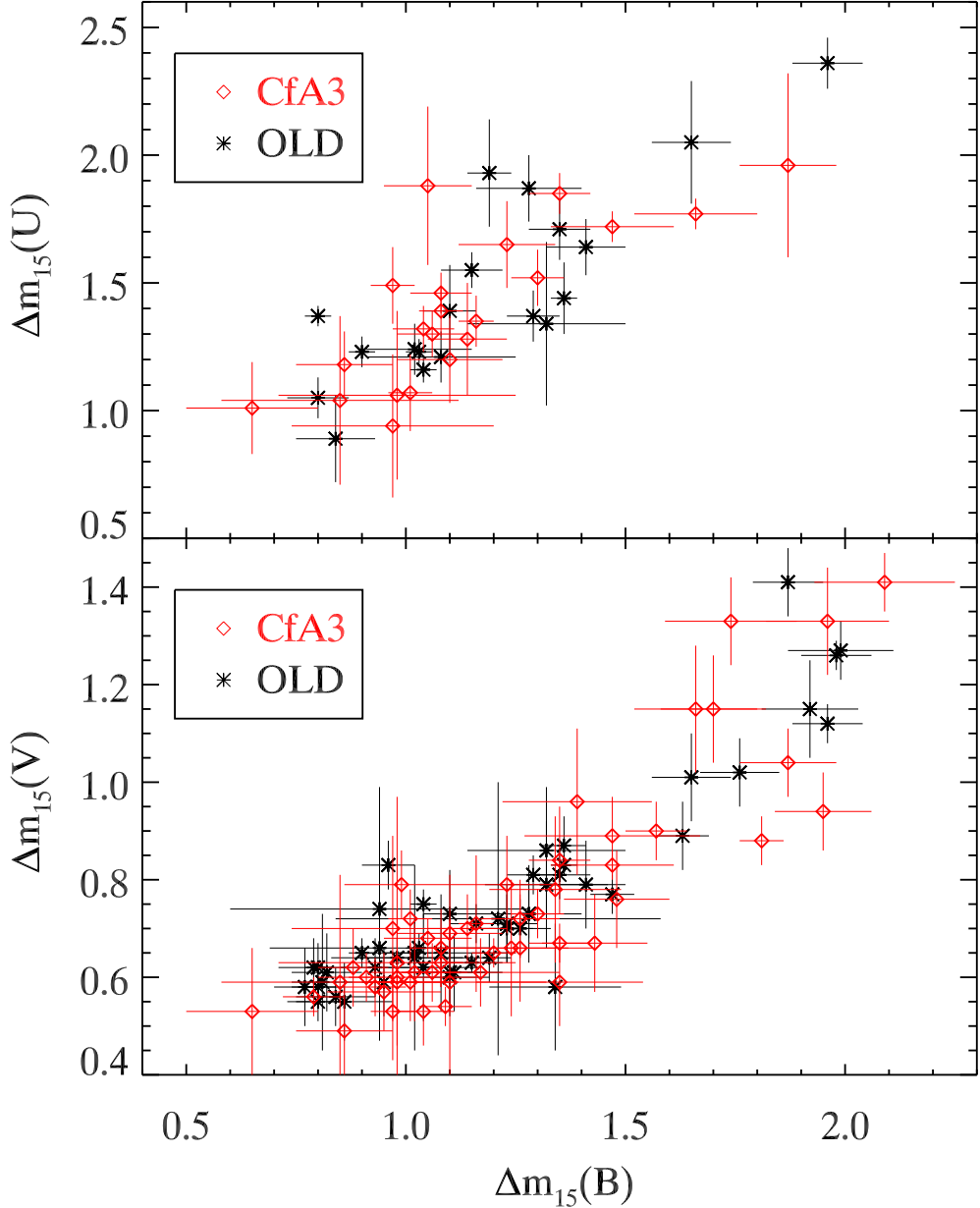}
}
\caption{Plots of $\Delta m_{15}(U)$ and  $\Delta m_{15}(V)$ versus \dm,
measured directly from suitable SN Ia light curves.  A linear correlation
is seen in the $U$ and $B$ data.  A tight correlation exists in $B$ and \emph{V}
between the slow and normal decliners while the faster decliners (many
of which are 1991bg-like objects) show
larger scatter.
}
\label{fig_dm15ubv}
\end{figure}

In Figure~\ref{fig_dm15ubv}, the 15-day decline rates in $U$ and \emph{V} are
plotted against that of $B$.  There is a fairly tight cluster of points, with a
few outliers, in the bottom-left (slower decline) portion of the \emph{V}-versus-$B$
panel with the fast decliners in $B$ showing a wide range of \emph{V} decline rates
in the upper right portion.  The fastest \emph{V} decliners ($\Delta m_{15}(V) >
1.2$) are all 1991bg-like SN Ia (which can be identified as such by strong
\ion{Ti}{2} lines in their spectra).  A linear relation between $B$ and \emph{V}
decline rates cannot describe the data well.  A quadratic relation does a
better job.  The upper panel shows a roughly linear relation between the $U$
and $B$ decline rates, with a larger amount of scatter than in the
\emph{V}-versus-$B$ comparison.  This same qualitative effect was seen in the
stretch relations of J06.

\begin{figure}
\scalebox{0.90}[0.90]{
\plotone{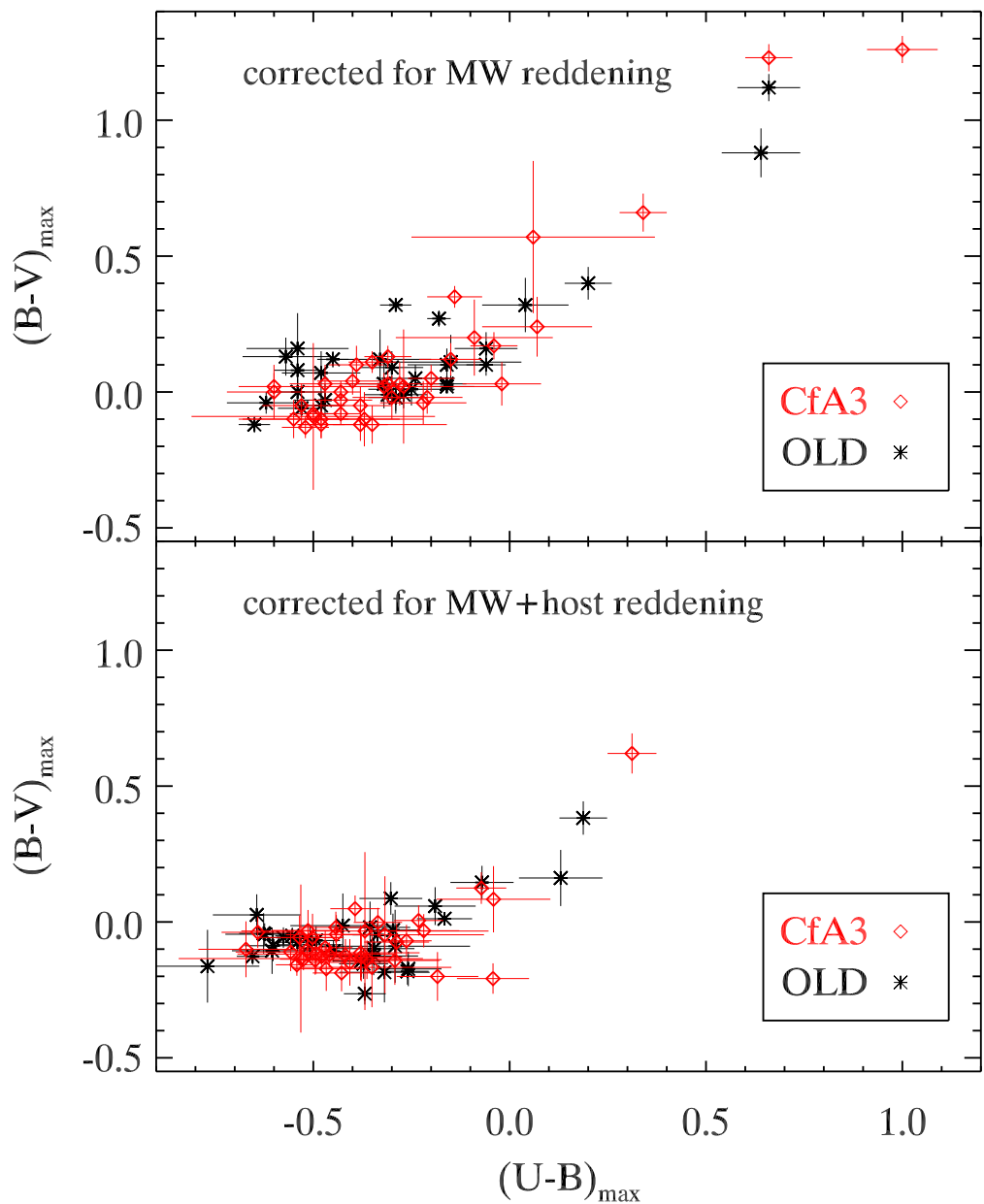}
}
\caption{Peak SN Ia color, before and after MLCS2k2 ($R_V=1.7$) correction for
host reddening.  Milky Way reddening is removed from both panels.  The
bottom panel gives a good idea of intrinsic $B-V$ and $U-B$ color at $B$
maximum.  The three reddest objects in the bottom panel are, in order of 
increasing redness:  SN 1986G, 1999by and 2005ke.
}
\label{fig_bvub}
\end{figure}

The $U-B$ and $B-V$ colors at $B$ maximum, directly measured from the light
curves with Milky Way reddening correction but no host-galaxy reddening
correction, are plotted in the top panel of Figure~\ref{fig_bvub}.  In the
bottom panel, they are corrected for the host galaxy reddening as measured by
MLCS2k2 ($R_V=1.7$), and mostly form a tight cluster with
$-0.2<(B-V)_{max}<0.1$ and $-0.7<(U-B)_{max} <0.0$.  This behavior is similar
to that seen in J06.

\begin{figure}
\scalebox{0.90}[0.90]{
\plotone{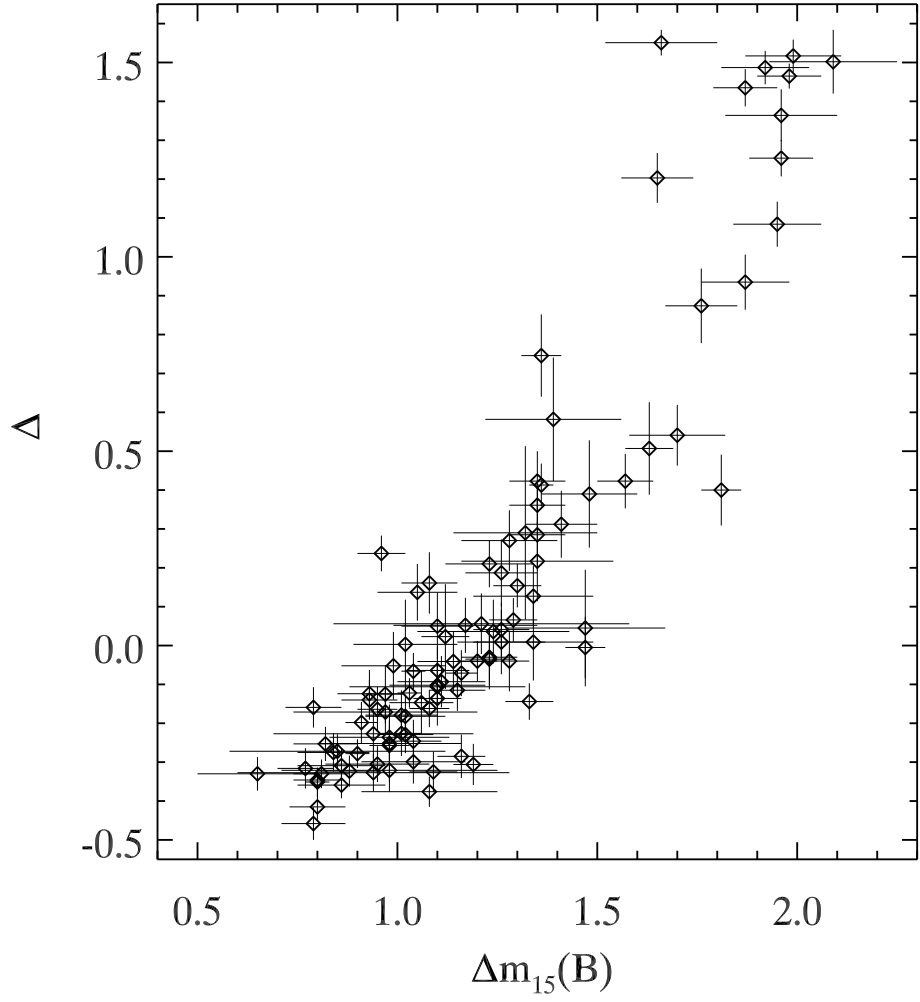}
}
\caption{
MLCS2k2 $\Delta$ versus \dm.  Fairly good correlation between
the two except at the largest values of \dm, where many of the objects
are 1991bg-like.
}
\label{fig_deltadm15}
\end{figure}

In Figure~\ref{fig_deltadm15}, the MLCS2k2 light-curve shape and luminosity 
parameter, $\Delta$, is plotted against
the directly-measured $\Delta m_{15}(B)$.  There is a linear
correlation for the slow and moderate decliners.  The fast decliners have
a wide range in $\Delta$.

In Figure~\ref{fig_bvdelta}, the host-galaxy corrected peak colors are plotted
versus $\Delta$ and $\Delta m_{15}(B)$.  Immediately standing out is the large
range of red intrinsic $(B-V)_{max}$ colors amongst the fast decliners, similar to what
was found by \citet{garnavich04}.  The redder fast decliners are all
1991bg-like objects 
while many of the relatively-bluer (though still red) fast
decliners are the more intermediate 1986G-like objects.  The $(B-V)_{max}$
versus $\Delta$ panel does suggest three interesting (though arbitrary)
groupings of points.  The slow and normal decliners form one group, with
typical peak color of $(B-V)_{max}\sim-0.1$.  The moderately-high $\Delta$
objects, including 1986G-like and some seemingly spectroscopically ``normal"
objects, form a second grouping with a typical color of 0.1.  Finally, the
1991bg-like objects have a typical $(B-V)_{max}$ color of 0.5.  Another way of
grouping the objects in this plot is by their slope in color versus $\Delta$.
Those with $\Delta < 0.1$ have a flat slope while those with $0.1 < \Delta <
1.3$ have a slightly positive slope and those with $\Delta > 1.4$ have a very
steep slope.  The 1991bg-like objects are clearly distinct in $B-V$ color
at maximum.

\subsection{Histograms of Color, Extinction and $\Delta$}

\begin{figure}
\scalebox{0.90}[0.90]{
\plotone{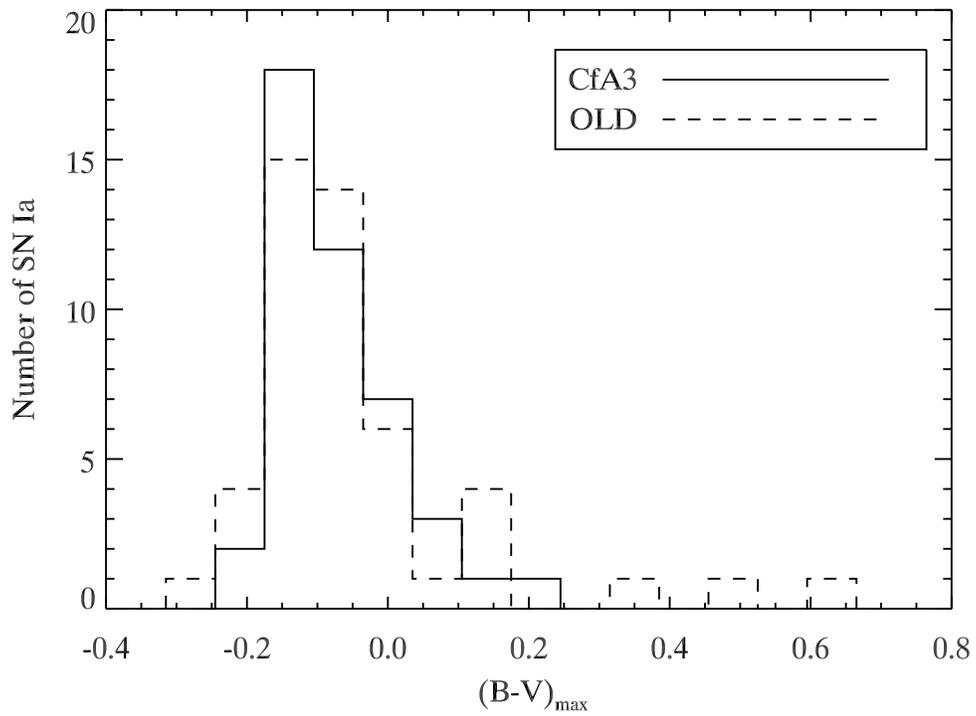}
}
\caption{44 CfA3 and 48 OLD SN Ia with $z_{CMB}\geq0.01$ had reliable, direct 
$(B-V)_{max}$ measurements.  The two samples show excellent agreement--a KS
test gives $87\%$ probability that the two samples are drawn from the 
same distribution.
}
\label{fig_bvhist}
\end{figure}

Our intention is to use light-curve fitters that were trained on objects
from the OLD sample, apply it to the CfA3 sample, and use the combined
OLD+CfA3 sample to infer useful cosmological properties.
To compare how similar the OLD and the CfA3 samples are, 
we take the nearby SN Ia with $z_{CMB}\geq0.01$.  We plot a histogram of
the directly-measured, intrinsic peak color, $(B-V)_{max}$, corrected for host
reddening (see Figure~\ref{fig_bvhist}).  There are 44 CfA3 and 48 Old objects
and their distributions are fairly close--the two-sample Kolmogorov-Smirnov 
(KS)
test gives a probability of $87\%$ that they come from the same distribution.

\begin{figure}
\scalebox{0.90}[0.90]{
\plotone{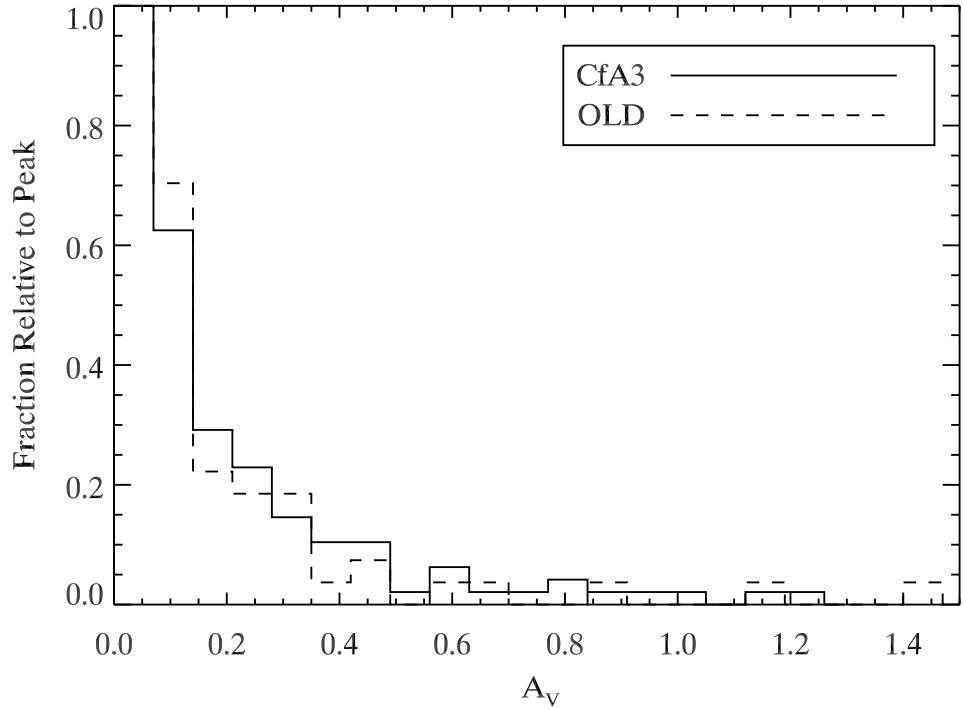}
}
\caption{Histograms of the CfA3 and OLD SN Ia \emph{V} extinction as calculated
by MLCS2k2 ($R_V=1.7$).  133 CfA3 and 70 OLD, useful for cosmological
measurements, with $z_{CMB}\geq0.01$ and good
MLCS2k2 fits are included.  The distributions are normalized to their
respective peaks and good agreement is seen-a KS test gives $74\%$ probability
that the two samples are drawn from the
same distribution.
}
\label{fig_avhist}
\end{figure}

A histogram of $A_V$, as measured by MLCS2k2, is shown in
Figure~\ref{fig_avhist}.  There are 133 CfA3 and 70 OLD objects, all with
$z_{CMB}\geq0.01$ and with good MLCS2k2 fits.  Their distributions are quite
similar--the KS test gives a probability of $74\%$ that they are from the
same distribution.

\begin{figure}
\scalebox{0.90}[0.90]{
\plotone{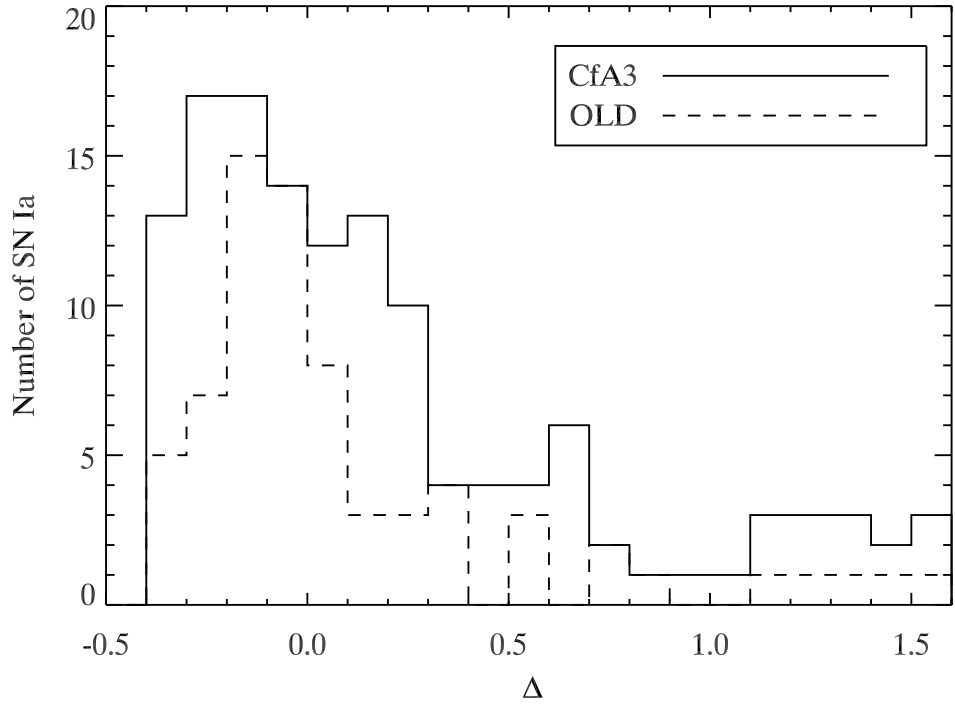}
}
\caption{Histograms of the 133 CfA3 and 70 OLD SN Ia values of $\Delta$, all
at $z_{CMB}\geq0.01$.  The CfA3 sample shows a wider distribution in $\Delta$,
probably due to our prioritization of slow and fast decliners.  
}
\label{fig_delhist}
\end{figure}

Finally, in Figure~\ref{fig_delhist}, a histogram of $\Delta$ for the two
samples is shown.  In part, due to our prioritization of fast and slow
decliners in our observing strategy, CfA3 has a wider distribution, most
noticeably in the most-negative end and in the range around $\Delta=0.2$.  The
increased number of highly-negative $\Delta$ SN Ia is especially helpful for
dark energy studies where these brighter objects are preferentially found due
to magnitude-limited high-z searches.  In H09, the three high-redshift
samples used (ESSENCE, SNLS and Higher-Z) do not have any objects with
$\Delta>0.75$.  The KS test gives a probability of
$9.4\%$ that the two samples are from the same distribution.  This should not
be interpreted as an intrinsic difference in the underlying populations from
which the CfA3 and OLD samples were drawn but, rather, as evidence of different
selection effects, mostly related to which objects were chosen to be observed.

\subsection{Intrinsic Absolute Magnitude}

Using a $\Lambda$CDM universe with ($\Omega_m=0.3$, $\Omega_{\Lambda}=0.7$,
$h=0.7$), we calculate distance moduli based on the CMB redshifts for all the
CfA3 and OLD SN Ia with $z_{CMB}\geq0.01$ and good MLCS2k2 ($R_V=1.7$) fits for
which we could also directly measure the peak \emph{UBV} magnitudes.  We subtract
off the distance moduli in the top panels of Figures~\ref{fig_babsmag},
\ref{fig_vabsmag} and \ref{fig_uabsmag} to show the absolute magnitudes before
correction for host-galaxy extinction.  In the bottom panels, we also subtract
off the MLCS2k2-calculated host extinction to give a good estimate of the
intrinsic absolute magnitudes of our sample.  

The two most noticeable things are that the extinction correction does a good
job of reducing the scatter and that $\Delta$ is more tightly correlated 
with intrinsic absolute magnitude than is $\Delta m_{15}(B)$.  Part of this
may be due to the larger uncertainties in our \dm~measurements.

Focusing on the lower right panel of Figure~\ref{fig_babsmag}, there appears
to be a linear relation between $B$-band intrinsic absolute magnitude and
$\Delta$, out to $\Delta\approx1.2$.  In \emph{V}, a linear relation with a negative
quadratic term in $\Delta$ would do well.  The three faint and high-$\Delta$
objects are 1991bg-like objects.  We also note that if the $z=0.01$ cut were
not made that there would be several more 1991bg-like objects from our direct
light curve measurements with $\Delta \approx 1.5$ and $M_B\approx-17$,
confirming the faint and separate nature of 1991bg-like SN Ia.  In order to
include these, MLCS2k2 uses a positive term in $\Delta$ but this comes at the
expense of underestimating the luminosity of the SN Ia in the range
$0.7<\Delta<1.2$ (see Figure~\ref{fig_vabsmag} where the solid line shows the
dependence of intrinsic absolute magnitude versus $\Delta$ for the MLCS2k2
model light curves).  It may be better to avoid 1991bg-like objects altogether
for use in the light-curve-shape/magnitude relation, or to at least treat them
separately.  Removing them can be accomplished by spectroscopic identification
of 1991bg-like features or by simply removing all objects above a certain
$\Delta$ and/or intrinsic peak color.

\section{Conclusion}

The goal of our research was to produce a large sample of nearby SN Ia
light curves that would better sample the whole range of SN Ia and serve
to reduce the statistical and systematic uncertainties in dark energy
calculations using SN Ia.

\begin{figure}
\begin{center}
\scalebox{0.75}[0.75]{
\plotone{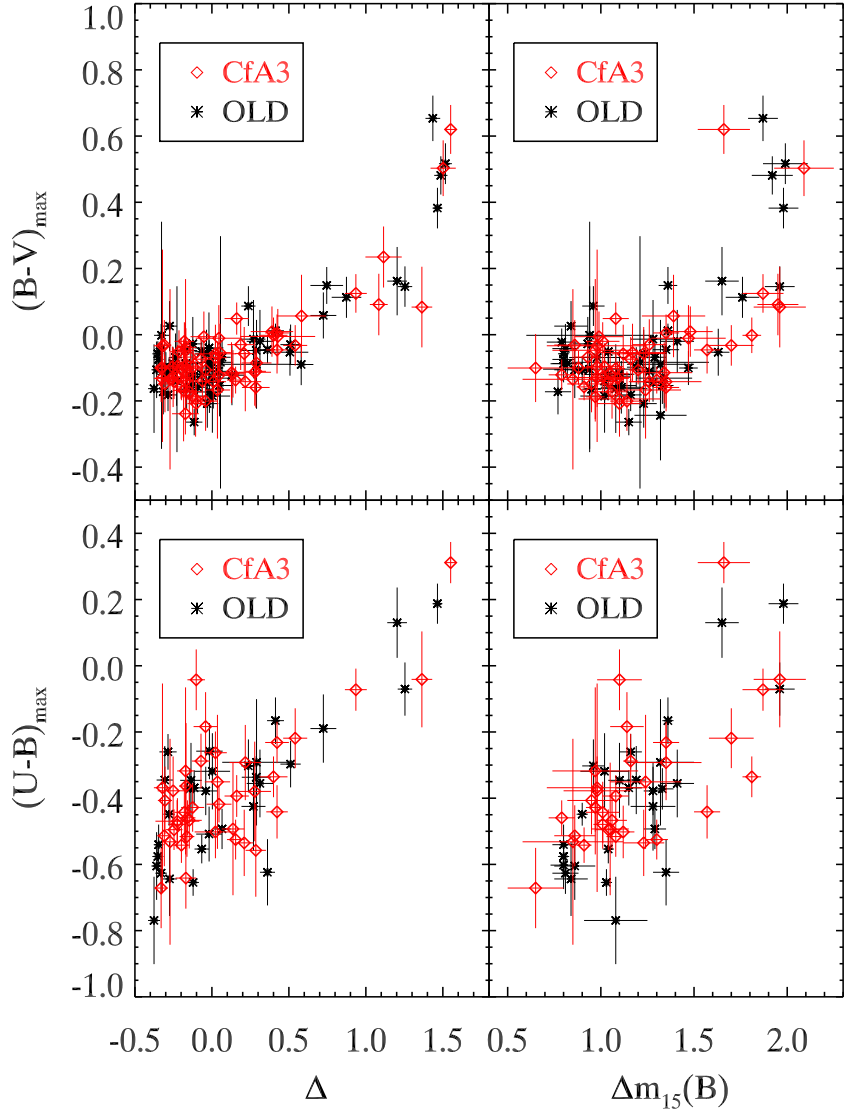}
}
\end{center}
\caption{The $B-V$ and $U-B$ peak colors, corrected for MLCS2k2-calculated
reddening is shown for all well-measured objects, with no cut on redshift.
The upper-left panel shows an upward curving distribution that can
be broken into three, somewhat arbitrary, groupings.  The slow and normal 
decliners with $-0.4\leq\Delta<0.7$ have a typical color of 
$(B-V)_{max}\approx-0.1$, although
the upward trend in color starts in the right portion of this group.
The second group has a typical color of 0.1 with $-0.7<\Delta<1.4$ and
includes objects similar to both 1992A and 1986G.  This second group may
be a ``transitional" group (or ``intersectional" if there are two underlying
groups), both photometrically and spectroscopically,
to the third group, consisting of 1991bg-like SN Ia, with $(B-V)_{max}\approx
0.5$ and $\Delta\approx1.5$.
}
\label{fig_bvdelta}
\end{figure}

\begin{figure}
\begin{center}
\scalebox{0.75}[0.75]{
\plotone{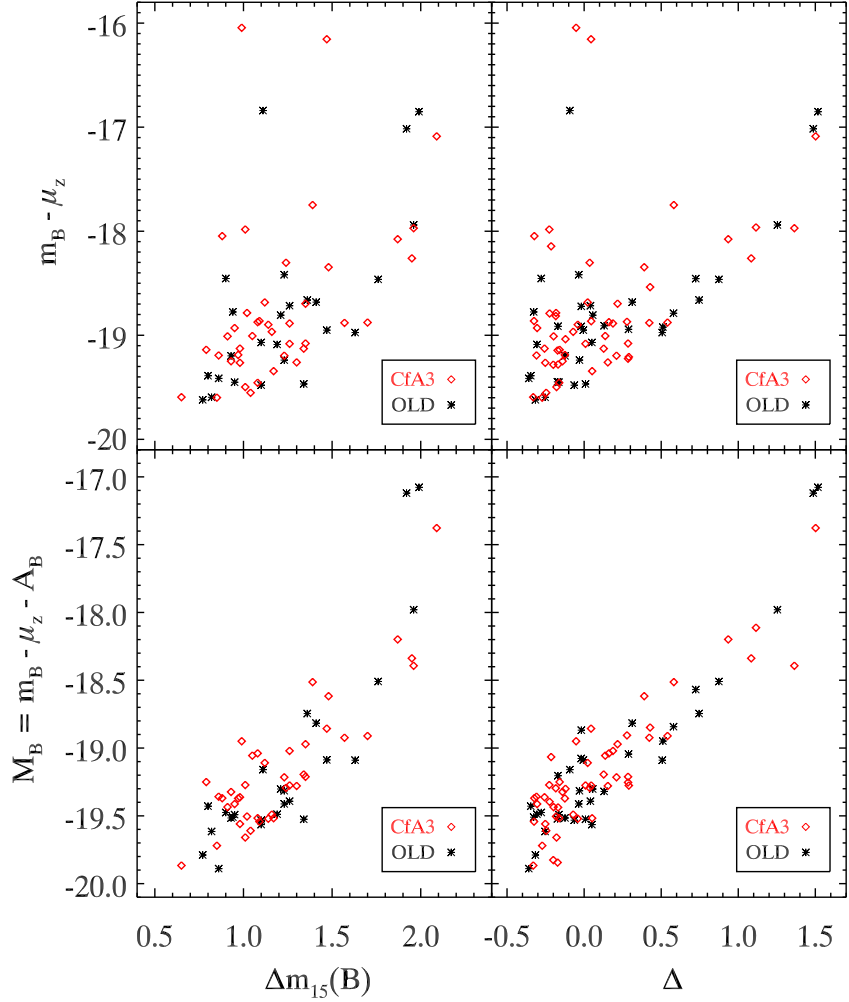}
}
\end{center}
\caption{SN Ia absolute magnitude versus \dm and $\Delta$.  The top 
panels show SN Ia peak apparent magnitude in $B$, 
directly measured from the suitable light curves, after
K-corrections and correction for MW extinction, minus the distance
modulus ($\Omega_M=0.3$, $\Omega_\Lambda=0.7$, $h=0.7$). 
All objects are at $z_{CMB}\geq0.01$ and error bars are omitted to not obscure
the data points.  The lower panels further subtract off the
host-galaxy extinction, $A_B$, as calculated by MLCS2k2 ($R_V=1.7$), giving
a good estimate of SN Ia intrinsic absolute magnitude, $M_B$.  This is plotted
against \dm and $\Delta$.  A linear trend is evident in both lower panels,
except for the faintest objects which are all 1991bg-like.  The relation
between $M_B$ and $\Delta$ is tighter than between $M_B$ and \dm.  If objects
below $z_{CMB}=0.01$ were included then three more 1991bg-like SN Ia would be
in the vicinity of ($\Delta=1.5$, $M_B=-17$), but with higher uncertainty
due to peculiar velocities.
}
\label{fig_babsmag}
\end{figure}

\begin{figure}
\scalebox{0.90}[0.90]{
\plotone{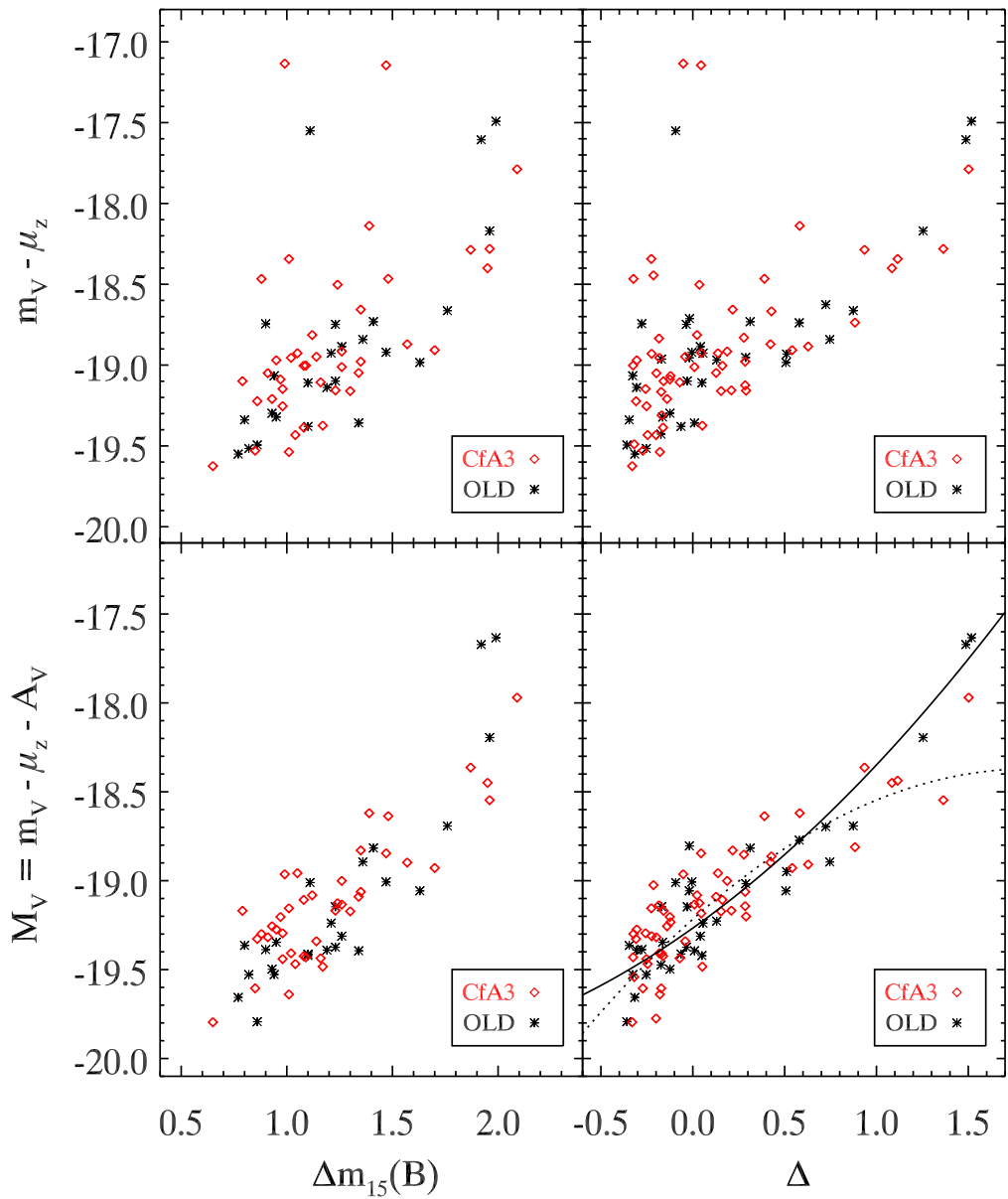}
}
\caption{SN Ia \emph{V} absolute magnitude versus \dm and $\Delta$.  Same as 
in Figure \ref{fig_babsmag} but for \emph{V}.  The solid line is the MLCS2k2 
model intrinsic absolute magnitude, M$_V(\Delta)$, from \citep{jha07} while the 
dotted line shows that a negative quadratic
fit may be better in \emph{V} if the three faintest, 1991bg-like, objects are not 
included.
}
\label{fig_vabsmag}
\end{figure}

\begin{figure}
\scalebox{0.90}[0.90]{
\plotone{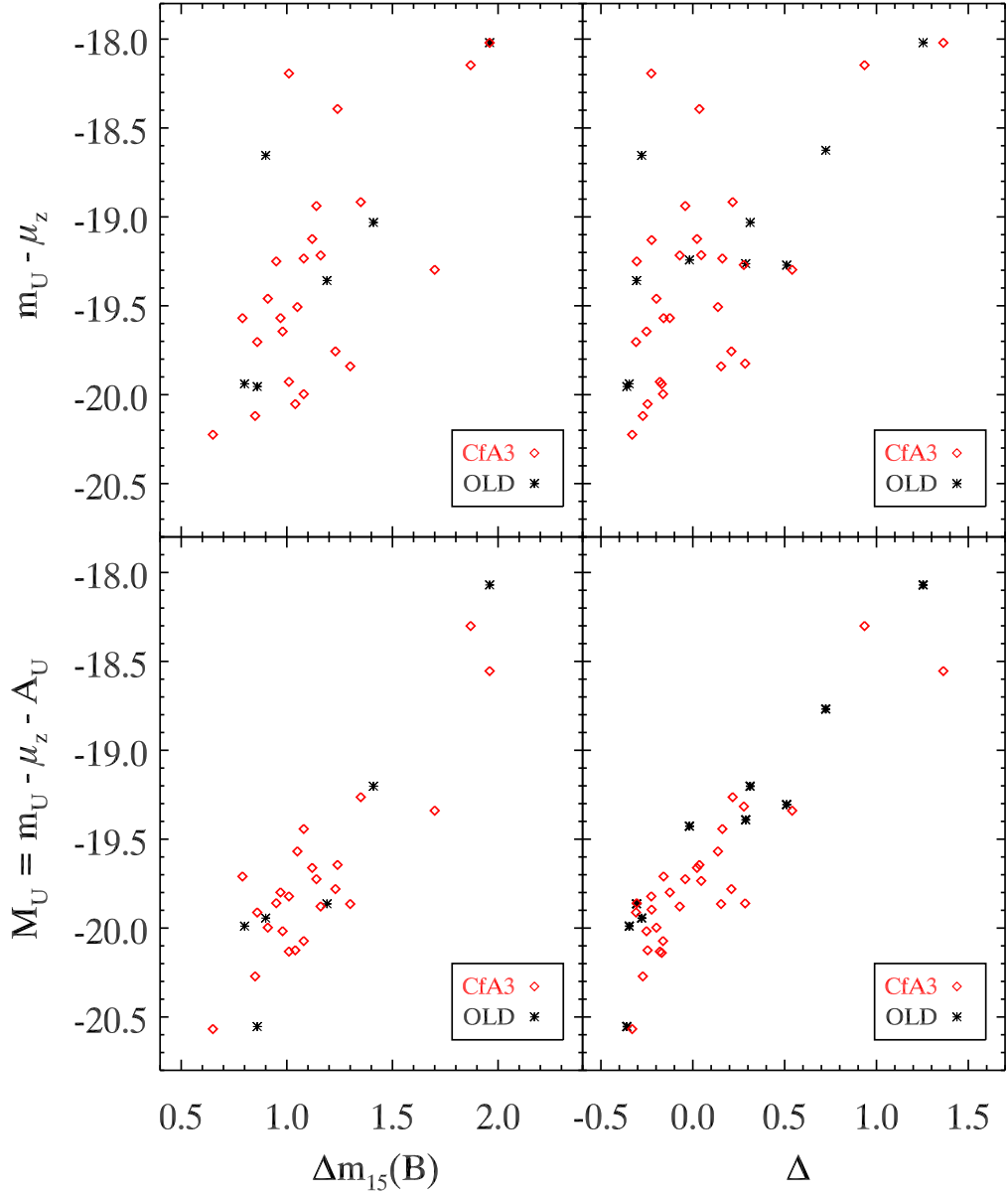}
}
\caption{SN Ia $U$ absolute magnitude versus \dm and $\Delta$.  Same as 
in Figure \ref{fig_babsmag} but for $U$.  There is fairly good correlation
between $M_U$ and light curve shape.
}
\label{fig_uabsmag}
\end{figure}

We have presented 185 nearby CfA3 SN Ia light curves, adding a significant
number of fast and slow decliners.  We have shown that the photometry is
internally consistent, and that it is externally consistent at roughly the same
level as seen in other nearby SN Ia photometry.  The intrinsic properties of SN
Ia have been explored in \emph{UBV}, confirming previous studies.  One of the
most important findings is that $(B-V)_{max}$ and the peak intrinsic magnitude
in $B$ and \emph{V} do not correlate well with light curve shape amongst the
fast decliners (see Figures~\ref{fig_bvdelta}, \ref{fig_babsmag}, and
\ref{fig_vabsmag}).  However, when the 1991bg-like objects are removed, the
remaining fast decliners still seem to be well correlated
with color and intrinsic absolute magnitude.  This suggests that 1991bg-like SN
Ia should be excluded from light-curve/distance fitter training samples and
from dark energy studies.  We believe that this will improve the performance of
fitters, like MLCS2k2, that have used them in their training samples.

The CfA3 sample is an important contribution to dark energy studies because it
is the largest homogeneously-reduced nearby sample, doubling the
cosmologically-useful sample.  The addition of the CfA3 sample to the
literature SN Ia and its effects on the dark energy calculations is explored in
H09.  H09 show that the statistical uncertainty in $w$ is reduced by a factor
of 1.2-1.3 by adding the CfA3 sample.  The CfA3 sample lowers the statistical
uncertainty on static dark energy measurements to the point where systematic
uncertainties begin to dominate.  The CfA3 sample can be used to lower
systematic uncertainties in dark energy studies in two ways.  First, either the
complete sample or the CfA3-Keplercam subsample can be used as a stand-alone
nearby sample that reduces or eliminates the uncertainties that arise from
different reduction pipelines, filters, cameras, and comparison-star
calibration amongst nearby SN Ia.  Second, it will be very useful for
retraining light curve fitters and making them more precise.  The two CfA3
subsamples, CfA3-Keplercam and CfA3-4Shooter, are the largest and
second-largest homogeneously-observed and reduced nearby samples to date.  
A large fraction of the CfA3 objects have spectra.  A
few dozen CfA3 objects also have near infrared photometry and will help
disentangle host-galaxy reddening from intrinsic SN Ia color.  This will lead
to a large decrease in the systematic uncertainty associated with host-galaxy
dust absorption.  Other large optical samples will soon be published too, with
dozens of objects overlapping the CfA3 sample, providing a good opportunity to
search for systematic errors and better combine data sets from different
groups.  We plan on publishing another 70-100 SN Ia (CfA4 sample) 
when the host-galaxy
reference images are obtained and comparison star calibration is
completed.  These will have been observed and reduced in the same way as the
CfA3-Keplercam subsample, and when added together will number roughly 200.  

Future nearby SN Ia studies should focus on reducing their statistical and
systematic photometric uncertainties by obtaining higher SNR measurements and
improved understanding of their passbands and SN Ia calibration.  The goal of
$1\%$ SN Ia photometry should be pursued so that more precise SN Ia measurments
can produce tighter limits on both static and especially dynamic models of dark
energy.  The largest hurdles to achieving $1\%$ photometry are in understanding
atmospheric transmission and instrumental passbands.  Efforts along the lines
of \citet{stubbs06} should be pursued and improved.  Use of calibrated
photodiodes and monochromatic light allows for a precise determination of the
system's transmission function.  The pixel-by-pixel CCD response can be
precisely determined by taking monochromatic flat fields at a
sufficiently-sampled range of wavelengths and measuring the intensity of this
light with the calibrated photodiode.  Additionally, improved treatment of
host-galaxy contamination of SN light should be developed, through improved
image-subtraction software and/or improved galaxy+SN models that measure both
without image subtraction.

\acknowledgments

We thank the staff at FLWO for their dedicated work in maintaining the 1.2m
telescope and instruments.  This work has been supported, in part, by NSF
grant AST0606772 to Harvard University.  AR thanks the Goldberg Fellowship Program
for its support.

\facility{\emph{Facility}:  FLWO:1.2m}


\clearpage
\begin{deluxetable}{llllll}
\tabletypesize{\scriptsize}
\tablecolumns{6}
\tablewidth{0pc}
\tablecaption{SN Ia Discovery Data}
\tablecomments{
Positions are a weighted mean of our measured SN positions, usually in $R/r'$ but occasionally from \emph{V}
when insufficient $R/r'$ data was available.  These are generally an improvement over the positions
reported by the discoverer.
}
\tablehead{ \colhead{SN Ia} & \colhead{Galaxy} &  \colhead{Discovery Date} & \colhead{Position} & \colhead{CBET/IAUC} & \colhead{Discoverer} 
}
\startdata
2001C&  Anon Gal       & 2001 01 04&06:59:36.138 +59:31:01.21&IAUC 7555&Puckett, Sehgal \\
2001G&  MCG +08-17-43  & 2001 01 08&09:09:33.215 +50:16:50.83&IAUC 7560&Armstrong \\
2001N&  NGC 3327       & 2001 01 21&10:39:58.060 +24:05:25.68&IAUC 7568&Chornock \\
2001V&  NGC 3987       & 2001 02 19&11:57:24.910 +25:12:09.49&IAUC 7585&Berlind \\
2001ah& UGC 6211       & 2001 03 27&11:10:29.838 +55:09:39.03&IAUC 7603&Puckett, Peoples \\
2001ay& IC 4423        & 2001 04 18&14:26:16.943 +26:14:55.24&IAUC 7611&LOTOSS \\
2001az& UGC 10483      & 2001 04 27&16:34:27.476 +76:01:46.34&IAUC 7614&Puckett, Peoples \\
2001bf& MCG +04-42-22  & 2001 05 03&18:01:34.059 +26:15:01.82&IAUC 7620&Armstrong \\
2001cp& UGC 10738      & 2001 06 19&17:11:02.600 +05:50:27.04&IAUC 7645&LOTOSS \\
2001da& NGC 7780       & 2001 07 09&23:53:32.741 +08:07:02.20&IAUC 7658&LOTOSS \\
2001eh& UGC 1162       & 2001 09 09&01:38:12.056 +41:39:18.95&IAUC 7712&Armstrong \\
2001en& NGC 523        & 2001 09 26&01:25:22.856 +34:01:30.06&IAUC 7724&LOTOSS; BAO \\
2001ep& NGC 1699       & 2001 10 03&04:57:00.349 $-$04:45:40.04&IAUC 7727&LOTOSS \\
2001fe& UGC 5129       & 2001 11 02&09:37:57.021 +25:29:40.84&IAUC 7742&Armstrong \\
2001fh& Anon Gal       & 2001 11 03&21:20:42.538 +44:23:53.14&IAUC 7744&LOTOSS \\
2001gb& IC 582         & 2001 11 20&09:59:00.960 +17:49:12.32&IAUC 7758&LOTOSS \\
2001gc& UGC 3375       & 2001 11 21&05:55:26.111 +51:54:34.22&IAUC 7759&LOTOSS \\
2001ic& NGC 7503       & 2001 12 07&23:10:43.298 +07:34:10.25&IAUC 7770&LOTOSS \\
2001ie& UGC 5542       & 2001 12 09&10:16:50.954 +60:16:53.32&IAUC 7771&Bincoletto \\
2002G&  Anon Gal       & 2002 01 18&13:07:55.285 +34:05:07.09&IAUC 7797&LOTOSS \\
2002ar& NGC 3746       & 2002 02 03&11:37:43.863 +22:00:34.47&IAUC 7819&LOTOSS \\
2002bf& Anon Gal       & 2002 02 22&10:15:42.314 +55:40:07.35&IAUC 7836&LOTOSS \\
2002bo& NGC 3190       & 2002 03 09&10:18:06.515 +21:49:41.63&IAUC 7847&Cacella; Hirose \\
2002bz& MCG +05-34-33  & 2002 04 03&14:24:40.524 +26:37:35.29&IAUC 7866&Puckett, Gauthier \\
2002cd& NGC 6916       & 2002 04 08&20:23:34.402 +58:20:47.30&IAUC 7871&Armstrong \\
2002ck& UGC 10030      & 2002 04 23&15:47:00.762 $-$00:59:24.92&IAUC 7884&LOTOSS \\
2002cr& NGC 5468       & 2002 05 01&14:06:37.652 $-$05:26:21.34&IAUC 7890&Kushida \\
2002de& NGC 6104       & 2002 06 01&16:16:30.334 +35:42:30.09&IAUC 7914&LOTOSS \\
2002dj& NGC 5018       & 2002 06 12&13:13:00.414 $-$19:31:08.56&IAUC 7918&LOTOSS \\
2002do& MCG +07-41-1   & 2002 06 17&19:56:12.853 +40:26:10.47&IAUC 7923&LOTOSS \\
2002dp& NGC 7678       & 2002 06 18&23:28:30.103 +22:25:38.05&IAUC 7924&Klotz \\
2002es& UGC 2708       & 2002 08 23&03:23:47.196 +40:33:53.56&IAUC 7959&LOTOSS \\
2002eu& Anon Gal       & 2002 08 30&01:49:43.549 +32:37:42.31&IAUC 7963&LOTOSS \\
2002fb& NGC 759        & 2002 09 06&01:57:48.869 +36:20:26.00&IAUC 7967&LOTOSS \\
2002fk& NGC 1309       & 2002 09 17&03:22:05.706 $-$15:24:02.99&IAUC 7973&Kushida; BAO \\
2002ha& NGC 6962       & 2002 10 21&20:47:18.592 +00:18:45.36&IAUC 7997&LOTOSS \\
2002hd& MCG -01-23-8   & 2002 10 24&08:54:03.366 $-$07:11:21.48&IAUC 7999&LOTOSS \\
2002he& UGC 4322       & 2002 10 28&08:19:58.804 +62:49:13.71&IAUC 8002&LOTOSS \\
2002hu& MCG +06-6-12   & 2002 11 07&02:18:20.027 +37:27:58.58&IAUC 8012&Boles \\
2002hw& UGC 52         & 2002 11 09&00:06:49.025 +08:37:48.64&IAUC 8014&LOTOSS \\
2002jy& NGC 477        & 2002 12 17&01:21:16.231 +40:29:55.27&IAUC 8035&Arbour; Vanmunster \\
2002kf& Anon Gal       & 2002 12 27&06:37:15.283 +49:51:10.87&IAUC 8040&Brady \\
2003D&  MCG -01-25-9   & 2003 01 06&09:38:53.551 $-$04:51:05.61&IAUC 8043&Puckett, Langoussis \\
2003K&  IC 1129        & 2003 01 11&15:32:01.832 +68:14:36.12&IAUC 8048&Puckett, Langoussis \\
2003U&  NGC 6365       & 2003 01 27&17:22:45.626 +62:09:50.67&IAUC 8059&Boles \\
2003W&  UGC 5234       & 2003 01 28&09:46:49.496 +16:02:37.77&IAUC 8061&LOTOSS \\
2003ae& Anon Gal       & 2003 01 23&09:28:22.570 +27:26:41.29&IAUC 8066&NEAT/Wood-Vasey et al. \\
2003ai& IC 4062        & 2003 02 08&13:00:58.699 +39:51:24.66&IAUC 8068&LOTOSS \\
2003cg& NGC 3169       & 2003 03 21&10:14:16.016 +03:28:02.01&IAUC 8097&Itagaki; Arbour \\
2003ch& UGC 3787       & 2003 03 21&07:17:57.890 +09:41:34.84&IAUC 8097&LOTOSS \\
2003cq& NGC 3978       & 2003 03 30&11:56:14.156 +60:31:19.67&IAUC 8103&Arbour \\
2003du& UGC 9391       & 2003 04 22&14:34:35.728 +59:20:03.93&IAUC 8121&LOTOSS \\
2003fa& Anon Gal       & 2003 06 01&17:44:07.733 +40:52:51.08&IAUC 8140&LOTOSS \\
2003hu& Anon Gal       & 2003 09 06&19:11:31.121 +77:53:34.91&IAUC 8196&Armstrong \\
2003ic& MCG -02-2-86   & 2003 09 16&00:41:50.334 $-$09:18:19.11&IAUC 8201&LOTOSS \\
2003it& UGC 40         & 2003 10 13&00:05:48.523 +27:27:08.62&IAUC 8225&Puckett, Cox \\
2003iv& MCG +02-8-14   & 2003 10 17&02:50:07.244 +12:50:45.29&IAUC 8226&LOTOSS \\
2003kc& MCG +05-23-37  & 2003 11 21&09:46:34.293 +30:39:19.27&IAUC 8242&LOSS \\
2003kf& MCG -02-16-2   & 2003 11 27&06:04:35.484 $-$12:37:42.87&IAUC 8245&LOSS \\
2004K&  ESO 579-G22    & 2004 01 19&14:23:39.802 $-$19:26:50.13&IAUC 8273&LOSS \\
2004L&  MCG +03-27-38  & 2004 01 21&10:27:04.125 +16:01:07.80&IAUC 8274&LOSS \\
2004ap& Anon Gal       & 2004 03 08&10:05:43.813 +10:16:16.68&IAUC 8300&LOSS \\
2004as& Anon Gal       & 2004 03 11&11:25:39.185 +22:49:49.05&IAUC 8302&LOSS \\
2004bg& UGC 6363       & 2004 04 07&11:21:01.542 +21:20:22.95&IAUC 8317&Armstrong \\
2004ef& UGC 12158      & 2004 09 04&22:42:10.021 +19:59:39.89&IAUC 8399&Boles; Armstrong \\
2004fu& NGC 6949       & 2004 11 04&20:35:11.608 +64:48:26.41&IAUC 8428&Arbour \\
2005M&  NGC 2930       & 2005 01 19&09:37:32.356 +23:12:02.20&IAUC 8470&Puckett, George \\
2005am& NGC 2811       & 2005 02 22&09:16:13.087 $-$16:18:15.97&IAUC 8490&Martin \\
2005cf& MCG -01-39-3   & 2005 05 28&15:21:32.225 $-$07:24:47.66&CBET 158 &LOSS \\
2005dv& NGC 5283       & 2005 09 04&13:41:04.478 +67:40:19.53&CBET 217 &Dainese, Dimai \\
2005el& NGC 1819       & 2005 09 25&05:11:48.744 +05:11:39.19&CBET 233 &LOSS \\
2005eq& MCG -01-9-6    & 2005 09 30&03:08:49.357 $-$07:02:00.24&IAUC 8608&LOSS \\
2005eu& Anon Gal       & 2005 10 04&02:27:43.239 +28:10:36.71&CBET 242 &LOSS \\
2005ew& Anon Gal       & 2005 10 04&03:39:23.747 +35:02:49.38&CBET 244 &Nearby SN Factory \\
2005hc& MCG +00-6-3    & 2005 10 12&01:56:47.950 $-$00:12:49.42&CBET 259 &SDSS-II \\
2005hf& Anon Gal       & 2005 10 25&01:27:05.991 +19:07:00.83&IAUC 8622&Quimby et al. \\
2005hj& Anon Gal       & 2005 10 26&01:26:48.397 $-$01:14:17.30&CBET 266 &Quimby et al. \\
2005hk& UGC 272        & 2005 10 30&00:27:50.879 $-$01:11:53.32&IAUC 8625&SDSS-II; LOSS \\
2005iq& MCG -03-1-8    & 2005 11 05&23:58:32.422 $-$18:42:32.97&IAUC 8628&LOSS \\
2005ir& Anon Gal       & 2005 10 28&01:16:43.796 +00:47:40.89&CBET 277 &SDSS II; Quimby et al. \\
2005kc& NGC 7311       & 2005 11 09&22:34:07.308 +05:34:06.04&IAUC 8629&Puckett, Sostero \\
2005ke& NGC 1371       & 2005 11 13&03:35:04.356 $-$24:56:38.93&IAUC 8630&LOSS \\
2005ki& NGC 3332       & 2005 11 18&10:40:28.219 +09:12:08.21&IAUC 8632&LOSS \\
2005ls& MCG +07-7-1    & 2005 12 09&02:54:15.914 +42:43:29.15&IAUC 8643&Armstrong \\
2005lu& MCG -03-07-40  & 2005 12 11&02:36:03.753 $-$17:15:49.50&IAUC 8645&LOSS \\
2005lz& UGC 1666       & 2005 12 24&02:10:49.727 +34:58:57.84&CBET 329 &Puckett, Gagliano \\
2005mc& UGC 4414       & 2005 12 23&08:27:06.277 +21:38:46.61&CBET 331 &THCA Supernova Survey  \\
2005ms& UGC 4614       & 2005 12 27&08:49:14.320 +36:07:46.72&CBET 343 &Puckett, Kroes \\
2005mz& NGC 1275       & 2005 12 31&03:19:49.910 +41:30:18.86&CBET 347 &Newton, Peoples, Puckett \\
2005na& UGC 3634       & 2005 12 31&07:01:36.659 +14:07:58.75&CBET 350 &Newton, Ceravolo, Puckett \\
2006B&  UGC 12538      & 2006 01 08&23:21:09.803 +33:24:00.74&CBET 356 &Puckett, Sostero \\
2006D&  MCG -01-33-34  & 2006 01 11&12:52:33.871 $-$09:46:30.56&CBET 362 &Colesanti, et al. \\
2006H&  Anon Gal       & 2006 01 15&03:26:01.533 +40:41:41.69&CBET 367 &Puckett, Sostero \\
2006N&  MCG +11-8-12   & 2006 01 21&06:08:31.268 +64:43:24.82&CBET 375 &Armstrong \\
2006S&  UGC 7934       & 2006 01 26&12:45:39.033 +35:05:12.16&CBET 379 &Puckett, Gagliano \\
2006X&  NGC 4321       & 2006 02 04&12:22:53.911 +15:48:31.65&IAUC 8667&Suzuki; Migliardi \\
2006ac& NGC 4619       & 2006 02 09&12:41:44.894 +35:04:07.93&IAUC 8669&LOSS \\
2006ah& Anon Gal       & 2006 02 09&13:46:13.540 $-$09:07:51.92&CBET 402 &Nearby SN Factory \\
2006ak& Anon Gal       & 2006 02 17&11:09:32.640 +28:37:51.63&CBET 408 &Tyurina, Lipunov et al. \\
2006al& Anon Gal       & 2006 02 19&10:39:28.238 +05:11:00.39&IAUC 8677&Holmes, Devore \\
2006an& Anon Gal       & 2006 02 21&12:14:38.749 +12:13:47.75&CBET 413 &Quimby, Castro \\
2006ar& MCG +11-13-36  & 2006 03 05&10:37:30.616 +65:00:57.78&CBET 420 &Boles \\
2006ax& NGC 3663       & 2006 03 20&11:24:03.432 $-$12:17:29.52&CBET 435 &LOSS \\
2006az& NGC 4172       & 2006 03 23&12:12:14.650 +56:10:47.11&IAUC 8691&Newton, Cox, Puckett \\
2006bb& UGC 4468       & 2006 03 25&08:33:31.096 +41:31:04.20&CBET 444 &Puckett, Gagliano \\
2006bd& UGC 6609       & 2006 03 26&11:38:28.420 +20:31:34.45&CBET 448 &Puckett, Cox \\
2006bk& MCG +06-23-20  & 2006 04 03&15:04:33.606 +35:57:50.53&CBET 462 &Boles \\
2006bq& NGC 6685       & 2006 04 23&18:39:58.941 +39:58:56.34&CBET 479 &Puckett, Pelloni \\
2006br& NGC 5185       & 2006 04 25&13:30:01.716 +13:24:56.61&CBET 482 &Puckett, Sostero \\
2006bt& Anon Gal       & 2006 04 26&15:56:30.526 +20:02:45.34&CBET 485 &LOSS \\
2006bu& Anon Gal       & 2006 04 27&13:52:47.736 +05:18:48.41&CBET 490 &Holmes, Devore, Graves \\
2006bw& Anon Gal       & 2006 04 27&14:33:56.806 +03:47:55.82&CBET 497 &LOSS \\
2006bz& Anon Gal       & 2006 05 04&13:00:43.362 +27:57:41.28&IAUC 8707&LOSS \\
2006cc& UGC 10244      & 2006 05 06&16:09:56.460 +43:07:35.89&CBET 505 &LOSS \\
2006cf& UGC 6015       & 2006 05 11&10:54:02.585 +46:01:36.44&IAUC 8710&LOSS; Puckett, Toth \\
2006cg& Anon Gal       & 2006 05 06&13:05:02.382 +28:44:25.11&CBET 509 &Quimby, Mondol \\
2006cj& Anon Gal       & 2006 05 17&12:59:24.519 +28:20:51.36&CBET 515 &Quimby, Mondol, Castro \\
2006cm& UGC 11723      & 2006 05 24&21:20:17.423 $-$01:41:02.08&CBET 521 &Puckett, Langoussis \\
2006cp& UGC 7357       & 2006 05 28&12:19:14.890 +22:25:37.89&CBET 524 &LOSS \\
2006cq& IC 4239        & 2006 05 29&13:24:25.040 +30:57:22.32&CBET 527 &Newton, Briggs, Puckett \\
2006cs& MCG +06-30-79  & 2006 06 03&13:45:33.879 +35:36:36.58&CBET 536 &LOSS \\
2006cz& MCG -01-38-2   & 2006 06 14&14:48:36.643 $-$04:44:30.91&IAUC 8721&LOSS \\
2006ef& NGC 809        & 2006 08 18&02:04:19.529 $-$08:43:42.50&CBET 597 &LOSS \\
2006ej& NGC 191        & 2006 08 23&00:38:59.812 $-$09:00:57.43&CBET 603 &LOSS \\
2006em& NGC 911        & 2006 08 25&02:25:44.313 +41:56:31.55&CBET 605 &LOSS \\
2006en& MCG +05-54-41  & 2006 08 26&23:10:05.053 +30:13:23.82&CBET 606 &Puckett, Peoples \\
2006et& NGC 232        & 2006 09 03&00:42:45.779 $-$23:33:29.80&CBET 616 &Itagaki \\
2006eu& MCG +08-36-16  & 2006 09 03&20:02:51.147 +49:19:02.18&CBET 618 &LOSS \\
2006ev& UGC 11758      & 2006 09 12&21:30:59.329 +13:59:21.30&IAUC 8747&Ory \\
2006gj& UGC 2650       & 2006 09 18&03:17:35.718 $-$01:41:30.18&CBET 631 &Puckett, Toth \\
2006gr& UGC 12071      & 2006 08 21&22:32:22.677 +30:49:43.80&CBET 638 &LOSS \\
2006gt& Anon Gal       & 2006 09 18&00:56:17.318 $-$01:37:46.66&CBET 641 &Quimby, Mondol \\
2006ha& IC 1461        & 2006 09 27&22:58:34.280 +15:10:25.53&CBET 649 &LOSS \\
2006hb& MCG -04-12-34  & 2006 09 27&05:02:01.302 $-$21:07:55.18&CBET 649 &LOSS \\
2006hn& UGC 6154       & 2006 09 28&11:07:18.439 +76:41:50.52&CBET 653 &Sehgal, Gagliano, Puckett \\
2006is& Anon Gal       & 2006 09 18&05:17:34.372 $-$23:46:54.67&CBET 659 &LOSS \\
2006je& IC 1735        & 2006 10 15&01:50:53.264 +33:05:53.27&CBET 675 &LOSS \\
2006ke& UGC 3365       & 2006 10 19&05:52:37.391 +66:49:00.78&CBET 682 &LOSS \\
2006kf& UGC 2829       & 2006 10 21&03:41:50.472 +08:09:24.93&CBET 686 &LOSS \\
2006le& UGC 3218       & 2006 10 26&05:00:41.934 +62:15:18.98&CBET 700 &LOSS \\
2006lf& UGC 3108       & 2006 10 26&04:38:29.511 +44:02:01.82&CBET 704 &LOSS \\
2006mo& MCG +06-02-17  & 2006 11 01&00:46:38.479 +36:19:57.70&CBET 719 &LOSS \\
2006mp& MCG +08-31-29  & 2006 11 03&17:12:00.194 +46:33:21.54&CBET 720 &Puckett, Gagliano \\
2006mq& ESO 494-G26    & 2006 10 22&08:06:12.378 $-$27:33:45.38&CBET 721 &LOSS \\
2006nz& Anon Gal       & 2006 11 08&00:56:29.206 $-$01:13:35.90&CBET 743 &SDSS-II \\
2006oa& Anon Gal       & 2006 11 11&21:23:42.939 $-$00:50:36.50&CBET 743 &SDSS-II \\
2006ob& Anon Gal       & 2006 11 13&01:51:48.133 +00:15:48.46&CBET 743 &SDSS-II \\
2006on& Anon Gal       & 2006 11 11&21:55:58.482 $-$01:04:12.79&CBET 745 &SDSS-II \\
2006or& NGC 3891       & 2006 11 18&11:48:03.469 +30:21:23.02&CBET 749 &Puckett, Kroes\\
2006os& UGC 2384       & 2006 11 21&02:55:00.998 +16:00:35.26&CBET 751 &Quimby, Castro\\
2006ot& ESO 544-G31    & 2006 11 22&02:15:04.800 $-$20:45:58.97&CBET 754 &LOSS (Joubert, Li) \\
2006qo& UGC 4133       & 2006 11 29&08:00:08.422 +56:22:07.25&CBET 763 &Joubert, Li (LOSS) \\
2006sr& UGC 14         & 2006 12 12&00:03:35.024 +23:11:45.67&IAUC 8784&Rich \\
2006td& Anon Gal       & 2006 12 24&01:58:15.761 +36:20:57.76&CBET 787 &Kloehr \\
2006te& Anon Gal       & 2006 12 28&08:11:42.963 +41:33:16.80&CBET 791 &Trondal, Luckas, Schwartz \\
2007F&  UGC 8162       & 2007 01 11&13:03:15.059 +50:37:07.53&CBET 803 &Puckett, Gagliano \\
2007H&  Anon Gal       & 2007 01 10&08:35:02.009 $-$08:20:16.00&CBET 806 &Joubert, Li (LOSS) \\
2007N&  MCG -01-33-12  & 2007 01 21&12:49:01.212 $-$09:27:10.77&CBET 818 &Lee, Li (LOSS) \\
2007O&  UGC 9612       & 2007 01 21&14:56:05.161 +45:24:17.37&CBET 818 &Lee, Li (LOSS) \\
2007R&  UGC 4008       & 2007 01 26&07:46:37.513 +44:47:22.51&CBET 823 &Puckett, Gray \\
2007S&  UGC 5378       & 2007 01 29&10:00:31.237 +04:24:25.26&CBET 825 &Puckett, Gorelli \\
2007ae& UGC 10704      & 2007 02 19&17:01:52.067 +79:01:54.26&CBET 856 &Nissinen, Hentunen \\
2007af& NGC 5584       & 2007 03 01&14:22:21.064 $-$00:23:37.92&CBET 863 &Itagaki \\
2007ai& MCG -04-38-4   & 2007 03 06&16:12:53.740 $-$21:37:48.57&CBET 870 &Li (LOSS) \\
2007al& Anon Gal       & 2007 03 10&09:59:18.467 $-$19:28:25.39&CBET 875 &Madison, Li (LOSS) \\
2007ap& MCG +03-41-3   & 2007 03 13&15:56:23.035 +16:30:57.92&CBET 883 &Puckett, Kroes \\
2007ar& MCG +10-19-62  & 2007 03 12&13:21:01.797 +58:33:02.80&CBET 886 &Duszanowicz \\
2007au& UGC 3725       & 2007 03 18&07:11:46.095 +49:51:13.08&CBET 895 &Lee, Li (LOSS) \\
2007ax& NGC 2577       & 2007 03 21&08:22:43.242 +22:33:16.91&CBET 904 &Arbour \\
2007ba& UGC 9798       & 2007 03 29&15:16:42.581 +07:23:47.91&CBET 911 &Winslow, Li (LOSS) \\
2007bc& UGC 6332       & 2007 04 04&11:19:14.566 +20:48:32.26&CBET 913 &Prasad, Li (LOSS) \\
2007bd& UGC 4455       & 2007 04 04&08:31:33.375 $-$01:11:57.73&CBET 914 &Prasad, Li (LOSS) \\
2007bm& NGC 3672       & 2007 04 20&11:25:02.309 $-$09:47:53.96&CBET 936 &Martin \\
2007bz& IC 3918        & 2007 04 22&12:56:53.764 +22:22:23.12&CBET 941 &Nearby SN Factory \\
2007ca& MCG -02-34-61  & 2007 04 25&13:31:05.840 $-$15:06:06.52&CBET 945 &Prasad, Li \\
2007cg& ESO 508-G75    & 2007 05 11&13:25:33.588 $-$24:39:08.29&CBET 960 &Thrasher, Li (LOSS) \\
2007ci& NGC 3873       & 2007 05 15&11:45:45.851 +19:46:13.74&CBET 966 &Puckett, Crowley \\
2007co& MCG +05-43-16  & 2007 06 04&18:23:03.599 +29:53:49.39&CBET 977 &Nicolas \\
2007cp& IC 807         & 2007 06 13&12:42:12.748 $-$17:24:07.45&CBET 980 &Khandrika, Li (LOSS) \\
2007cq& Anon Gal       & 2007 06 21&22:14:40.423 +05:04:48.57&CBET 983 &Orff, Newton \\
2007qe& Anon Gal       & 2007 11 13&23:54:12.958 +27:24:33.02&CBET 1138&Yuan et al. (ROTSE)\\
2007sr& NGC 4038       & 2007 12 18&12:01:52.800 $-$18:58:21.83&CBET 1172&Drake et al. \\
2008L&  NGC 1259       & 2008 01 14&03:17:16.596 +41:22:56.23&CBET 1212&Fujita \\
2008af& UGC 9640       & 2008 02 09&14:59:28.493 +16:39:11.77&CBET 1248&Boles \\
2008bf& NGC 4055       & 2008 03 18&12:04:02.877 +20:14:42.29&CBET 1307&Parisky (LOSS) \\
\enddata
\label{table_sndiscovery}
\end{deluxetable}

\clearpage
\begin{deluxetable}{lrcc}
\tablecolumns{4}
\tablewidth{0pc}
\tablecaption{Photometric Color Terms}
\tablecomments{
~Lower-case $ubvri$ refer to the instrumental magnitudes while
\emph{UBVRIr'i'} refer to the standard magnitudes.  All color terms
implicitly contain an additive constant.  For example, for the
Keplercam:  $(v-V) = 0.0185(B-V)$ + const; $(u-b) = 1.0279(U-B)$ + const.
}
\tablehead{
\colhead{Detector/Filters} & \colhead{Color Term} &  \colhead{Value} & \colhead{Nights}
}
\startdata
Keplercam/\emph{UBVr'i'}& $(u-b)/(U-B)$  &  $1.0279 \pm 0.0069$ & 20\\
Keplercam/\emph{UBVr'i'}& $(b-v)/(B-V)$  &  $0.9212 \pm 0.0029$ & 37\\
Keplercam/\emph{UBVr'i'}& $(v-V)/(B-V)$  &  $0.0185 \pm 0.0023$ & 37\\
Keplercam/\emph{UBVr'i'}& $(v-r)/(V-r')$ &  $1.0508 \pm 0.0029$ & 37\\
Keplercam/\emph{UBVr'i'}& $(v-i)/(V-i')$ &  $1.0185 \pm 0.0020$ & 37\\
\hline
&&&\\
Minicam/\emph{UBVr'i'}  & $(u-b)/(U-B)$  &  $1.0060 \pm 0.0153$ & 4\\
Minicam/\emph{UBVr'i'}  & $(b-v)/(B-V)$  &  $0.9000 \pm 0.0095$ & 4\\
Minicam/\emph{UBVr'i'}  & $(v-V)/(B-V)$  &  $0.0380 \pm 0.0030$ & 4\\
Minicam/\emph{UBVr'i'}  & $(v-r)/(V-r')$ &  $1.0903 \pm 0.0140$ & 4\\
Minicam/\emph{UBVr'i'}  & $(v-i)/(V-i')$ &  $1.0375 \pm 0.0088$ & 4\\
\hline
&&&\\
4Shooter/\emph{UBVRI}     & $(u-b)/(U-B)$  &  $0.9912 \pm 0.0078$ & 16\\
4Shooter/\emph{UBVRI}     & $(b-v)/(B-V)$  &  $0.8928 \pm 0.0019$ & 16\\
4Shooter/\emph{UBVRI}     & $(v-V)/(B-V)$  &  $0.0336 \pm 0.0020$ & 16\\
4Shooter/\emph{UBVRI}     & $(v-r)/(V-R)$  &  $1.0855 \pm 0.0058$ & 16\\
4Shooter/\emph{UBVRI}     & $(v-i)/(V-I)$  &  $1.0166 \pm 0.0067$ & 16\\
\enddata
\label{table_colorterms}
\end{deluxetable}

\begin{deluxetable}{lccccccccc}
\tablecolumns{10}
\tablewidth{0pc}
\rotate
\tabletypesize{\scriptsize}
\tablecaption{Direct-fit and MLCS2k2 Light Curve Properties}
\tablecomments{The peak magnitudes, $m_U$, $m_B$, and $m_V$, are measured at the time of maximum light in each band while $(B-V)_{Bmax}$ and $(U-B)_{Bmax}$ are measured at the time of maximum
light in $B$.  For these measurements, the light curves were $K$-corrected, corrected for 
Milky Way extinction, and corrected for time dilation.  The host-galaxy extinction, $A_V$, is from MLCS2k2 ($R_V=1.7$) and has not been removed from the peak magnitudes listed.  The host-galaxy
color excesses, $E_{BV}$ and $E_{UB}$, are derived from $A_V$.
}
\tablehead{
\colhead{SN} & \colhead{$z_{CMB}$} & \colhead{$m_U$} & \colhead{$m_B$} & \colhead{$m_V$} & \colhead{$(B-V)_{Bmax}$} & \colhead{$(U-B)_{Bmax}$} & \colhead{$A_V$} & \colhead{$E_{BV}$} & \colhead{$E_{UB}$} 
}
\startdata
80N& 0.0055& 12.10(0.06)& 12.40(0.03)& 12.38(0.03)& -0.02(0.05)& -0.29(0.07)& 0.2110(0.0510)& 0.1241(0.0300)& 0.0881(0.0213)\\
81B& 0.0072& 11.62(0.10)& 11.93(0.04)& 11.89(0.05)& 0.01(0.06)& -0.25(0.11)& 0.2300(0.0630)& 0.1353(0.0371)& 0.0961(0.0263)\\
81D& 0.0055& 12.28(0.16)& 12.49(0.09)& 12.34(0.05)& 0.11(0.10)& -0.15(0.18)& 0.3390(0.1520)& 0.1994(0.0894)& 0.1416(0.0635)\\
86G& 0.0027& 12.58(0.10)& 12.03(0.08)& 11.11(0.06)& 0.88(0.09)& 0.64(0.10)& 1.2210(0.0860)& 0.7182(0.0506)& 0.5099(0.0359)\\
89B& 0.0035& 12.20(0.07)& 12.23(0.09)& 11.88(0.05)& 0.32(0.10)& 0.04(0.11)& 0.8590(0.0810)& 0.5053(0.0476)& 0.3588(0.0338)\\
90N& 0.0043&      -     & 12.67(0.05)& 12.65(0.04)& 0.01(0.06)&      -     & 0.2210(0.0510)& 0.1300(0.0300)& 0.0923(0.0213)\\
90af& 0.0502&      -     & 17.77(0.04)& 17.76(0.04)& -0.01(0.06)&      -     & 0.0730(0.0640)& 0.0429(0.0376)& 0.0305(0.0267)\\
91T& 0.0069& 11.16(0.03)& 11.60(0.02)& 11.45(0.02)& 0.12(0.03)& -0.45(0.04)& 0.3020(0.0390)& 0.1776(0.0229)& 0.1261(0.0163)\\
91bg& 0.0046&      -     & 14.60(0.05)& 13.85(0.04)& 0.71(0.06)&      -     & 0.0960(0.0570)& 0.0565(0.0335)& 0.0401(0.0238)\\
92A& 0.0059& 12.36(0.07)& 12.53(0.02)& 12.48(0.01)& 0.02(0.02)& -0.16(0.07)& 0.0140(0.0140)& 0.0082(0.0082)& 0.0058(0.0058)\\
92ag& 0.0259&      -     & 16.20(0.08)& 16.16(0.06)& 0.03(0.08)&      -     & 0.3120(0.0810)& 0.1835(0.0476)& 0.1303(0.0338)\\
92al& 0.0141&      -     & 14.45(0.04)& 14.55(0.04)& -0.10(0.05)&      -     & 0.0330(0.0270)& 0.0194(0.0159)& 0.0138(0.0113)\\
92bc& 0.0198&      -     & 15.08(0.04)& 15.16(0.04)& -0.08(0.05)&      -     & 0.0120(0.0120)& 0.0071(0.0071)& 0.0050(0.0050)\\
92bh& 0.0451&      -     & 17.59(0.04)& 17.54(0.04)& 0.02(0.06)&      -     & 0.1830(0.0790)& 0.1076(0.0465)& 0.0764(0.0330)\\
92bo& 0.0181&      -     & 15.69(0.04)& 15.74(0.04)& -0.07(0.06)&      -     & 0.0340(0.0290)& 0.0200(0.0171)& 0.0142(0.0121)\\
92bp& 0.0789&      -     & 18.30(0.07)& 18.41(0.06)& -0.11(0.09)&      -     & 0.0360(0.0310)& 0.0212(0.0182)& 0.0151(0.0129)\\
93H& 0.0248&      -     & 16.71(0.05)& 16.51(0.04)& 0.13(0.06)&      -     & 0.0290(0.0260)& 0.0171(0.0153)& 0.0121(0.0109)\\
93O& 0.0519&      -     & 17.58(0.05)& 17.72(0.05)& -0.18(0.07)&      -     & 0.0480(0.0340)& 0.0282(0.0200)& 0.0200(0.0142)\\
93ag& 0.0500&      -     & 17.82(0.09)& 17.78(0.07)& 0.02(0.10)&      -     & 0.1020(0.0660)& 0.0600(0.0388)& 0.0426(0.0275)\\
94D& 0.0031& 11.14(0.09)& 11.78(0.04)& 11.82(0.02)& -0.04(0.04)& -0.62(0.10)& 0.0090(0.0090)& 0.0053(0.0053)& 0.0038(0.0038)\\
94S& 0.0160&      -     & 14.76(0.05)& 14.78(0.06)& -0.02(0.08)&      -     & 0.0470(0.0340)& 0.0276(0.0200)& 0.0196(0.0142)\\
94T& 0.0357&      -     & 17.32(0.03)& 17.14(0.04)& 0.18(0.05)&      -     & 0.0530(0.0420)& 0.0312(0.0247)& 0.0222(0.0175)\\
94ae& 0.0055&      -     & 12.95(0.06)& 12.99(0.03)& -0.04(0.07)&      -     & 0.0490(0.0320)& 0.0288(0.0188)& 0.0204(0.0133)\\
95D& 0.0077&      -     & 13.17(0.06)& 13.25(0.05)& -0.10(0.07)&      -     & 0.0680(0.0440)& 0.0400(0.0259)& 0.0284(0.0184)\\
95E& 0.0117&      -     & 16.68(0.05)& 15.97(0.05)& 0.70(0.07)&      -     & 1.4600(0.0640)& 0.8588(0.0376)& 0.6097(0.0267)\\
95ac& 0.0488&      -     & 17.06(0.04)& 17.13(0.04)& -0.11(0.06)&      -     & 0.1060(0.0550)& 0.0624(0.0324)& 0.0443(0.0230)\\
95ak& 0.0220&      -     & 16.00(0.06)& 15.94(0.06)& 0.03(0.08)&      -     & 0.2590(0.0720)& 0.1524(0.0424)& 0.1082(0.0301)\\
95al& 0.0059& 12.72(0.10)& 13.33(0.05)& 13.19(0.05)& 0.13(0.07)& -0.57(0.11)& 0.1770(0.0490)& 0.1041(0.0288)& 0.0739(0.0204)\\
95bd& 0.0144&      -     & 15.20(0.33)& 14.91(0.25)& 0.27(0.33)&      -     & 0.4620(0.1590)& 0.2718(0.0935)& 0.1930(0.0664)\\
96X& 0.0078& 12.36(0.06)& 12.98(0.05)& 13.02(0.04)& -0.05(0.05)& -0.48(0.06)& 0.0310(0.0240)& 0.0182(0.0141)& 0.0129(0.0100)\\
96bo& 0.0163&      -     & 15.83(0.05)& 15.50(0.04)& 0.31(0.06)&      -     & 0.6260(0.0710)& 0.3682(0.0418)& 0.2614(0.0297)\\
97E& 0.0133& 14.77(0.10)& 15.12(0.08)& 15.07(0.07)& 0.03(0.09)& -0.32(0.10)& 0.0850(0.0510)& 0.0500(0.0300)& 0.0355(0.0213)\\
97bp& 0.0094& 13.81(0.05)& 13.91(0.03)& 13.73(0.03)& 0.10(0.04)& -0.06(0.05)& 0.4790(0.0480)& 0.2818(0.0282)& 0.2001(0.0200)\\
97br& 0.0080& 13.04(0.09)& 13.63(0.12)& 13.42(0.08)& 0.16(0.13)& -0.54(0.13)& 0.5490(0.0540)& 0.3229(0.0318)& 0.2293(0.0226)\\
97dg& 0.0297& 16.33(0.3)& 16.85(0.06)& 16.86(0.04)& -0.03(0.06)& -0.47(0.08)& 0.0920(0.0520)& 0.0541(0.0306)& 0.0384(0.0217)\\
98aq& 0.0045& 11.62(0.03)& 12.31(0.02)& 12.43(0.02)& -0.12(0.03)& -0.65(0.04)& 0.0110(0.0110)& 0.0065(0.0065)& 0.0046(0.0046)\\
98bp& 0.0102& 15.20(0.08)& 15.28(0.05)& 15.05(0.04)& 0.16(0.06)& -0.06(0.08)& 0.0250(0.0200)& 0.0147(0.0118)& 0.0104(0.0084)\\
98bu& 0.0040& 11.78(0.04)& 12.12(0.02)& 11.78(0.02)& 0.32(0.02)& -0.29(0.04)& 0.6310(0.0400)& 0.3712(0.0235)& 0.2636(0.0167)\\
98de& 0.0156&      -     & 17.30(0.05)& 16.66(0.04)& 0.60(0.05)&      -     & 0.1420(0.0610)& 0.0835(0.0359)& 0.0593(0.0255)\\
98es& 0.0096& 13.26(0.06)& 13.83(0.04)& 13.75(0.07)& 0.08(0.07)& -0.54(0.06)& 0.2070(0.0420)& 0.1218(0.0247)& 0.0865(0.0175)\\
99aa& 0.0153& 14.17(0.06)& 14.72(0.03)& 14.77(0.02)& -0.06(0.03)& -0.53(0.06)& 0.0250(0.0210)& 0.0147(0.0124)& 0.0104(0.0088)\\
99ac& 0.0098& 13.77(0.06)& 14.09(0.04)& 14.05(0.03)& -0.01(0.05)& -0.27(0.06)& 0.2440(0.0420)& 0.1435(0.0247)& 0.1019(0.0175)\\
99aw& 0.0392&      -     & 16.73(0.04)& 16.74(0.03)& -0.01(0.04)&      -     & 0.0210(0.0160)& 0.0124(0.0094)& 0.0088(0.0067)\\
99by& 0.0028& 13.73(0.02)& 13.54(0.06)& 13.10(0.02)& 0.40(0.06)& 0.20(0.06)& 0.0300(0.0220)& 0.0176(0.0129)& 0.0125(0.0092)\\
99cc& 0.0315& 16.44(0.05)& 16.76(0.02)& 16.75(0.02)& -0.01(0.03)& -0.31(0.06)& 0.0640(0.0490)& 0.0376(0.0288)& 0.0267(0.0204)\\
99cl& 0.0087& 15.51(0.07)& 14.87(0.04)& 13.72(0.04)& 1.12(0.05)& 0.66(0.08)& 2.1980(0.0660)& 1.2929(0.0388)& 0.9180(0.0275)\\
99da& 0.0125&      -     & 16.65(0.04)& 16.06(0.04)& 0.52(0.05)&      -     & 0.0660(0.0490)& 0.0388(0.0288)& 0.0275(0.0204)\\
99dk& 0.0139& 14.54(0.09)& 14.81(0.05)& 14.76(0.04)& 0.05(0.05)& -0.24(0.10)& 0.2520(0.0580)& 0.1482(0.0341)& 0.1052(0.0242)\\
99dq& 0.0135& 13.88(0.10)& 14.42(0.08)& 14.34(0.06)& 0.07(0.08)& -0.48(0.10)& 0.2990(0.0510)& 0.1759(0.0300)& 0.1249(0.0213)\\
99ee& 0.0106& 14.65(0.03)& 14.85(0.02)& 14.56(0.02)& 0.27(0.03)& -0.18(0.03)& 0.6430(0.0410)& 0.3782(0.0241)& 0.2685(0.0171)\\
99ek& 0.0176&      -     & 15.61(0.37)& 15.49(0.28)& 0.10(0.37)&      -     & 0.3120(0.1560)& 0.1835(0.0918)& 0.1303(0.0652)\\
99gp& 0.0260& 15.40(0.06)& 15.99(0.05)& 15.97(0.03)& -0.00(0.05)& -0.54(0.07)& 0.1490(0.0440)& 0.0876(0.0259)& 0.0622(0.0184)\\
00E& 0.0042&      -     & 12.86(0.24)& 12.68(0.19)& 0.17(0.24)&      -     & 0.4660(0.1220)& 0.2741(0.0718)& 0.1946(0.0510)\\
00cf& 0.0365&      -     & 17.08(0.03)& 17.11(0.03)& -0.05(0.04)&      -     & 0.0860(0.0550)& 0.0506(0.0324)& 0.0359(0.0230)\\
00cn& 0.0232& 16.40(0.09)& 16.57(0.05)& 16.40(0.03)& 0.10(0.06)& -0.16(0.10)& 0.0710(0.0600)& 0.0418(0.0353)& 0.0297(0.0251)\\
00dk& 0.0164& 14.99(0.07)& 15.34(0.05)& 15.33(0.04)& -0.02(0.06)& -0.29(0.07)& 0.0170(0.0150)& 0.0100(0.0088)& 0.0071(0.0062)\\
01ba& 0.0305&      -     & 16.18(0.05)& 16.31(0.05)& -0.15(0.06)&      -     & 0.0250(0.0210)& 0.0147(0.0124)& 0.0104(0.0088)\\
01bt& 0.0144&      -     & 15.26(0.05)& 15.09(0.04)& 0.14(0.05)&      -     & 0.4260(0.0630)& 0.2506(0.0371)& 0.1779(0.0263)\\
01cz& 0.0163&      -     & 15.05(0.06)& 14.95(0.05)& 0.09(0.07)&      -     & 0.2000(0.0700)& 0.1176(0.0412)& 0.0835(0.0293)\\
01el& 0.0037& 12.56(0.04)& 12.75(0.03)& 12.70(0.01)& 0.03(0.03)& -0.16(0.05)& 0.5000(0.0440)& 0.2941(0.0259)& 0.2088(0.0184)\\
01ep& 0.0129& 14.52(0.04)& 14.87(0.04)& 14.81(0.03)& 0.03(0.05)& -0.31(0.05)& 0.2590(0.0540)& 0.1524(0.0318)& 0.1082(0.0226)\\
01fe& 0.0143& 14.02(0.10)& 14.68(0.04)& 14.65(0.03)& 0.02(0.05)& -0.60(0.09)& 0.0990(0.0490)& 0.0582(0.0288)& 0.0413(0.0204)\\
01fh& 0.0114& 13.87(0.59)& 14.19(0.50)& 14.31(0.38)& -0.15(0.50)& -0.25(0.59)& 0.0770(0.0620)& 0.0453(0.0365)& 0.0322(0.0259)\\
01V& 0.0162& 14.01(0.09)& 14.64(0.08)& 14.61(0.05)& -0.00(0.10)& -0.60(0.12)& 0.1710(0.0410)& 0.1006(0.0241)& 0.0714(0.0171)\\
02bo& 0.0054&      -     & 13.94(0.08)& 13.59(0.07)& 0.34(0.11)&      -     & 0.9080(0.0500)& 0.5341(0.0294)& 0.3792(0.0209)\\
02cd& 0.0097& 15.57(0.32)& 15.53(0.27)& 14.93(0.22)& 0.57(0.28)& 0.06(0.31)& 1.0260(0.1320)& 0.6035(0.0776)& 0.4285(0.0551)\\
02cr& 0.0103&      -     & 14.16(0.04)& 14.23(0.04)& -0.07(0.05)&      -     & 0.1220(0.0630)& 0.0718(0.0371)& 0.0510(0.0263)\\
02cx& 0.0250&      -     & 17.54(0.10)& 17.34(0.08)& 0.17(0.13)&      -     & 0.7030(0.0680)& 0.4135(0.0400)& 0.2936(0.0284)\\
02de& 0.0281& 16.32(0.06)& 16.66(0.03)& 16.52(0.02)& 0.13(0.04)& -0.31(0.06)& 0.3820(0.0840)& 0.2247(0.0494)& 0.1595(0.0351)\\
02dj& 0.0104&      -     & 13.98(0.07)& 13.83(0.06)& 0.10(0.08)&      -     & 0.3420(0.0780)& 0.2012(0.0459)& 0.1429(0.0326)\\
02dp& 0.0105& 14.16(0.06)& 14.60(0.05)& 14.47(0.05)& 0.10(0.07)& -0.39(0.07)& 0.2680(0.0900)& 0.1576(0.0529)& 0.1119(0.0376)\\
02er& 0.0085& 13.91(0.13)& 14.24(0.11)& 14.10(0.09)& 0.12(0.11)& -0.33(0.13)& 0.2270(0.0740)& 0.1335(0.0435)& 0.0948(0.0309)\\
02fk& 0.0070&      -     & 13.11(0.05)& 13.23(0.04)& -0.12(0.06)&      -     & 0.0340(0.0230)& 0.0200(0.0135)& 0.0142(0.0096)\\
02ha& 0.0134&      -     & 14.69(0.08)& 14.77(0.07)& -0.09(0.09)&      -     & 0.0420(0.0320)& 0.0247(0.0188)& 0.0175(0.0133)\\
02hu& 0.0382& 16.08(0.05)& 16.58(0.04)& 16.70(0.03)& -0.12(0.05)& -0.48(0.06)& 0.0360(0.0300)& 0.0212(0.0176)& 0.0151(0.0125)\\
03W& 0.0211& 15.60(0.08)& 15.85(0.04)& 15.71(0.05)& 0.12(0.06)& -0.15(0.08)& 0.3300(0.0500)& 0.1941(0.0294)& 0.1378(0.0209)\\
03cg& 0.0053& 16.38(0.05)& 15.79(0.05)& 14.56(0.02)& 1.23(0.05)& 0.66(0.06)& 2.2090(0.0530)& 1.2994(0.0312)& 0.9226(0.0222)\\
03du& 0.0066& 13.07(0.16)& 13.43(0.06)& 13.54(0.02)& -0.12(0.07)& -0.35(0.19)& 0.0320(0.0220)& 0.0188(0.0129)& 0.0133(0.0092)\\
03iv& 0.0335& 16.57(0.15)& 16.97(0.09)& 17.01(0.08)& -0.10(0.10)& -0.37(0.15)& 0.0230(0.0240)& 0.0135(0.0141)& 0.0096(0.0100)\\
03kf& 0.0077& 12.93(0.25)& 13.28(0.21)& 13.25(0.16)& 0.02(0.21)& -0.27(0.25)& 0.1140(0.0800)& 0.0671(0.0471)& 0.0476(0.0334)\\
04as& 0.0321&      -     & 16.93(0.15)& 16.91(0.03)& 0.01(0.15)&      -     & 0.3030(0.0580)& 0.1782(0.0341)& 0.1265(0.0242)\\
05am& 0.0095&      -     & 13.62(0.04)& 13.60(0.03)& 0.01(0.05)& -0.27(0.06)& 0.0370(0.0330)& 0.0218(0.0194)& 0.0155(0.0138)\\
05cf& 0.0070& 12.84(0.08)& 13.24(0.07)& 13.28(0.05)& -0.05(0.07)& -0.38(0.08)& 0.2080(0.0700)& 0.1224(0.0412)& 0.0869(0.0293)\\
05el& 0.0148& 14.28(0.10)& 14.84(0.08)& 14.88(0.06)& -0.05(0.08)& -0.53(0.10)& 0.0120(0.0130)& 0.0071(0.0076)& 0.0050(0.0054)\\
05eq& 0.0284& 15.77(0.08)& 16.28(0.06)& 16.25(0.05)& 0.03(0.07)& -0.47(0.09)& 0.1040(0.0470)& 0.0612(0.0276)& 0.0435(0.0196)\\
05eu& 0.0341&      -     &      -     & 16.39(0.08)&      -     &      -     & 0.0520(0.0380)& 0.0306(0.0224)& 0.0217(0.0159)\\
05hc& 0.0450& 16.93(0.06)& 17.31(0.04)& 17.41(0.04)& -0.12(0.06)& -0.38(0.07)& 0.1150(0.0520)& 0.0676(0.0306)& 0.0480(0.0217)\\
05hk& 0.0118& 15.44(0.03)& 15.84(0.05)& 15.71(0.02)& 0.04(0.05)& -0.27(0.06)& 0.8100(0.0440)& 0.4765(0.0259)& 0.3383(0.0184)\\
05iq& 0.0330& 16.30(0.19)& 16.80(0.07)& 16.88(0.03)& -0.10(0.07)& -0.48(0.20)& 0.0310(0.0260)& 0.0182(0.0153)& 0.0129(0.0109)\\
05kc& 0.0138& 15.49(0.11)& 15.58(0.09)& 15.38(0.07)& 0.20(0.14)& -0.09(0.20)& 0.6240(0.0740)& 0.3671(0.0435)& 0.2606(0.0309)\\
05ke& 0.0045& 15.15(0.04)& 14.88(0.05)& 14.14(0.04)& 0.66(0.07)& 0.34(0.06)& 0.0680(0.0400)& 0.0400(0.0235)& 0.0284(0.0167)\\
05ki& 0.0208& 14.96(0.12)& 15.56(0.07)& 15.66(0.02)& -0.10(0.07)& -0.55(0.14)& 0.0180(0.0150)& 0.0106(0.0088)& 0.0075(0.0062)\\
05lz& 0.0402& 17.33(0.10)& 17.55(0.07)& 17.59(0.06)& -0.04(0.08)& -0.22(0.11)& 0.1730(0.0680)& 0.1018(0.0400)& 0.0723(0.0284)\\
05mc& 0.0261& 17.14(0.06)& 17.21(0.04)& 17.00(0.04)& 0.17(0.05)& -0.04(0.06)& 0.0770(0.0510)& 0.0453(0.0300)& 0.0322(0.0213)\\
05ms& 0.0259& 15.70(0.04)& 16.13(0.03)& 16.17(0.03)& -0.08(0.04)& -0.43(0.05)& 0.0700(0.0400)& 0.0412(0.0235)& 0.0293(0.0167)\\
05mz& 0.0170& 16.32(0.13)& 16.37(0.11)& 16.06(0.09)& 0.24(0.11)& 0.07(0.14)& 0.2660(0.0890)& 0.1565(0.0524)& 0.1111(0.0372)\\
06ac& 0.0236& 15.83(0.05)& 16.19(0.03)& 16.06(0.03)& 0.11(0.04)& -0.35(0.06)& 0.1040(0.0470)& 0.0612(0.0276)& 0.0435(0.0196)\\
06ar& 0.0229&      -     & 16.46(0.03)& 16.33(0.03)& 0.11(0.04)&      -     & 0.1960(0.1240)& 0.1153(0.0729)& 0.0819(0.0518)\\
06ax& 0.0180& 14.47(0.05)& 15.01(0.04)& 15.08(0.03)& -0.08(0.04)& -0.50(0.06)& 0.0380(0.0290)& 0.0224(0.0171)& 0.0159(0.0121)\\
06az& 0.0316& 15.87(0.06)& 16.45(0.03)& 16.55(0.03)& -0.13(0.04)& -0.52(0.06)& 0.0120(0.0120)& 0.0071(0.0071)& 0.0050(0.0050)\\
06br& 0.0255&      -     & 19.08(0.07)& 18.09(0.04)& 0.99(0.08)&      -     & 1.7010(0.1020)& 1.0006(0.0600)& 0.7104(0.0426)\\
06bt& 0.0325&      -     & 16.91(0.04)& 16.77(0.03)& 0.12(0.04)&      -     & 0.4280(0.0530)& 0.2518(0.0312)& 0.1788(0.0222)\\
06bz& 0.0277&      -     & 18.33(0.08)& 17.63(0.03)& 0.61(0.05)&      -     & 0.1820(0.1150)& 0.1071(0.0676)& 0.0760(0.0480)\\
06cc& 0.0328& 17.60(0.06)& 17.81(0.03)& 17.45(0.02)& 0.35(0.04)& -0.14(0.07)& 0.8120(0.0510)& 0.4776(0.0300)& 0.3391(0.0213)\\
06cm& 0.0152&      -     & 18.05(0.05)& 16.96(0.04)& 1.07(0.06)&      -     & 1.8290(0.0790)& 1.0759(0.0465)& 0.7639(0.0330)\\
06cp& 0.0233&      -     & 15.89(0.14)& 15.87(0.09)& 0.02(0.05)&      -     & 0.4400(0.0640)& 0.2588(0.0376)& 0.1837(0.0267)\\
06D& 0.0097& 13.90(0.05)& 14.13(0.04)& 14.06(0.04)& 0.05(0.05)& -0.20(0.06)& 0.0760(0.0420)& 0.0447(0.0247)& 0.0317(0.0175)\\
06gj& 0.0277&      -     & 17.67(0.08)& 17.28(0.07)& 0.34(0.10)&      -     & 0.4820(0.1260)& 0.2835(0.0741)& 0.2013(0.0526)\\
06gr& 0.0335& 16.59(0.09)& 16.91(0.07)& 16.87(0.05)& 0.03(0.08)& -0.28(0.10)& 0.3040(0.0520)& 0.1788(0.0306)& 0.1269(0.0217)\\
06kf& 0.0208&      -     &      -     & 15.90(0.12)&      -     &      -     & 0.0240(0.0240)& 0.0141(0.0141)& 0.0100(0.0100)\\
06le& 0.0173& 14.26(0.32)& 14.78(0.27)& 14.85(0.21)& -0.09(0.27)& -0.50(0.31)& 0.0760(0.0600)& 0.0447(0.0353)& 0.0317(0.0251)\\
06lf& 0.0130&      -     & 13.70(0.63)& 13.88(0.49)& -0.18(0.63)&      -     & 0.0950(0.0740)& 0.0559(0.0435)& 0.0397(0.0309)\\
06mp& 0.0229&      -     & 15.96(0.03)& 15.93(0.04)& -0.00(0.04)&      -     & 0.1660(0.0680)& 0.0976(0.0400)& 0.0693(0.0284)\\
06N& 0.0143&      -     & 15.08(0.07)& 15.09(0.05)& -0.03(0.07)& -0.43(0.08)& 0.0270(0.0230)& 0.0159(0.0135)& 0.0113(0.0096)\\
06nz& 0.0372&      -     & 18.11(0.06)& 17.73(0.05)& 0.29(0.08)&      -     & 0.0940(0.0780)& 0.0553(0.0459)& 0.0393(0.0326)\\
06oa& 0.0589& 17.46(0.08)& 17.84(0.06)& 17.85(0.06)& -0.02(0.08)& -0.30(0.10)& 0.1860(0.0670)& 0.1094(0.0394)& 0.0777(0.0280)\\
06ob& 0.0582& 17.78(0.08)& 18.20(0.05)& 18.17(0.04)& -0.02(0.06)& -0.21(0.09)& 0.0210(0.0210)& 0.0124(0.0124)& 0.0088(0.0088)\\
06qo& 0.0300&      -     & 16.81(0.04)& 16.64(0.04)& 0.17(0.06)&      -     & 0.4530(0.0630)& 0.2665(0.0371)& 0.1892(0.0263)\\
06S& 0.0329& 16.34(0.05)& 16.79(0.02)& 16.75(0.02)& -0.00(0.03)& -0.43(0.05)& 0.2680(0.0460)& 0.1576(0.0271)& 0.1119(0.0192)\\
06sr& 0.0232&      -     & 16.14(0.07)& 16.11(0.05)& 0.02(0.07)&      -     & 0.0850(0.0530)& 0.0500(0.0312)& 0.0355(0.0222)\\
06td& 0.0150&      -     & 15.72(0.06)& 15.60(0.05)& 0.11(0.06)&      -     & 0.1710(0.0790)& 0.1006(0.0465)& 0.0714(0.0330)\\
06X& 0.0063& 16.28(0.07)& 15.28(0.04)& 13.97(0.02)& 1.26(0.05)& 1.00(0.09)& 2.4960(0.0430)& 1.4682(0.0253)& 1.0424(0.0180)\\
07af& 0.0063&      -     & 13.13(0.03)& 13.10(0.02)& 0.02(0.03)&      -     & 0.2150(0.0540)& 0.1265(0.0318)& 0.0898(0.0226)\\
07au& 0.0202&      -     & 16.46(0.06)& 16.32(0.07)& 0.12(0.09)&      -     & 0.0490(0.0390)& 0.0288(0.0229)& 0.0204(0.0163)\\
07bc& 0.0219&      -     & 15.82(0.04)& 15.92(0.03)& -0.11(0.05)&      -     & 0.0840(0.0590)& 0.0494(0.0347)& 0.0351(0.0246)\\
07bd& 0.0320&      -     & 16.53(0.03)& 16.58(0.03)& -0.06(0.04)&      -     & 0.0430(0.0330)& 0.0253(0.0194)& 0.0180(0.0138)\\
07ca& 0.0152&      -     & 15.95(0.05)& 15.65(0.04)& 0.29(0.05)&      -     & 0.5800(0.0690)& 0.3412(0.0406)& 0.2423(0.0288)\\
07ci& 0.0191&      -     &      -     & 15.86(0.02)&      -     &      -     & 0.0740(0.0630)& 0.0435(0.0371)& 0.0309(0.0263)\\
07co& 0.0266& 16.39(0.10)& 16.43(0.08)& 16.38(0.06)& 0.03(0.08)& -0.02(0.10)& 0.3920(0.0690)& 0.2306(0.0406)& 0.1637(0.0288)\\
07cq& 0.0247&      -     & 15.82(0.07)& 15.79(0.06)& 0.01(0.08)&      -     & 0.1090(0.0590)& 0.0641(0.0347)& 0.0455(0.0246)\\
07F& 0.0242&      -     & 15.87(0.03)& 15.91(0.02)& -0.04(0.04)&      -     & 0.0470(0.0380)& 0.0276(0.0224)& 0.0196(0.0159)\\
07qe& 0.0229&      -     & 16.01(0.04)& 15.99(0.03)& -0.01(0.05)&      -     & 0.1480(0.0590)& 0.0871(0.0347)& 0.0618(0.0246)\\
07S& 0.0151&      -     & 15.82(0.03)& 15.40(0.03)& 0.39(0.04)&      -     & 0.8330(0.0540)& 0.4900(0.0318)& 0.3479(0.0226)\\
08bf& 0.0257& 15.29(0.08)& 15.72(0.04)& 15.68(0.04)& 0.04(0.05)& -0.40(0.09)& 0.1020(0.0490)& 0.0600(0.0288)& 0.0426(0.0204)\\

\enddata
\label{table_lcproperties1}
\end{deluxetable}

\begin{deluxetable}{lcccrccr}
\tablecolumns{8}
\tablewidth{0pc}
\tabletypesize{\scriptsize}
\tablecaption{$\Delta m_{15}$, $\Delta$, and Time Between Points at Maximum Light}
\tablecomments{$\Delta t$ is the time between the closest point before and the closest point after maximum light.  $\Delta$ is the light-curve shape parameter from MLCS2k2 with $R_V=1.7$. 
}
\tablehead{
\colhead{SN} & \colhead{$\Delta m_{15}(U)$} & \colhead{\dm} & \colhead{$\Delta m_{15}(V)$} & \colhead{$\Delta$} & \colhead{$\Delta t (U)$} & \colhead{$\Delta t (B)$} & \colhead{$\Delta t (V)$} 
}
\startdata
80N&      -     & 1.28(0.05)& 0.73(0.05)& -0.0400(0.0780)& 2.02& 2.02& 0.98\\
81B& 1.39(0.18)& 1.10(0.06)& 0.73(0.09)& -0.1360(0.0700)& 0.66& 0.36& 3.50\\
81D& 1.34(0.32)& 1.32(0.18)& 0.86(0.13)& 0.2900(0.2230)& 1.00& 5.99& 5.99\\
86G& 2.05(0.24)& 1.65(0.09)& 1.01(0.09)& 1.2030(0.0640)& 5.09& 0.99& 3.94\\
89B& 1.24(0.10)& 1.02(0.13)& 0.64(0.19)& 0.0030(0.1150)& 0.99& 3.99& 3.99\\
90N&      -     & 1.04(0.13)& 0.62(0.11)& -0.2990(0.0560)&  -   & 7.90& 7.02\\
90af&      -     & 1.63(0.06)& 0.89(0.07)& 0.5070(0.1190)&      -     & 0.99& 1.90\\
91T& 1.37(0.04)& 0.80(0.03)& 0.62(0.05)& -0.3510(0.0360)& 1.00& 1.00& 3.05\\
91bg&      -     & 1.87(0.08)& 1.41(0.07)& 1.4350(0.0480)&      -     & 0.99& 0.15\\
92A& 1.44(0.14)& 1.36(0.03)& 0.83(0.02)& 0.4130(0.0550)& 7.04& 0.96& 1.11\\
92ag&      -     & 1.10(0.09)& 0.60(0.08)& 0.0500(0.0860)&      -     & 0.98& 1.05\\
92al&      -     & 1.10(0.08)& 0.61(0.08)& -0.0640(0.0610)&      -     & 3.78& 3.78\\
92bc&      -     & 0.82(0.08)& 0.61(0.08)& -0.2530(0.0440)&      -     & 3.85& 3.85\\
92bh&      -     &      -     &      -     & -0.1700(0.0860)&      -     & 2.81& 2.81\\
92bo&      -     &      -     &      -     & 0.5800(0.0790)&      -     & 4.02& 2.94\\
92bp&      -     & 1.34(0.15)& 0.58(0.13)& 0.0090(0.0990)&      -     & 6.48& 6.48\\
93H&      -     & 1.76(0.09)& 1.02(0.07)& 0.8740(0.0960)&      -     & 1.11& 0.87\\
93O&      -     & 1.23(0.07)& 0.71(0.08)& -0.0300(0.0720)&      -     & 2.13& 3.61\\
93ag&      -     &      -     &      -     & -0.0190(0.0940)&      -     & 6.67& 6.67\\
94D& 1.71(0.12)& 1.35(0.07)& 0.81(0.05)& 0.3610(0.0490)& 0.81& 0.65& 0.49\\
94S&      -     &      -     & 0.68(0.09)& -0.1730(0.0780)&      -     & 1.97& 1.97\\
94T&      -     & 1.36(0.05)& 0.87(0.06)& 0.7460(0.1060)&      -     & 0.87& 0.87\\
94ae&      -     & 0.98(0.15)& 0.64(0.04)& -0.2360(0.0440)&      -     & 7.96& 0.99\\
95D&      -     & 1.02(0.07)& 0.65(0.07)& -0.2290(0.0480)&      -     & 1.98& 2.02\\
95E&      -     & 1.11(0.11)& 0.61(0.08)& -0.0930(0.0660)&      -     & 2.87& 2.87\\
95ac&      -     & 0.77(0.07)& 0.58(0.08)& -0.3160(0.0520)&      -     & 2.78& 3.83\\
95ak&      -     &     -   & 0.86(0.07)& 0.1300(0.0800)&      -     & 1.01& 0.86\\
95al& 0.89(0.17)& 0.84(0.09)& 0.56(0.06)& -0.2750(0.0490)& 3.97& 3.98& 1.00\\
95bd&      -     & 0.94(0.34)& 0.74(0.25)& -0.3270(0.0490)&      -     & 5.97& 0.98\\
96X& 1.37(0.10)& 1.29(0.06)& 0.81(0.04)& 0.0660(0.0560)& 0.99& 1.09& 1.00\\
96bo&      -     & 1.23(0.06)& 0.70(0.07)& -0.0350(0.0780)&      -     & 1.96& 3.93\\
97E& 1.64(0.11)& 1.41(0.09)& 0.79(0.09)& 0.3120(0.0860)& 1.07& 1.86& 3.94\\
97bp&      -     & 1.16(0.06)& 0.71(0.04)& -0.2850(0.0560)& 1.00& 0.98& 2.05\\
97br& 1.21(0.10)& 1.08(0.17)& 0.65(0.12)& -0.3760(0.0390)& 1.90& 6.79& 5.96\\
97dg&      -     &      -     &      -     & -0.0180(0.0840)& 9.0& 9.0& 9.0\\
98aq& 1.23(0.05)& 1.03(0.03)& 0.66(0.03)& -0.1220(0.0380)& 2.01& 1.03& 0.94\\
98bp& 2.36(0.10)& 1.96(0.08)& 1.12(0.04)& 1.2540(0.0470)& 2.98& 2.98& 1.04\\
98bu& 1.16(0.05)& 1.04(0.03)& 0.75(0.03)& -0.0660(0.0470)& 0.91& 0.82& 0.82\\
98de&      -     & 1.99(0.12)& 1.27(0.06)& 1.5170(0.0420)&      -     & 0.98& 1.02\\
98es&      -     & 0.81(0.07)& 0.59(0.14)& -0.3300(0.0370)& 1.03& 0.89& 7.99\\
99aa&      -     & 0.80(0.06)& 0.58(0.03)& -0.3460(0.0320)& 4.82& 2.02& 0.86\\
99ac&      -     & 1.33(0.06)&      -     & -0.1440(0.0470)& 4.02& 0.98& 1.97\\
99aw&      -     & 0.79(0.08)& 0.62(0.06)& -0.4580(0.0420)&      -     & 4.72& 1.18\\
99by&      -     & 1.98(0.08)& 1.26(0.03)& 1.4650(0.0320)& 0.99& 0.97& 1.00\\
99cc&      -     &      -     &      -     & 0.2880(0.0940)& 4.91& 0.81& 0.98\\
99cl&      -     &      -     &      -     & -0.0160(0.0860)& 1.98& 1.95& 2.99\\
99da&      -     & 1.92(0.11)& 1.15(0.10)& 1.4870(0.0430)&      -     & 1.84& 2.08\\
99dk& 1.93(0.21)& 1.19(0.05)& 0.64(0.05)& -0.3060(0.0530)& 7.89& 1.24& 1.24\\
99dq&      -     & 0.86(0.11)& 0.55(0.08)& -0.3590(0.0340)& 1.01& 0.95& 0.94\\
99ee& 1.23(0.06)& 0.90(0.03)& 0.65(0.03)& -0.2780(0.0370)& 2.94& 0.95& 0.87\\
99ek&      -     & 1.21(0.37)& 0.72(0.28)& 0.0560(0.0780)&      -     & 2.03& 0.85\\
99gp& 1.05(0.08)& 0.80(0.07)& 0.55(0.04)& -0.4150(0.0350)& 1.86& 2.85& 0.91\\
00E&      -     & 0.94(0.25)& 0.66(0.19)& -0.2270(0.0630)&      -     & 4.96& 3.94\\
00cf&      -     & 1.47(0.05)& 0.77(0.04)& -0.0050(0.0790)&      -     & 1.13& 1.13\\
00cn&      -     &      -     &      -     & 0.7240(0.0840)& 5.95& 5.95& 0.92\\
00dk& 1.86(0.08)&      -     &      -     & 0.5110(0.0660)& 0.91& 1.84& 4.91\\
01ba&      -     & 0.95(0.05)& 0.59(0.07)& -0.1640(0.0580)&      -     & 0.98& 2.90\\
01bt&      -     & 1.26(0.07)& 0.70(0.06)& 0.0410(0.0690)&      -     & 2.83& 3.01\\
01cz&      -     & 0.93(0.08)& 0.62(0.06)& -0.1240(0.0610)&      -     & 2.89& 1.97\\
01el& 1.55(0.07)& 1.15(0.07)& 0.63(0.02)& -0.1150(0.0530)& 4.03& 4.02& 1.02\\
01ep&      -     &      -     &      -     & 0.0460(0.0810)& 0.84& 3.97& 3.98\\
01fe&      -     &      -     &      -     & -0.1690(0.0670)& 6.88& 5.88& 4.03\\
01fh&      -     &      -     &      -     & 0.6320(0.1240)& 0.97& 3.01& 2.93\\
01V& 1.01(0.18)& 0.65(0.15)& 0.53(0.13)& -0.3300(0.0430)& 8.70& 7.74& 7.74\\
02bo&      -     &     -     &     -     & -0.1060(0.0640)&  -   &  -   &  -   \\
02cd& 1.06(0.33)& 0.98(0.27)&     -     & -0.3210(0.0560)& 3.95& 3.96&  -   \\
02cr&      -     & 1.26(0.11)& 0.66(0.11)& 0.0090(0.0830)&  -   & 6.83& 6.82\\
02cx&      -     & 1.32(0.14)& 0.79(0.13)& -0.5320(0.0580)&      -     & 5.86& 6.87\\
02de&      -     &      -     &      -     & -0.2240(0.1430)& 2.00& 2.94& 2.94\\
02dj&      -     &      -     &      -     & -0.2000(0.1200)&      -     & 1.96& 7.94\\
02dp&      -     & 1.12(0.06)&      -     & 0.0230(0.1350)& 1.97& 1.97& 6.86\\
02er& 1.87(0.13)& 1.28(0.12)& 0.73(0.10)& 0.2700(0.0780)& 0.99& 1.03& 1.94\\
02fk&      -     &      -     & 0.65(0.10)& -0.0620(0.0560)&      -     & 6.98& 6.02\\
02ha&      -     & 1.34(0.15)& 0.78(0.15)& 0.1270(0.0790)&  -   & 8.84& 8.84\\
02hu& 1.32(0.09)& 1.04(0.07)& 0.53(0.07)& -0.2460(0.0550)& 3.84& 3.85& 3.84\\
03W& 1.35(0.10)& 1.16(0.04)& 0.71(0.14)& -0.0710(0.0590)& 1.04& 0.94& 8.90\\
03cg&      -     &      -     & 0.68(0.05)& 0.0230(0.0740)& 3.08& 5.91& 2.84\\
03du&      -     &      -     &      -     & -0.1680(0.0450)& 6.94& 2.86& 2.86\\
03iv&      -     &      -     &      -     & 0.2780(0.1060)&  -   & 1.97& 4.80\\
03kf& 0.94(0.28)& 0.97(0.23)& 0.70(0.19)& -0.1710(0.0550)& 9.02& 6.10& 6.11\\
04as&      -     &      -     & 0.64(0.06)& -0.1840(0.0700)&      -     &  -   & 2.81\\
05am&      -     & 1.73(0.05)& 0.89(0.05)& 0.4000(0.0910)&      -     & 1.01& 0.95\\
05cf& 1.30(0.09)& 1.06(0.08)& 0.61(0.06)& -0.1470(0.0810)& 3.05& 3.03& 1.00\\
05el& 1.65(0.17)& 1.23(0.11)& 0.79(0.10)& 0.2100(0.0600)& 8.84& 4.85& 4.85\\
05eq& 1.18(0.13)& 0.86(0.11)& 0.49(0.10)& -0.3090(0.0450)& 5.78& 5.79& 5.78\\
05eu&      -     &      -     & 0.67(0.15)& -0.3190(0.0560)&      -     &      -     & 7.89\\
05hc& 1.49(0.15)& 0.97(0.05)& 0.53(0.10)& -0.1250(0.0750)& 1.04& 0.86& 5.69\\
05hk& 1.72(0.06)& 1.47(0.14)& 0.83(0.03)& -0.3100(0.0460)& 0.88& 8.87& 0.91\\
05iq& 1.88(0.31)& 1.05(0.10)& 0.68(0.04)& 0.1370(0.0730)& 5.84& 4.72& 0.89\\
05kc&      -     & 1.24(0.19)& 0.66(0.14)& 0.0360(0.0820)& 2.88& 1.93& 1.93\\
05ke& 1.77(0.06)& 1.66(0.14)& 1.15(0.13)& 1.5510(0.0330)& 0.01& 8.86& 7.86\\
05ki&     -     &     -     &     -     & 0.2850(0.0660)& 8.80& 8.80& 1.02\\
05lz&      -     & 1.35(0.19)& 0.59(0.09)& 0.2170(0.1030)& 3.02& 3.02& 3.02\\
05mc& 1.96(0.36)& 1.87(0.11)& 1.04(0.07)& 0.9350(0.0710)& 1.02& 3.92& 3.91\\
05ms&      -     & 0.79(0.07)& 0.56(0.04)& -0.1590(0.0520)& 3.90& 3.91& 1.83\\
05mz&      -     & 1.96(0.14)& 1.33(0.11)& 1.3640(0.0670)& 1.98& 1.94& 3.97\\
06ac& 1.46(0.08)& 1.08(0.07)& 0.66(0.08)& 0.1610(0.0790)& 1.91& 3.86& 4.97\\
06ar&      -     &      -     &      -     & 0.4280(0.2430)&      -     & 3.93& 2.85\\
06ax& 1.39(0.10)& 1.08(0.05)& 0.63(0.05)& -0.1620(0.0480)& 4.93& 1.95& 1.94\\
06az& 1.52(0.11)& 1.30(0.06)& 0.73(0.05)& 0.1540(0.0560)& 4.83& 2.08& 1.91\\
06br&      -     & 1.47(0.20)& 0.89(0.08)& 0.0450(0.1500)&      -     & 0.89& 0.90\\
06bt&      -     & 1.09(0.06)& 0.54(0.04)& -0.3250(0.0520)&      -     & 1.04& 0.98\\
06bz&      -     & 2.09(0.16)& 1.41(0.06)& 1.5020(0.0820)&      -     & 0.92& 1.08\\
06cc& 1.07(0.15)& 1.01(0.05)& 0.72(0.06)& -0.2260(0.0580)& 3.75& 0.96& 2.81\\
06cm&      -     & 0.99(0.13)& 0.79(0.07)& -0.0520(0.0870)&      -     & 0.94& 1.97\\
06cp&      -     &      -     &      -     & -0.1720(0.0870)&  -   & 4.84& 4.84\\
06D& 1.85(0.08)& 1.35(0.07)& 0.84(0.11)& 0.4230(0.0770)& 3.88& 3.88& 6.96\\
06gj&      -     & 1.39(0.17)& 0.96(0.15)& 0.5820(0.1590)&      -     & 8.71& 8.71\\
06gr&      -     & 0.95(0.13)& 0.57(0.08)& -0.3050(0.0460)& 1.85& 6.91& 3.93\\
06kf&      -     &      -     & 0.77(0.12)& 0.6280(0.0970)&      -     &  -   & 0.87\\
06le& 1.04(0.33)& 0.85(0.27)& 0.59(0.22)& -0.2720(0.0440)& 2.01& 4.02& 4.03\\
06lf&      -     & 1.35(0.62)& 0.71(0.49)& 0.2920(0.0790)&      -     & 6.06& 6.06\\
06mp&      -     &      -     &      -     & -0.1210(0.0580)&      -     & 1.92& 5.87\\
06N&      -     & 1.57(0.07)& 0.90(0.06)& 0.4230(0.0700)&      -     & 1.85& 1.85\\
06nz&      -     &      -     & 1.18(0.14)& 1.1150(0.1180)&      -     & 7.58& 7.57\\
06oa&      -     & 0.98(0.18)& 0.60(0.14)& -0.2520(0.1000)& 0.93& 5.67& 3.77\\
06ob&      -     & 1.70(0.12)& 1.15(0.11)& 0.5410(0.0780)& 1.03& 1.80& 1.94\\
06qo&      -     & 1.02(0.10)& 0.61(0.10)& -0.1820(0.0550)&      -     & 5.74& 5.75\\
06S&      -     & 0.91(0.04)& 0.60(0.05)& -0.1980(0.0530)& 2.01& 0.93& 2.82\\
06sr&      -     & 1.26(0.09)& 0.72(0.08)& 0.1870(0.0860)&      -     & 3.87& 3.88\\
06td&      -     & 1.48(0.12)& 0.76(0.10)& 0.3900(0.1380)&      -     & 0.99& 0.98\\
06X&       -     & 1.10(0.12)& 0.69(0.03)& -0.1020(0.0570)&     -   & 7.04& 1.01\\
07af&      -     & 1.20(0.05)& 0.65(0.03)& -0.0400(0.0520)&      -     & 2.04& 0.94\\
07au&      -     & 1.95(0.11)& 0.94(0.08)& 1.0840(0.0580)&      -     & 5.88& 0.97\\
07bc&      -     & 1.35(0.07)& 0.67(0.04)& 0.2850(0.0900)&      -     & 2.92& 0.91\\
07bd&      -     &      -     &      -     & 0.2900(0.0970)&      -     & 3.93& 3.93\\
07ca&      -     &      -     &      -     & -0.2140(0.0570)&      -     & 2.92& 2.93\\
07ci&      -     &      -     & 0.86(0.03)& 0.8830(0.0780)&      -     &  -  & 0.99\\
07co& 1.28(0.22)& 1.14(0.09)& 0.70(0.07)& -0.0410(0.0780)& 2.69& 1.16& 1.01\\
07cq&      -     & 1.17(0.18)& 0.61(0.07)& 0.0520(0.0710)&      -     & 0.85& 2.00\\
07F&      -     & 0.93(0.07)& 0.58(0.06)& -0.1400(0.0510)&      -     & 3.84& 3.84\\
07qe&      -     & 0.98(0.05)& 0.59(0.06)& -0.2570(0.0490)&      -     & 1.93& 2.94\\
07S&      -     & 0.88(0.08)& 0.62(0.08)& -0.3230(0.0400)&      -     & 4.83& 4.83\\
08bf&      -     & 1.01(0.09)& 0.59(0.08)& -0.1790(0.0640)& 2.06& 4.91& 4.92\\

\enddata
\label{table_lcproperties2}
\end{deluxetable}


\clearpage

\clearpage

\clearpage


\clearpage

\clearpage

\clearpage

\clearpage

\clearpage

\clearpage

\clearpage

\clearpage

\clearpage

\clearpage

\clearpage

\clearpage

\clearpage

\clearpage


\clearpage

\clearpage

\clearpage

\clearpage

\clearpage

\clearpage

\clearpage

\clearpage

\clearpage

\clearpage

\end{document}